\documentclass[preprints,article,accept,moreauthors,pdftex]{Definitions/mdpi}
  
\usepackage{float}
\usepackage{bbm}
\firstpage{1} 
\pubvolume{1}
\issuenum{1}
\articlenumber{0}
\pubyear{2021}
\copyrightyear{2020}
\datereceived{} 
\dateaccepted{} 
\datepublished{}

\DeclareMathOperator{\Tr}{Tr}

\newcommand{\Mcal}{\mathcal{M}}

\newcommand{\vect}[1]{\boldsymbol{#1}_{\perp}}
\newcommand{\kt}{\vect{k}}
\newcommand{\pt}{\vect{p}}
\newcommand{\Pt}{\vect{P}}

\newcommand{\bt}{\vect{b}}

\newcommand{\xt}{\vect{x}}
\newcommand{\yt}{\vect{y}}
\newcommand{\zt}{\vect{z}}

\newcommand{\rt}{\vect{r}}

\newcommand{\der}{\mathrm{d}}

\hreflink{https://doi.org/} 

\Title{Mining for Gluon Saturation at Colliders}

\TitleCitation{Title}

\Author{Astrid Morreale $^{1, \ddagger}$\orcidA{}, Farid Salazar $^{2,3,4, \ddagger}$ }

\AuthorNames{Astrid Morreale, Farid Salazar }

\AuthorCitation{Morreale, A.; Salazar, F.}

\address{%
$^{1}$ \quad Los Alamos National Laboratory ; astrid@lanl.gov\\
$^{2}$ \quad Department of Physics and Astronomy, Stony Brook University, Stony Brook, NY 11794, USA\\
$^{3}$ \quad Physics Department, Brookhaven National Laboratory,  Bldg. 510A, Upton, NY 11973, USA \\
$^{4}$ \quad Center for Frontiers in Nuclear Science (CFNS), Stony Brook University, Stony Brook, NY 11794, USA \\
\quad farid.salazarwong@stonybrook.edu}

\secondnote{These authors contributed equally to this work.}

\abstract{}


\begin{document}


\section{Introduction}\label{Sec:intro}

Quantum chromodynamics (QCD) is the  theory of strong interactions of quarks and gluons collectively called partons, the basic constituents of all nuclear matter. Its non-abelian character manifests in nature in the form of two remarkable properties: color confinement and asymptotic freedom \cite{Gross:1973id,Politzer:1973fx, Hanada:2019czd}. Confinement forbids the existence of quarks and gluons in free form or separated by long distances, allowing only colorless bound states such as  mesons and baryons.  Asymptotic freedom on the other hand states that at short distances, quarks and gluons interact weakly due to smallness of the coupling constant $\alpha_{s}$ at asymptotically high energies. The latter property formed the theoretical basis behind the parton model \cite{Bjorken:1968dy,Bjorken:1969ja,Feynman:1969ej} allowing the use of perturbation theory while leading to the successful description of a plethora of experimental results from  fixed target experiments to high energy colliders. A multi-decade endeavor to quantify the structure of hadrons was launched with the introduction of the  universal parton distribution functions (PDFs) which obey the DGLAP (Dokshitzer-GribovLipatov-Altarelli-Parisi)  renormalization group equation~\cite{Gribov:1972ri,Lipatov:1974qm,Altarelli:1977zs,Dokshitzer:1977sg}. These PDF's describe the parton densities as functions of the longitudinal momentum fraction $x$. Beyond this one-dimensional picture, transverse momentum dependent (TMD) PDFs have been introduced to characterize the three-momenta of partons inside hadrons \cite{Collins:1981uw,Mulders:2000sh,Meissner:2007rx}. More recently, the complementary generalized parton distributions (GPDs) have been defined by furnishing PDFs with the two dimensional transverse spatial distribution of partons resulting in a tomographic picture of hadrons and nuclei \cite{Ji:1996ek,Radyushkin:1997ki,Mueller:1998fv}.

At high energies (or small $x$), gluon densities grow quickly resulting in a large parton occupation number. This growth is expected to be controlled by non-linear QCD effects at sufficiently small-$x$ \cite{Gribov:1984tu,Mueller:1985wy}. An appropriate description of the fundamental degrees of freedom of hadrons and nuclei in this regime is in terms of classical strong gluon fields, replacing the usual partonic description. A description of this  high density regime is given by the Color Glass Condensate (CGC), a semi-classical effective field theory (EFT) for small-$x$ gluons~\cite{McLerran:1993ni,McLerran:1993ka,McLerran:1994vd,Ayala:1995kg,Ayala:1995hx}.

High energy nuclear and particle physics experiments have spent the past decades quantifying the structure of protons and nuclei in terms of their fundamental constituents confirming extraordinary behaviour  of matter at extreme density and pressure conditions. In the process they have also measured seemingly unexpected phenomena which will need a new generation of theoretical efforts as well as pertinent collider experiments. These last decades have also resulted on  number of gluon saturation theoretical reviews which have paved the foundation for the current document~\cite{Iancu:2003xm,Gelis:2010nm,Kovchegov:2012mbw,Albacete:2014fwa,Blaizot:2016qgz}.

This paper is organized as follows, having given a brief introduction we then proceed to Sec.~\ref{Sec:CGC} which describes the principal underlying theoretical tools for the description of a gluon saturated state. Sec.~\ref{Sec:experiment} gives a selected overview of experimental signatures to date while Sec.~\ref{Sec:EIC} discusses future facilities pertinent for the discovery and quantification of gluon saturation. Sec~\ref{Sec:conclusions} gives an outlook for our field followed by acknowledgements (Sec.~\ref{Sec:Ack}) to the people and institutions that made this manuscript possible.

\section{Color Glass Condensate effective field theory}\label{Sec:CGC} 
This section will serve as a description to the underlying principles behind the Color Glass Condensate (CGC) as the effective field theory for saturated gluons. We will  discuss the separation of degrees of freedom into sources and fields and the computation of observables in the CGC in Sec.\,\ref{sec:sources_vs_fields}. Light-like Wilson lines and their correlators in the context of high energy scattering will follow in Sec.\,\ref{SSec:wilson} which will include a discussion on the dipole correlator and the saturation scale. Sec.\,\ref{sec:DIS_to_pA} will provide some examples of high energy processes in the CGC and discuss basic features of saturation. We briefly discuss the elements of small-$x$ quantum evolution in  Sec.\,\ref{sec:quantum_evolution_CGC}, where we introduce the Balitsky-Kovchegov equation and the JIMWLK evolution for the evolution of the sources. We note that a detailed exposition of the subject can be found in the reviews~\cite{Iancu:2003xm,Gelis:2010nm} as well as the  text book~\cite{Kovchegov:2012mbw}. 

\subsection{Separation of degrees of freedom: Sources and fields}
\label{sec:sources_vs_fields}

The CGC is an effective field theory for high energy QCD \cite{McLerran:1993ni,McLerran:1993ka,McLerran:1994vd,Ayala:1995kg,Ayala:1995hx}. For a hadron moving in the plus light-cone direction with large momentum $P^+$ probed at the scale $x_0 P^+$, with $x_0 \ll 1$, the CGC separates the partonic content of hadrons according to their longitudinal momentum $k^+ =x P^+$ where
$x$ refers to the longitudinal momentum fraction of the parton probed in the nucleus/nucleon.
Partons carrying large longitudinal momentum fraction $x \gtrsim x_0$ (large-$x$ partons) are treated as static and localized color sources $\rho$. Heisenberg's uncertainty principle justifies this view: the degree of localization of partons $\Delta z^-$ is much smaller than the longitudinal resolution $1/(x_0 P^+)$ of the probe:
\begin{align}
    \Delta z^- \sim \frac{1}{k^+} = \frac{1}{x P^+} \ll \frac{1}{x_0 P^+} \,.
\end{align}
Similarly, the time scale $\Delta z^+$ for the evolution of these large-$x$ partons is much larger than the time scale of the probe $\tau \sim \frac{2 x_0 P^+}{k_\perp^2}$, where $\kt$ is the transverse momentum of the produced quark:
\begin{align}
    \Delta z^+ \sim \frac{1}{k^-} = \frac{2k^+}{k_\perp^2} = \frac{2 x P^+}{k_\perp^2} \gg \frac{2 x_0 P^+}{k_\perp^2} \,.
\end{align}
From the point of view of a probe, large-$x$ partons are localized in the longitudinal direction $z^-$ and frozen in light-cone time $z^+$. Their color charge distribution is non-perturbative and will be dictated by a gauge invariant weight functional $W_{x_0}[\rho]$ as will be discussed hereafter.

For a hadron/nucleus moving close to the light-cone in the plus direction, these sources generate a current independent of the light-cone time $z^+$:
\begin{align}
    J^{\mu,a}(z) = \delta^{\mu+} \rho^a(z^-,\zt) \,,
    \label{eq:current_static sources}
\end{align}
where the support of $\rho$ along the minus light-cone direction is small.

The partons possessing a small momentum fraction $x \lesssim x_0 $ are treated as a delocalized dynamical field $A^{\mu,a}(z)$ (small-$x$ partons). This classical treatment of $A^{\mu,a}_{\rm cl}(z)$ is justified by noting that the occupation number of small-$x$ partons is large $\langle A_{\rm cl} A_{\rm cl} \rangle \sim 1/\alpha_s$. 

Sources and fields are related by the Yang-Mills equations $[D_\mu,F^{\mu\nu}] = J^{\nu}$, where $D_\mu = \partial_\mu - ig A_{\mu}$. The independence of $z^+$ of the current in Eq.\,\eqref{eq:current_static sources} is consistent with the conservation equation $\left[ D_\mu, J^\mu \right] = 0$ when working in an appropriate gauge ($A^{-}=0$). For this choice of gauge condition, the classical gauge field adopts a simple solution:
\begin{align}
    A^{\mu,a}_{\rm cl}(z) = \delta^{
    \mu+} \alpha^a(z^-,\zt) \,, \label{eq:A_classical_sol}
\end{align}
where $\alpha^a(z^-,\zt)$ satisfies the two-dimensional Poisson equation $\nabla_\perp^2 \alpha^a = - \rho^a$.

\begin{figure}[H]
    \centering
    \includegraphics[scale=0.33]{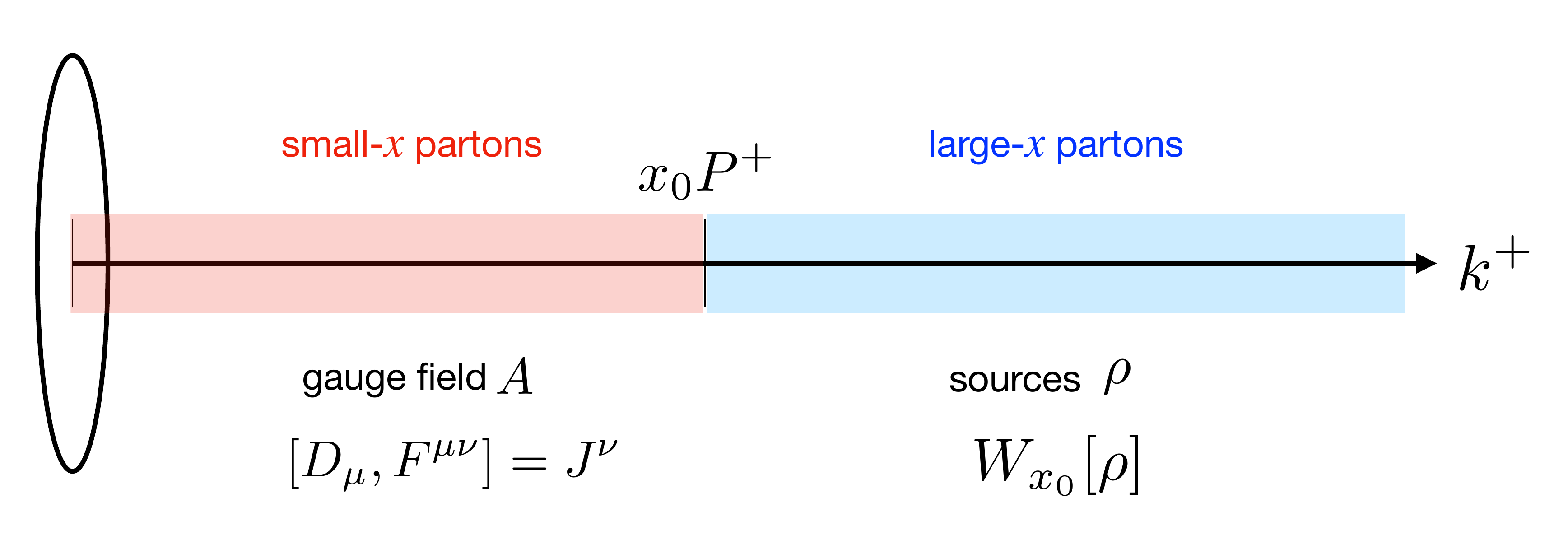}
    \caption{In the CGC EFT partons are organized as color sources or fields according to their longitudinal momentum fraction $x$  relative to the characteristic momentum fraction of the probe $x_0$. Sources are stochastic and their distribution is characterized by a gauge invariant weight functional $W_{x_0}[\rho]$ (represented in blue). The gauge field is a solution to Yang-Mills equations in the presence of the sources (represented in red).}
    \label{fig:CGC_seperation_dofs}
\end{figure}
As a consequence of the separation of degrees of freedom described above (see Fig.\,\ref{fig:CGC_seperation_dofs}), the expectation value of any observable $\mathcal{O}$ is computed in the CGC in a two step process:
\begin{enumerate}
    \item Compute the quantum expectation value/path integral $\mathcal{O}[\rho] = \langle \mathcal{O} \rangle_{\rho}$ in the presence of sources $\rho$ drawn from $W_{x_0}[\rho]$.
    
    \item Average over all possible configurations given by an appropriate gauge invariant weight functional $W_{x_0}[\rho]$.
\end{enumerate}
This procedure is summarized in the following expression:
\begin{align}
    \left \langle  \mathcal{O} \right \rangle_{x_0} = \int \left[\mathcal{D} \rho \right] W_{x_0}[\rho]  \mathcal{O} [\rho] \,.
    \label{eq:double_average_CGC}
\end{align}
For observables that involve longitudinal momentum fraction $x$ close to $x_0$, the path integral $\langle \mathcal{O} \rangle_{\rho}$ is dominated  by the classical solution in Eq.\,\eqref{eq:A_classical_sol}. When the observable is probed at significantly smaller values of $x \ll x_0$ one must account for quantum evolution. We will return to this point in Sec.\,\ref{sec:quantum_evolution_CGC}.

The most widely used choice for the weight function is the McLerran-Venugopalan (MV) model \cite{McLerran:1993ni,McLerran:1993ka}. For a sufficiently large nucleus, the MV model invokes the central limit theorem, thus constructing a distribution following Gaussian statistics (for a detailed exposition see \cite{Jeon:2004rk}):
\begin{align}
    W_{x_0}[\rho] = \mathcal{N} \exp \left\{-\frac{1}{2} \int \mathrm{d} x^- \mathrm{d}^2 \xt \frac{\rho^a(x^-,\xt) \rho^a(x^-,\xt)}{\lambda_{x_0}(x^-)} \right\}\,.
\end{align}
The function $\lambda_{x_0}(x^-)$ is related to the transverse color charge density distribution inside the nucleus. An energetic probe will interact coherently with the partons encountered along its longitudinal trajectory. Considering the contribution from the valence quarks only one finds the quantity
\begin{align}
    \mu^2 = \int \der x^- \lambda_{x_0}(x^-) = \frac{2\pi g^2 A}{R_A^2} \sim A^{1/3} \label{eq:mu2_MVmodel} \,,
\end{align}
where $A$ is the nuclear mass number, $R_A \sim A^{1/3}$ is the nuclear radius and $g$ is the strong coupling constant. This new quantity $\mu^2$ is closely related to the saturation scale $Q_s^2$ as we will see in the next section where we introduce the high energy correlators.

\subsection{High energy scattering: light-like Wilson lines and correlators}\label{SSec:wilson}
The interaction of a highly energetic color charged particle with large $k^-$ momentum, and small $k^+ = k_\perp^2/(2k^-)$, with the classical field $A_{\rm cl}$ created by a nucleus is more easily described in mixed space $(k^-,\xt)$, where $\xt$ is conjugate to $\kt$. In the eikonal approximation the scattering rotates the color state of the particle while keeping the longitudinal momentum $k^-$, transverse coordinates $\xt$ and any additional quantum numbers (e.g. polarization or helicities) unchanged. The effect of the rotation is encoded in the light-like Wilson lines which for quark and gluon read
\begin{align}
    V_{ij}(\xt) = \mathcal{P}\left( ig \int_{-\infty}^\infty A_{\rm cl}^{+,c}(z^- ,\xt) t^c_{ij} \ dz^- \right)\,, \\
    U_{ab}(\xt) = \mathcal{P}\left( ig \int_{-\infty}^\infty A_{\rm cl}^{+,c}(z^- ,\xt) T^c_{ab} \ dz^- \right) \,,
\end{align}
respectively, where $t^c$ and $T^c$ are generators of $SU(3)$ in the fundamental and adjoint representations respectively. Here $\mathcal{P}$ denotes path ordered exponential, $(i,j)$ and $(a,b)$ are color indices, and $\xt$ is the transverse location at which the color charged particle interacts with the background-field $A_{\rm cl}$.

\begin{figure}[H]
    \centering
    \includegraphics[scale=0.3]{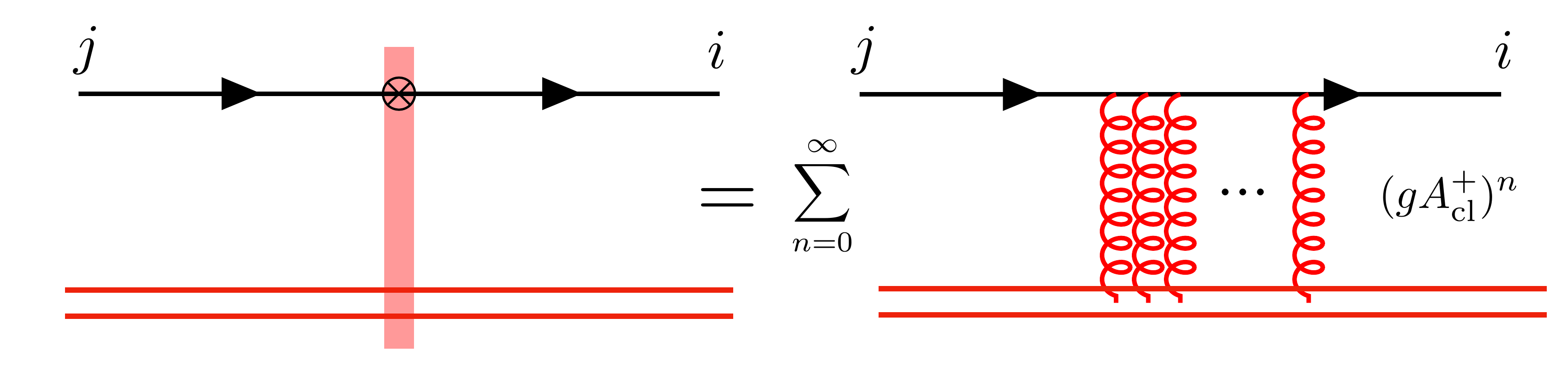}
    \caption{The interaction of a quark with the background field of the nucleus is encoded in a light-like Wilson line which resums multiple eikonal scatterings.}
\end{figure}

Light-like Wilson lines resum multiple interactions  $(gA_{\rm cl}^{+})^n$  with the background field. These are the fundamental degrees of freedom in high energy QCD scattering. Production cross-sections are expressed as convolutions of Wilson line correlators with perturbatively calculable and process-dependent impact factors, as we will see in Sec.\,\ref{sec:DIS_to_pA}.

The simplest and most important of such correlators is the two-point correlator or dipole:
\begin{align}
    S^{(2)}_{x_0}(\xt,\yt) = \frac{1}{N_c} \left \langle \mathrm{Tr} \left[ V(\xt) V^\dagger(\yt)\right] \right \rangle_{x_0} \,,
\end{align}
where $N_c=3$ is the number of colors. This correlator represents the scattering amplitude of a quark anti-quark dipole interacting with the background field of a nucleus at transverse locations $\xt$ and $\yt$. It is the building block of many processes in high energy QCD such as the total Deep Inelastic Scattering (DIS) cross-section at small-$x$.

In the MV model the dipole correlator only depends on the dipole separation $r_\perp = |\xt-\yt|$ and takes the form:
\begin{align}
    S^{(2)}_{x_0}(r_\perp) = \exp \left\{ -\frac{1}{4} \alpha_s C_F \mu^2 r_\perp^2  \log\left(\frac{1}{\Lambda r_\perp }+e\right) \right\} \,, \label{eq:dipole_MVmodel}
\end{align}
where $\mu^2$ was introduced in Eq.\eqref{eq:mu2_MVmodel}. $\alpha_s$ is the fine structure constant, $C_F = (N_c^2-1)/(2N_c)$ is the Casimir in the fundamental representation, and $\Lambda$ is an infrared cut-off.

For small separations $r_\perp $, the dipole correlator behaves as a color neutral object, and thus the scattering amplitude is close to unity (i.e. the scattering matrix $\mathcal{S} \approx \mathbbm{1}$, no scattering), this is known as color transparency. Mathematically, this follows from the unitarity of the Wilson lines. On the other hand, at large distances $r_\perp$ the dipole correlator vanishes as the Wilson lines decorrelate as expected from the black-disk limit.

The transition between these two regimes is delineated by defining the saturation scale $Q_s^2$ as
\begin{align}
    Q_s^2 = \frac{2}{r_{\perp,s}^2}\,, \quad \mathrm{where} \quad  S^{(2)}_{x_0}(r_{\perp,s}) = \exp(-c) \,, \label{eq:sat_scale}
\end{align}
where the constant $c$ is typically chosen to be $1/2$.
The inverse of the saturation scale provides a measure of the correlation length of the Wilson line pair. 

\begin{figure}[H]
    \centering
    \includegraphics[scale=0.25]{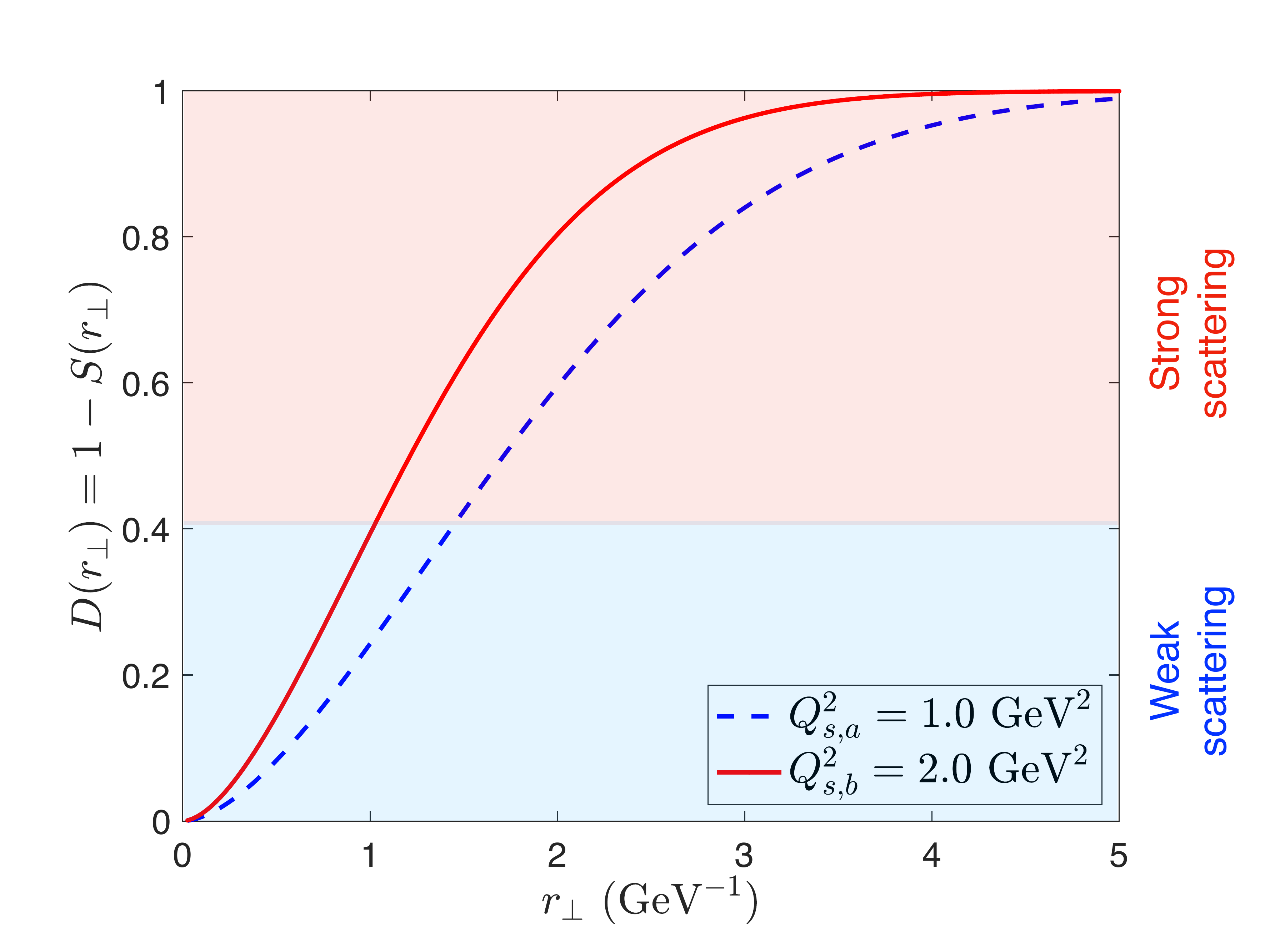}
    \caption{Dipole amplitude $D(r_\perp) = 1- S^{(2)}(r_\perp)$ in the MV model (see Eq.\,\eqref{eq:dipole_MVmodel}) with two different values of the saturation scale $Q_s^2$ defined in Eq.\,\eqref{eq:sat_scale} displaying the transition between weak and strong scattering regimes.}
    \label{fig:dipole_amplitude}
\end{figure}

By examining Eq.\,\eqref{eq:dipole_MVmodel}, it follows that the saturation scale is proportional to the color charge density $Q_s^2 \sim \mu^2 \sim A^{1/3}$; hence growing with larger nuclei, this is also refered to as the nuclear \emph{oomph} factor. In Sec.\,\ref{sec:quantum_evolution_CGC} we will argue that the saturation scale also grows with decreasing values of $x$ (or equivalently with increasing energies). This results in the relation
\begin{align}
    Q_s^2 \sim \frac{A^{1/3}}{x^{\lambda}}\,,
\end{align}
where $\lambda \approx 0.3$ and its arises from  estimates of the energy dependence of the saturation momentum from DIS and nucleus-nucleus (A-A)  scattering experiments~\cite{Gotsman:2015gba}.
In Fig.\,\ref{fig:dipole_amplitude} we plot the dipole amplitude $D(r_\perp) = 1- S(r_\perp)$ with two different values of the saturation scale, which can be interpreted as examining different nuclei species, or a nucleus at two different energies~\footnote{This is an oversimplified view point, as the small-$x$ evolution will not only change the value of $Q_s$ but also the functional form of the dipole. In the most general case, the saturation scale will also depend on the impact parameter $b_\perp$ as more color charge densities are expected in the center of the nucleus than in its periphery, modulo fluctuations.}. As expected the larger saturation scale leads to a more rapid transition to the strong scattering regime, where eventually the dipole amplitude approaches unity.

More complex correlators of light-like Wilson lines appear in less inclusive processes and in the small-$x$ evolution equations. Two notable examples are the double dipole correlator:
\begin{align}
    S^{(2,2)}_{x_0}(\xt,\yt;\yt',\xt') = \frac{1}{N_c^2} \left \langle \mathrm{Tr}  \left[V(\xt)V^\dagger(\yt) \right]\mathrm{Tr} \left[V(\yt')V^\dagger(\xt')\right] \right \rangle_{x_0}\,,
\end{align}
and the quadrupole correlator
\begin{align}
    S^{(4)}_{x_0}(\xt,\yt;\yt',\xt') = \frac{1}{N_c} \left \langle \mathrm{Tr}  \left[V(\xt)V^\dagger(\yt)V(\yt')V^\dagger(\xt')\right] \right \rangle_{x_0} \,.
\end{align}
As in the dipole case, these correlators implicitly contain the saturation scale $Q_s^2$. It is noted that this could be explicitly realized in the MV model, where the Gaussian approximation allows expressing any $n-$point Wilson line correlator as a non-linear function of the dipole. Other correlators involving Wilson lines in the adjoint representation $U(\zt)$ appear in the scattering/production of gluons. 

In the next section we will see the manifestations of these correlators of light-like Wilson lines in concrete high energy processes in QCD.

\subsection{From DIS to proton-nucleus (pA) collisions} 
\label{sec:DIS_to_pA}

As discussed in the previous section, high energy scattering processes in the CGC are expressed in terms of correlators of light-like Wilson lines with impact factors that can be systematically computed in perturbation theory. In this section we provide a few examples. 

\begin{figure}[H]
    \centering
    \includegraphics[scale=0.3]{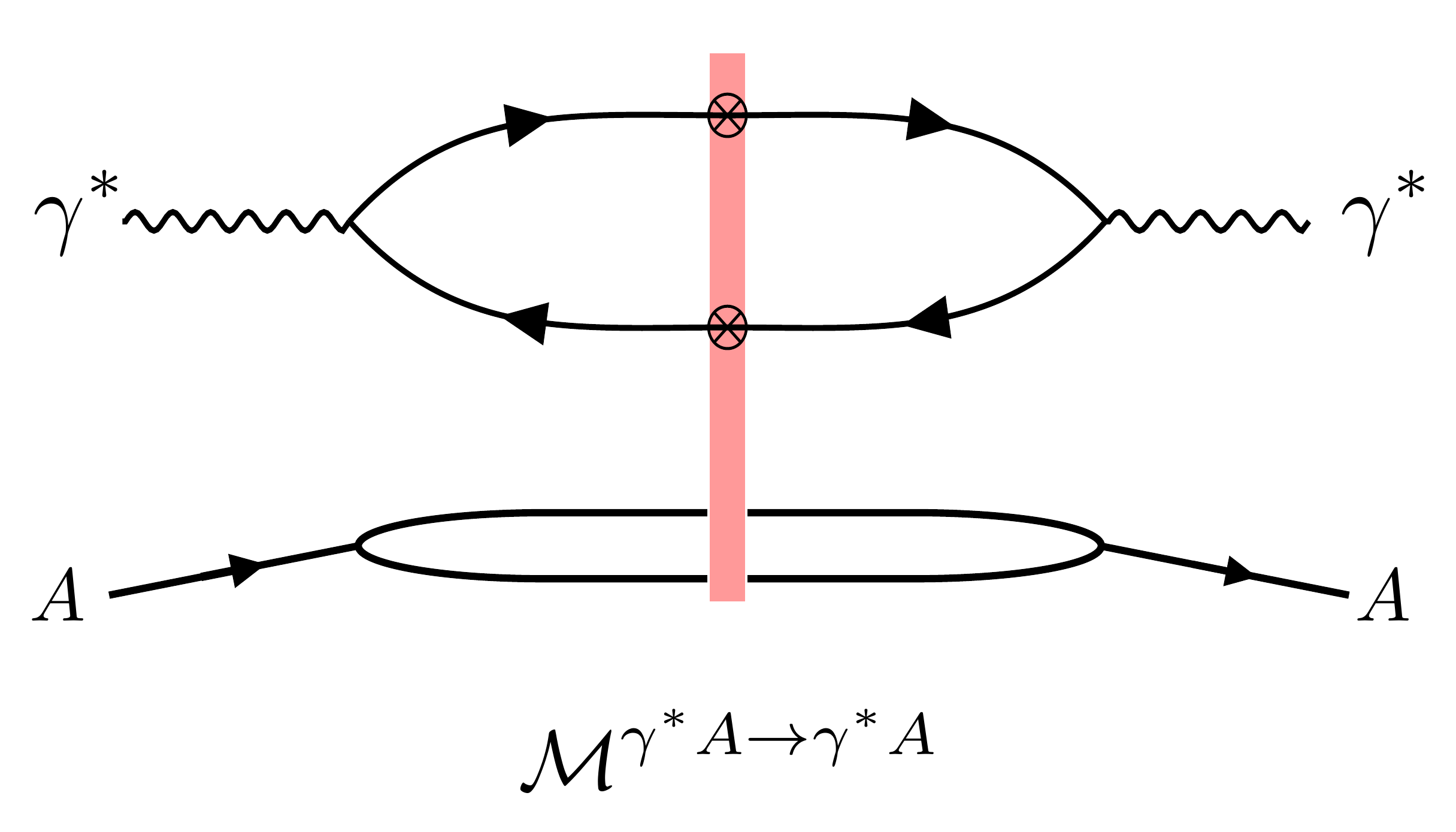}
    \caption{Feynman diagram for the forward scattering amplitude $\Mcal^{\gamma^* A \to \gamma^*A}$ of virtual photon-nucleus collision. The amplitude contains two light-like Wilson lines, which appear from the interaction of the quark anti-quark pair with the nucleus. This amplitude is related to the total DIS cross-section by virtue of the optical theorem $\sigma^{\gamma^*A} = 2 \mathrm{Im} (\Mcal^{\gamma^* A \to \gamma^*A})$. In the high energy limit, the forward amplitude is purely imaginary.}
    \label{fig:forward_scat_amplitude_DIS}
\end{figure}

The total DIS cross-section, at small-$x$, for a virtual photon scattering off a nucleus (see Fig.\,\ref{fig:forward_scat_amplitude_DIS}) can be expressed with the help of the optical theorem as \cite{Gelis:2002nn}:
    \begin{align}
    \sigma^{\gamma^*A}_{\lambda}(x,Q^2) = 2 \int \der^2 \rt \der^2 \bt \int_0^1 \der z \ \left |  \Psi^{\gamma^*}_{\lambda}(\rt,Q^2,z) \right |^2 \left[1 - S^{(2)}_{x}\left(\bt+\frac{\rt}{2}, \bt-\frac{\rt}{2}\right) \right] \,,
\end{align}
where $Q^2=-q^2$ and $\lambda$ are the virtuality and polarization of the photon respectively. Here $\Psi^{\gamma^*}_{\lambda}(\rt,Q^2,z)$ is the light-cone wave-function of the splitting of the virtual photon into a quark anti-quark pair which only depends on the dipole separation $\rt=\xt-\yt$ and it can be calculated from perturbation theory. The longitudinal momentum fraction of the quark relative to that of the photon is denoted as $z$ and that of the anti-quark is $1-z$ by momentum conservation. The dipole correlator arises from the interaction of the quark and anti-quark with the background field of the nucleus. In addition to the dependence on the dipole vector $\rt$, the dipole correlator can generally depend on the impact parameter vector defined as $\bt = \frac{1}{2}(\xt +\yt)$. 

The longitudinal momentum fraction $x$ is given by Bjorken $x=Q^2/W^2$, where $W$ is the center of mass energy per nucleon of the virtual photon-nucleus system.

In order to access the saturated regime one has to probe dipole sizes $r_\perp \sim 1/Q_s$ (see Fig.\,\ref{fig:dipole_amplitude}). The light-cone wave-functions $\Psi^{\gamma^*}_\lambda$ rapidly suppress dipoles with sizes $r^2_\perp \gtrsim 1/Q^2$ (more precisely $r^2_\perp \gtrsim 1/ \left[z(1-z)Q^2 \right]$). These two observations imply that saturation effects in DIS at small-$x$ are more visible at lower values of photon virtuality $\Lambda_{QCD}^2 \ll Q^2 \lesssim Q_s^2$. At high virtualities one probes the weak scattering regime where gluon saturation has not yet set in.

\begin{figure}[H]
    \centering
    \includegraphics[scale=0.3]{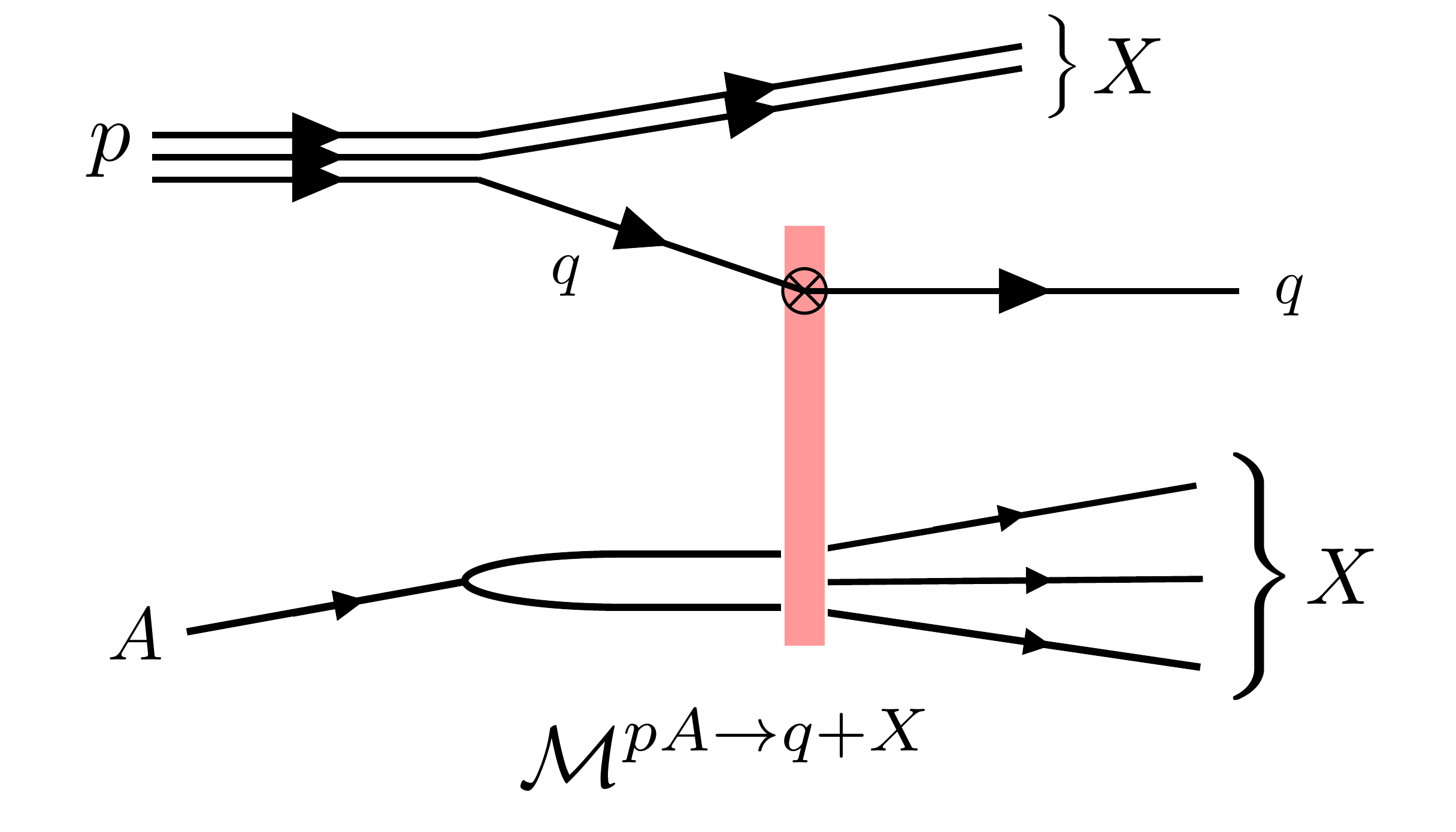}
    \caption{Feynman diagram for the amplitude $\Mcal^{pA \to q + X}$ for quark production in proton-nucleus collisions . A light-like Wilson line appears in the amplitude; thus the cross-section will feature a dipole correlator.}
\label{fig:amplitude_pA}\end{figure}

Another process which features the dipole correlator is the forward quark production in proton-nucleus scattering (Fig.~\ref{fig:amplitude_pA}), which can be studied via forward jet or hadron production. Within the \emph{hybrid factorization} approach, the differential cross-section reads \cite{Gelis:2002nn}
\begin{align}
    \frac{\der \sigma^{pA \to q X}}{\der \eta \der^2 \kt} = \frac{1}{(2\pi)^2} x_p q(x_p) C_{x_A}(\kt) \,,
\end{align}
where $\kt$ and $\eta$ are the transverse momentum and rapidity of the produced quark, $x_p q(x_p)$ is
the quark distribution in the proton for a collinear quark with longitudinal momentum
fraction $x_p$. In this case the dipole correlator appears from a light-like Wilson line $V(\xt)$ in the amplitude and another one $V^\dagger(\yt)$ in the complex conjugate amplitude. Here $x_A$ refers to the longitudinal momentum fraction of the gluon probed in the dense nucleus.

\begin{figure}[H]
    \centering
    \includegraphics[scale=0.4]{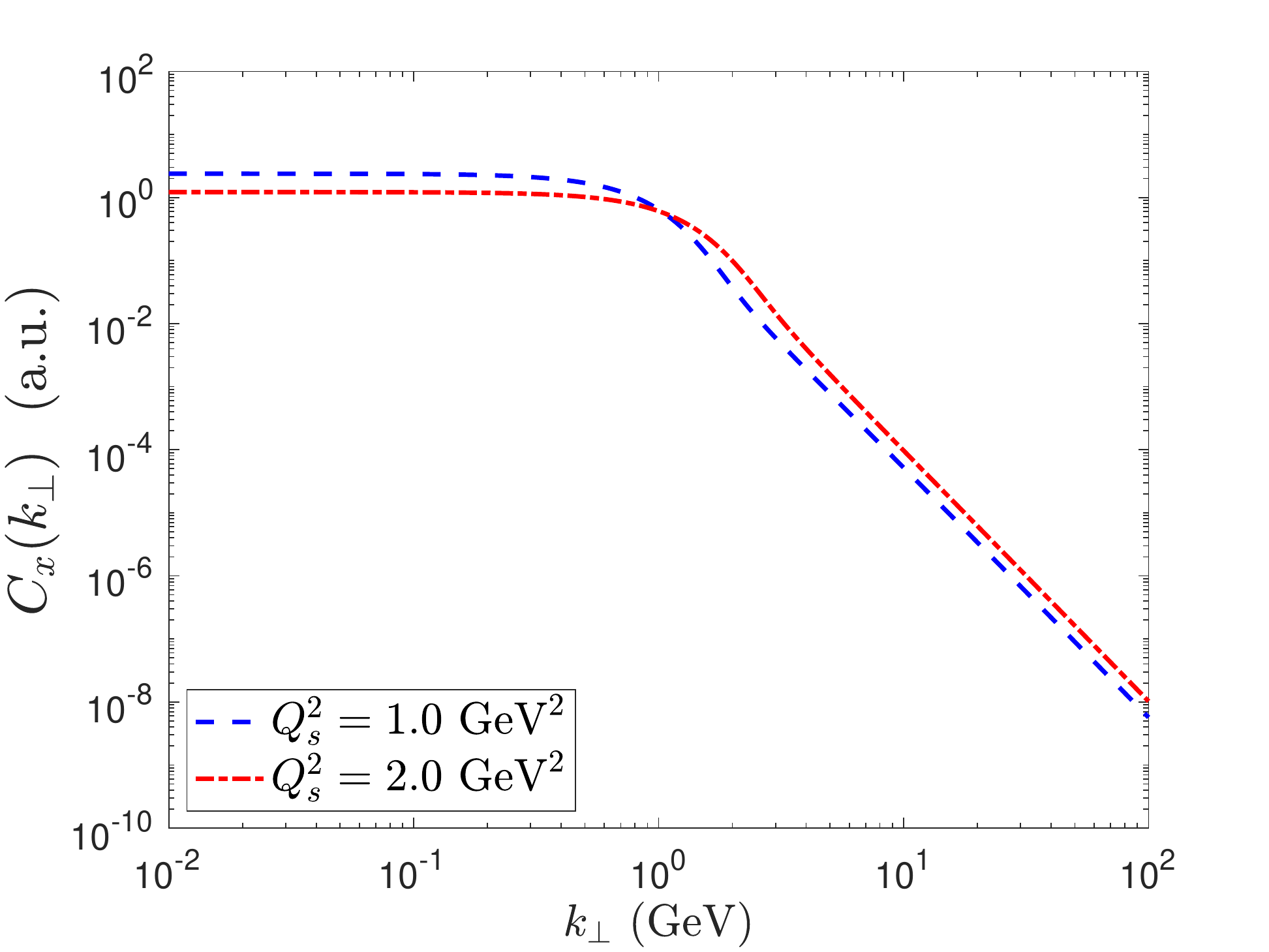}
    \caption{Fourier transform of the dipole correlator as a function of $k_\perp$ for two different values of the saturation scale. A transition between saturation and perturbative regime is observed near $k_\perp \sim Q_s$.   The $x$ dependence of the distribution is effectively accounted by the saturation scale. In a more careful treatment, the functional shape of the distribution also depends on $x$. }    \label{fig:dipole_FT_MV}
\end{figure}

The function $C_{x_A}(\kt)$ is the Fourier transform of the dipole amplitude:
\begin{align}
    C_{x_A}(\kt) = \int \der^2 \xt \der^2 \yt e^{-i \kt \cdot (\xt-\yt)} S^{(2)}_{x_A}(\xt,\yt) \,
\end{align}
This function determines the transverse momentum kick acquired by a collinear quark as it multiple-scatters from the nucleus. In Fig.\,\ref{fig:dipole_FT_MV} we plot the function $C(k_\perp)$ corresponding to the dipoles shown in Fig.\,\ref{fig:dipole_amplitude}, where we normalized by an overall factor of the transverse area as the MV model is translationally invariant. We observe a clear difference in the behavior of $C(k_\perp)$ between the small and large $k_\perp$ regions. In the perturbative limit it behaves as a power law
\begin{align}
    C_x(\kt) \sim \frac{Q_s^2(x)}{k_\perp^4}, \quad k_\perp \gtrsim Q_s \,
\end{align}
whereas in the saturated regime it approaches a constant 
\begin{align}
    C_x(\kt) \sim \frac{1}{Q_s^2(x)}, \quad k_\perp \lesssim Q_s \,
\end{align}
In Fig.\,\ref{fig:dipole_amplitude} we see  that as the saturation scale is increased, the distribution $C_x(\kt)$ is pushed to larger values of $k_\perp$.  This is one of the consequences of saturation: as the energy of the collision is increased, the saturation scale $Q_s$ grows and radiated gluons at small-$x$ are pushed to larger values of $k_\perp$ since the phase space $k_\perp \lesssim Q_s$ is overoccupied. 

The function $k_\perp C_x(k_\perp)$, where the additional factor of $k_\perp$ arises from the phase space, determines the transverse momentum acquired by the quark as it multiply scatters from the nucleus. It can be verified that this function peaks around $k_\perp \sim Q_s$. While it is possible to parametrize $C_x(k_\perp)$ as a function of both $k_\perp$ and $x$; however, in the CGC usually its Fourier conjugate, the dipole correlator is constrained by HERA data.

Appearing both in DIS and proton-nucleus collisions, the dipole correlator is a universal building block in high energy collisions. Evidently, its manifestation is different for each process, thus one can constrain different features of this object from independent measurements either by studying the $(x,Q^2)$ dependence in DIS or the $(\eta,\kt)$ distribution of quark jet production in proton-nucleus collisions.

\begin{figure}[H]
    \centering
    \includegraphics[scale=0.3]{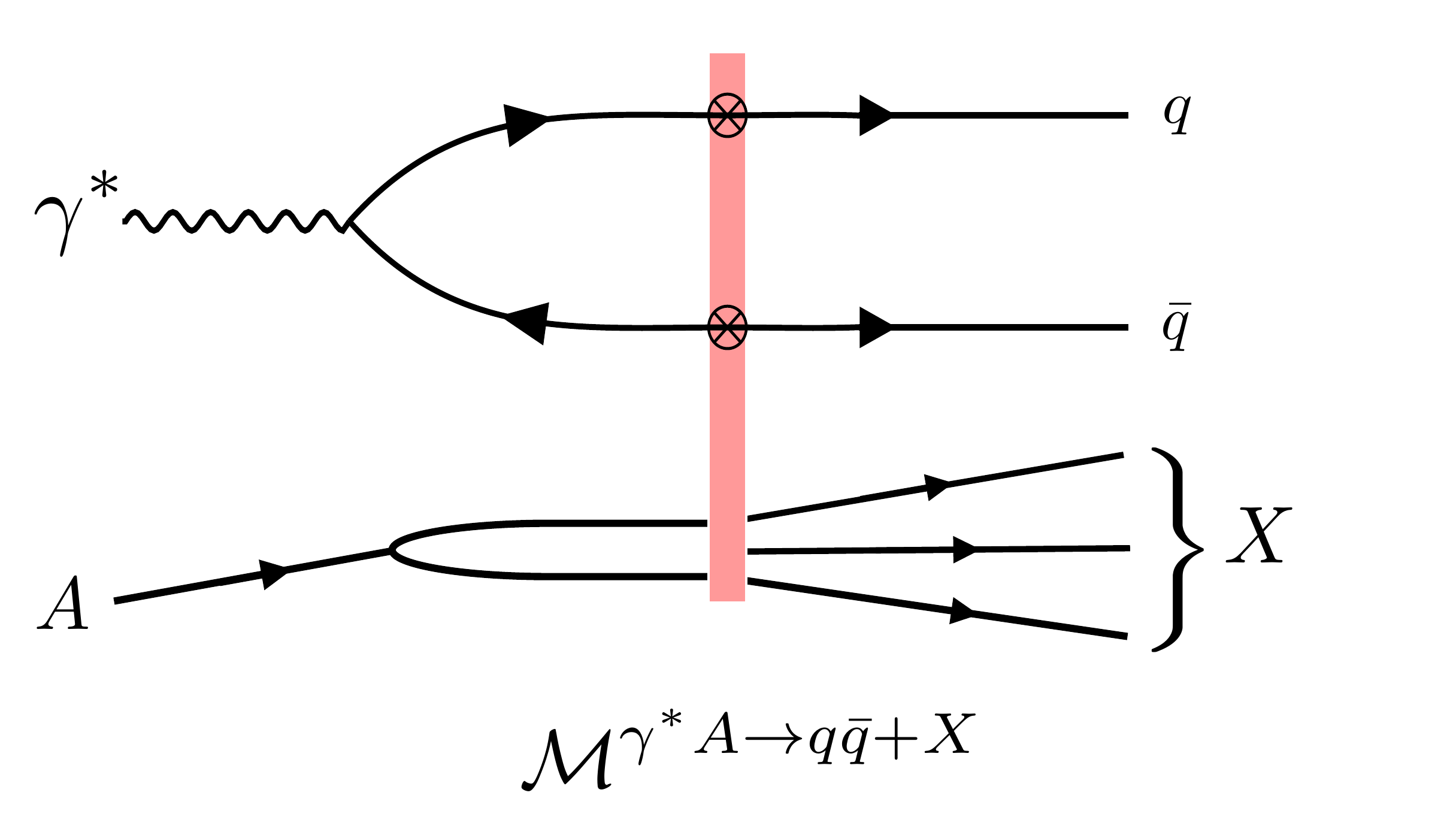}
    \caption{Feynman diagram for the amplitude $\Mcal^{\gamma^*A \to q\bar{q} + X}$ for the quark anti-quark dijet production in virtual photon-nucleus collisions. The amplitude contains the production of two light-like Wilson lines; thus the amplitude will feature a quadrupole (and dipole) correlator. }
    \label{fig:dijet_DIS_CGC_LO}
\end{figure}

We provide one more example of a high energy process, the production of a dijet or a dihadron pair in DIS as shown in Fig.\,\ref{fig:dijet_DIS_CGC_LO}. In the CGC for general small-$x$ kinematics, this process depends on the quadrupole correlator in a non-trivial way. However, in the limit where the dijets or dihadrons are back-to-back in transverse space, it is possible to establish a TMD factorization \cite{Dominguez:2011wm}. The corresponding Wilson line correlator is given by the small-$x$ Weizsäcker-Williams (WW) gluon TMD $x G_{WW}(x,\kt)$, where the transverse momentum $\kt$ refers to the imbalance of the dijet/dihadron system. The differential cross-section reads \cite{Dominguez:2011wm}:
\begin{align}
    \frac{\der \sigma_{\lambda}^{\gamma^*A \to q\bar{q}+X}}{\der z_1 \der z_2 \der^2 \kt \der^2 \Pt } = \delta(1-z_1-z_2) H_{\gamma^*g \to q\bar{q}}^{ij,\lambda}(Q^2,\Pt,z) x G^{ij}_{WW}(x,\kt) \,,
\end{align}
with $H_{\gamma^*g \to q\bar{q}}^{ij,\lambda}$ a perturbatively calculable factor, $z_{1,2}$ are the longitudinal momentum fraction of the jets/hadrons relative to the virtual photon and $\Pt$ denotes the mean transverse momenta of the jets/hadrons.

The WW gluon TMD is computed from a correlator of light-like Wilson lines and its derivatives:
\begin{align}
    x G^{ij}_{WW}(x,\kt) = \frac{4}{(2\pi)^3} \int \der^2 \bt \der^2 \bt' e^{-i \kt \cdot (\bt- \bt')}\left\langle \Tr\left[A^i(\bt) A^{j}(\bt') \right] \right  \rangle_x \,,
\end{align}
where $A^i(\bt) = \frac{i}{g} V(\bt) \partial^i V^\dagger(\bt)$ is the transverse gauge field in the light-cone gauge $A^+=0$.

\begin{figure}[H]
    \centering
    \includegraphics[scale=0.4]{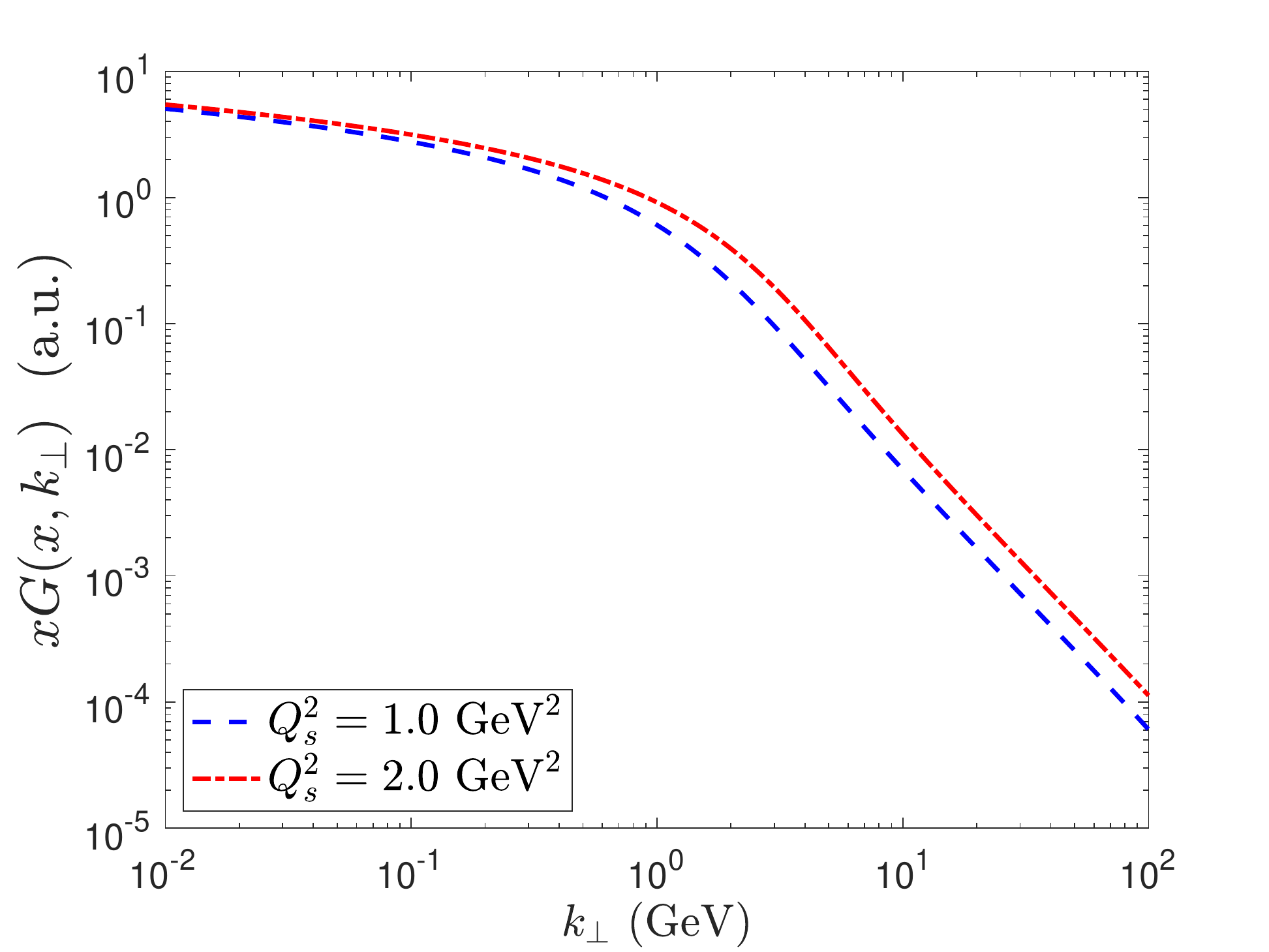}
    \caption{WW gluon TMD distribution as a function of $k_\perp$ for two different values of the saturation scale. A transition between saturation and perturbative regime is observed near $k_\perp \sim Q_s$. The $x$ dependence of the distribution is effectively accounted by the saturation scale. In a more careful treatment, the functional shape of the distribution also depends on $x$. }
    \label{fig:WW_TMD_MVmodel}
\end{figure}

Unlike the Fourier transform of the dipole correlator, this distribution has a probability density interpretation. In Fig.\,\ref{fig:WW_TMD_MVmodel} we plot the WW gluon TMD for two different values of the saturation scale. As expected we observe a transition in the behavior of this function near $k_\perp \sim Q_s$.  In the perturbative limit, this distribution has the following power law tail
\begin{align}
    xG^{ii}(x,k_\perp) \sim \frac{Q_s^2(x)}{k_\perp^2} , \quad k_\perp \gtrsim Q_s \,,
\end{align}
and a slow logarithmic growth in the saturated regime:
\begin{align}
    xG^{ii}(x,k_\perp) \sim \log\left(\frac{Q_s^2(x)}{k_\perp^2} \right), \quad k_\perp \gtrsim Q_s \,.
\end{align}
The transverse momentum imbalance of produced dihadrons/dijets which originated from the virtual photon (with zero transverse momentum, i.e. in the Breit frame) is dictated by the WW gluon TMD distribution. The comparison of the azimuthal angle distribution of dijets/dihadrons near the back-to-back configuration is one of the promising observables for the search of saturation as we will review in the later sections.

 It is worth pointing out that it might be possible to directly parametrize the WW gluon distribution $xG^{ii}(x,k_\perp)$ as a function of $x$ and $k_\perp$. In the CGC and within the Gaussian approximation the WW gluon distribution object is typically constructed from the dipole correlator which is constrained by HERA data.

\subsection{Quantum evolution}
\label{sec:quantum_evolution_CGC}
We close this section on the CGC EFT by very briefly reviewing the crucial aspect of quantum evolution and the renormalization group equations at small-$x$. Thus far we have focused on observables and light-like Wilson line correlators computed using the classical solutions to the Yang-Mills equations for color sources drawn from the MV model. This procedure is appropriate when the observables of interest are probed at a longitudinal momentum fraction $x$ close to $x_0$ at which the weight functional is constructed (for MV $x_0 \approx 0.01$). However, quantum fluctuations around the classical solution are enhanced by terms proportional to $\alpha_s \log(x_0/x)$. These terms can be of order $1$ for sufficiently small $x$, and thus require resummation. Physically, these contributions arise from gluon emissions in the interval $[x,x_0]$. 

\begin{figure}[H]
    \centering
    \includegraphics[scale=0.3]{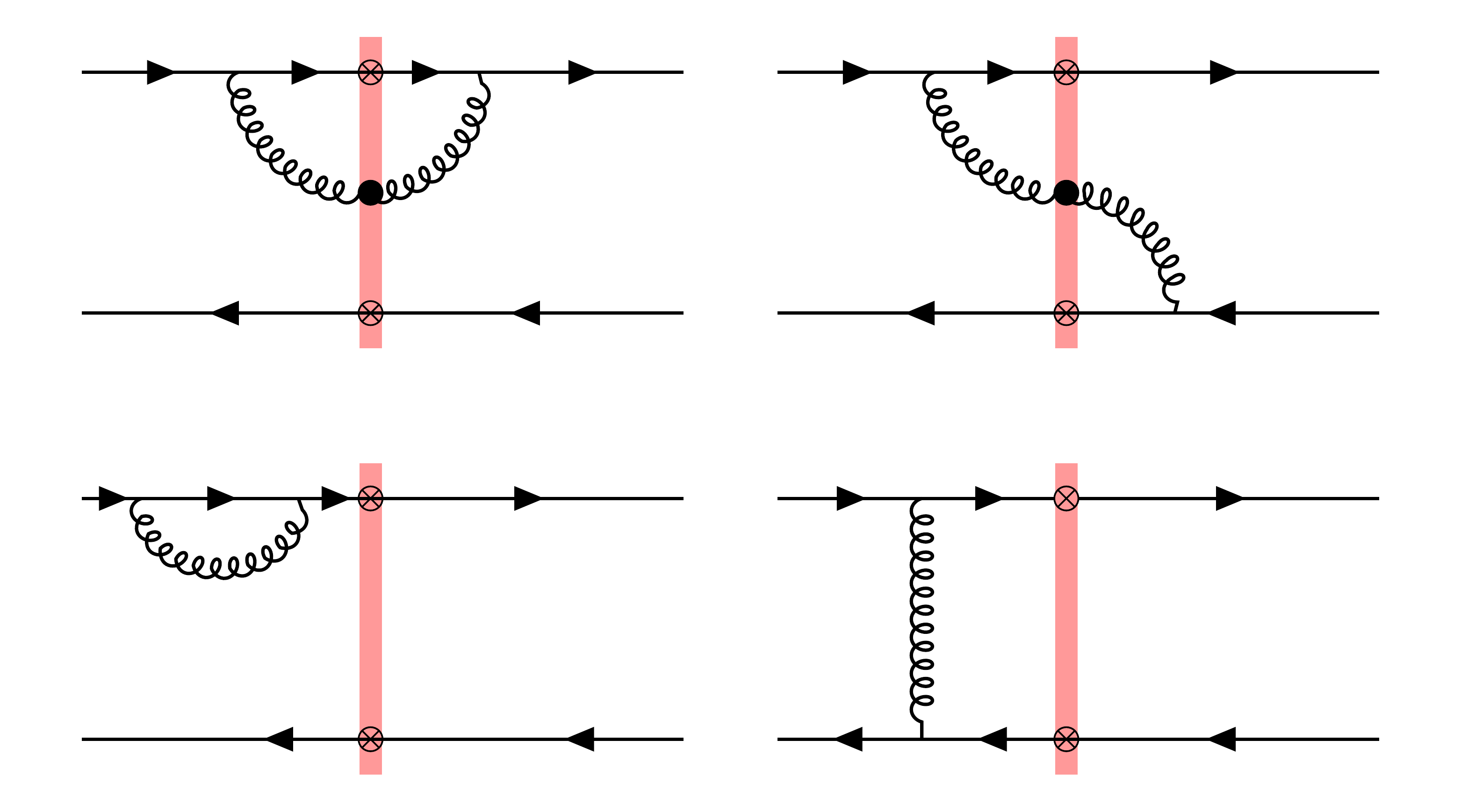}
    \caption{A subset of Feynman diagrams for the quantum evolution of the dipole correlator. Upper diagrams correspond to real gluon emission, while lower diagrams correspond to virtual contributions.}
    \label{fig:dipole_evolution_diagrams}
\end{figure}

\begin{figure}[H]
    \centering
    \includegraphics[scale=0.33]{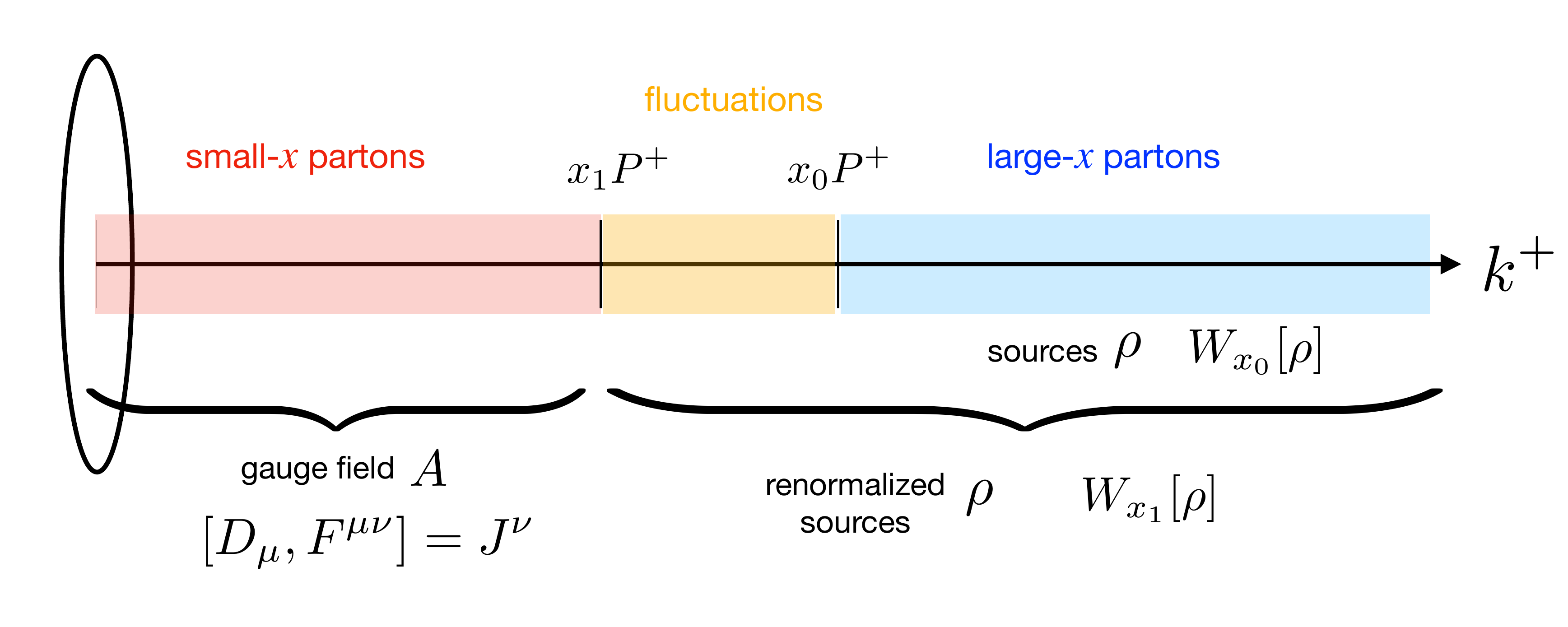}
    \caption{Schematic representation of the quantum evolution of the weight-functional. Quantum fluctuations in the interval $[x_1,x_0]$ (shown in yellow) are absorbed into stochastic fluctuations of the color sources by redefinition of the weight functional $W_{x_0}[\rho] \to W_{x_1}[\rho]$   (compare with Fig.\,\ref{fig:CGC_seperation_dofs}). The choice of the scale separating small-$x$ and large-$x$ partons is thus arbitrary and different choices are related by the JIMWLK renormalization group equations. The long right bracket represents how large-$x$ partons and fluctuations are considered as sources after properly evolving of weight functional.} 
    \label{fig:CGC_seperation_dofs_evolution}
\end{figure}

At large $N_c$, the resummation of these terms results in the Balitsky-Kovchegov (BK) equation for the small-$x$ evolution of the dipole correlator \cite{Balitsky:1995ub,Kovchegov:1999yj} (diagrams shown in Fig.\,\ref{fig:dipole_evolution_diagrams}):
\begin{align}
    \frac{\der S_x^{(2)}(r_\perp)}{d \log(1/x)} = \int \der^2 \rt' \frac{\rt^2}{\rt'^2 (\rt-\rt')^2}\left[S_x^{(2)}(r_\perp') S_x^{(2)}(|\rt-\rt'|) - S_x^{(2)}(r_\perp) \right] \,.
\end{align}
The terms quadratic in $S^{(2)}$ arise from the real emission diagrams in which the gluon crosses the shock-wave, while those linear in $S^{(2)}$ appear from virtual contributions.  The BK equation reduces to the Balitsky-Fadin-Kuraev-Lipatov (BFKL) equation \cite{Lipatov:1976zz,Kuraev:1977fs,Balitsky:1978ic} in the weak scattering regime $D_x(r_\perp)=1-S^{(2)}(r_\perp) \ll 1$ .
Remarkably, an alternative way to resum large logarithmic contributions is by the evolution of the weight-functional from the scale $x_0$ to $x$ following the equation:
\begin{align}
    \frac{\der W_x[\rho]}{\der \log(1/x)} = -\mathcal{H}_{\rm JIMWLK} W_x[\rho]
    \label{eq:evolution_weightfunctional}
\end{align}
where $\mathcal{H}_{\rm JIMWLK}$ is the Jalilian-Marian, Iancu, McLerran, Weigert, Leonidov, Kovner (JIMWLK) Hamiltonian \cite{JalilianMarian:1996xn,JalilianMarian:1997dw,Kovner:2000pt,Iancu:2000hn,Iancu:2001ad,Ferreiro:2001qy}. Physically, this procedure correspond to absorbing the quantum fluctuations in the interval $[x_0 - dx,x_0]$ into stochastic fluctuations of the color sources by redefinition of the sources $\rho$ (Fig.~\ref{fig:CGC_seperation_dofs_evolution}). Iterating this process through a self-similar Wilsonian renormalization group (RG) procedure results in Eq.\eqref{eq:evolution_weightfunctional}. This procedure is equivalent to an infinite hierarchy (known as the B-JIMWLK hierarchy) of non-linear coupled equations dictating the evolution of $n$-point Wilson line correlators \cite{Balitsky:1995ub,JalilianMarian:1996xn,JalilianMarian:1997dw,Kovner:2000pt,Iancu:2000hn,Iancu:2001ad,Ferreiro:2001qy,Weigert:2000gi}.

\begin{figure}[H]
    \centering
    \includegraphics[scale=0.18]{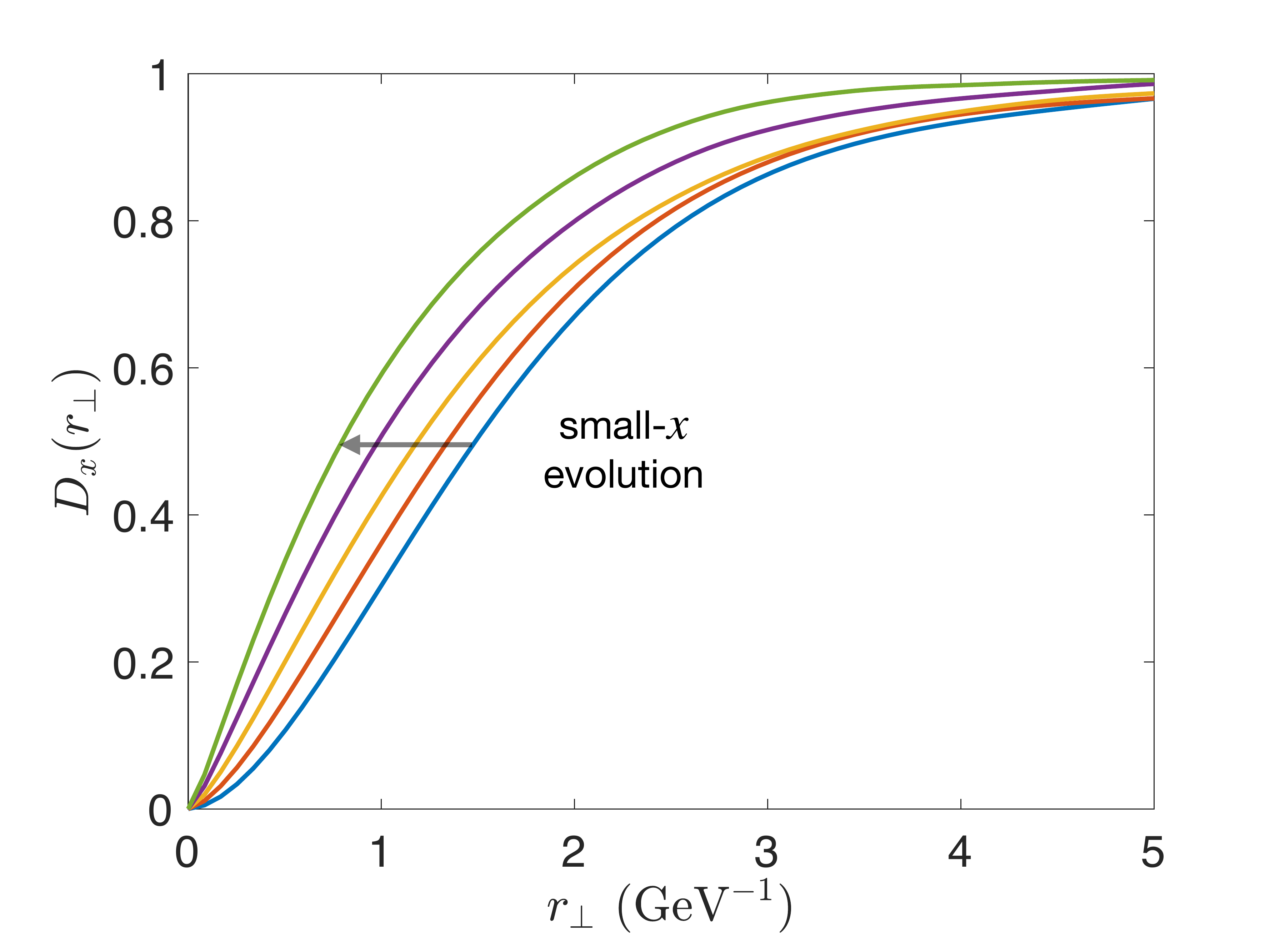}
    \includegraphics[scale=0.33]{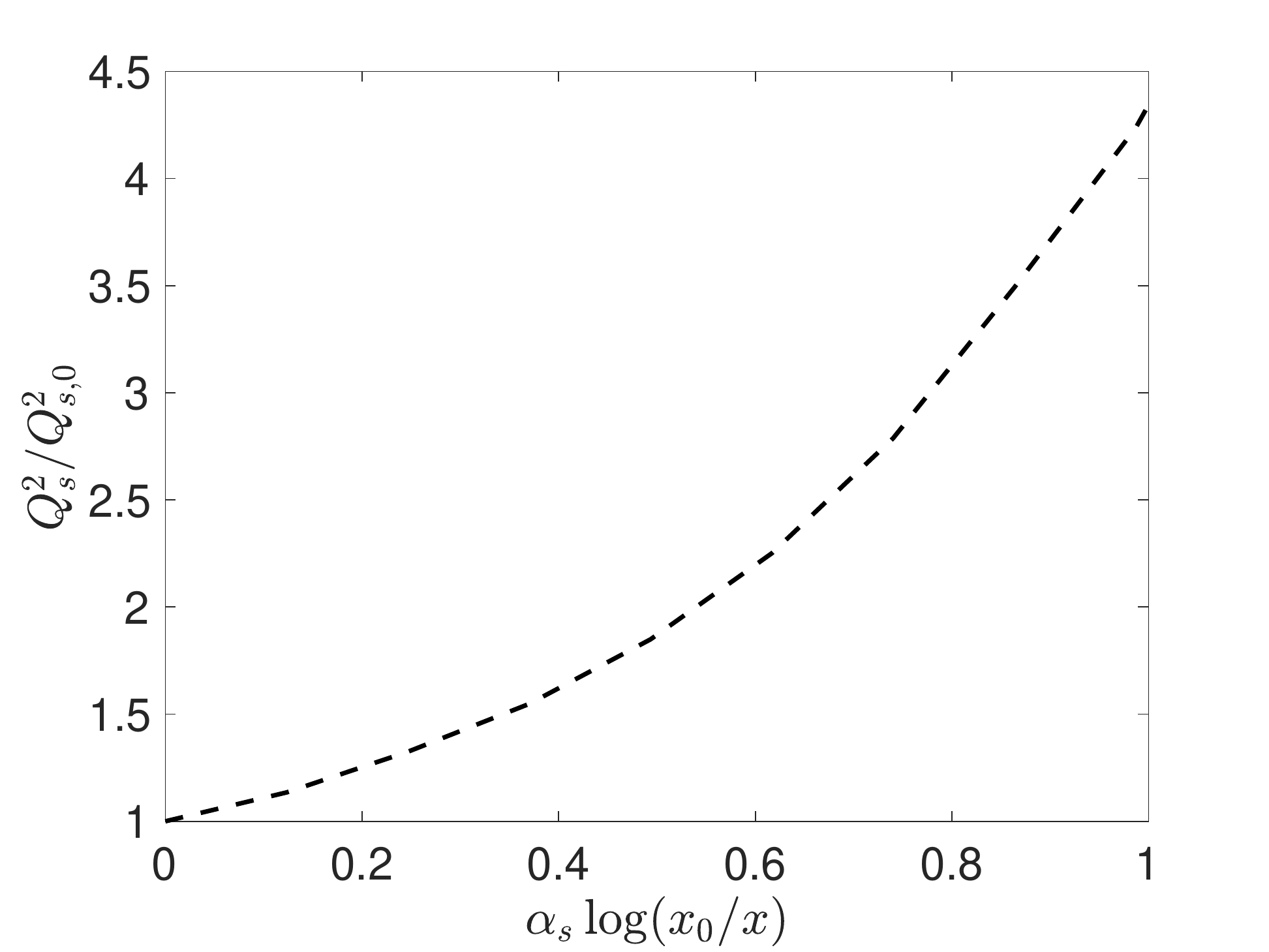}
    \caption{Left: Dipole amplitude $D_x(r_\perp)$ small-$x$ evolution drives a more rapid transition to the strong scattering regime. Right: small-$x$ evolution of the saturation scale $Q_s^2$ normalized by the saturation scale at $x_0$.}
    \label{fig:dipoleQs2__evolution}
\end{figure}

We end this section by illustrating the effect of small-$x$ JIMWLK evolution equations on the dipole amplitude. The quantum evolution effectively increases the color charge density as more partons are introduced as sources. This in turn implies an increase in the saturation scale which for fixed dipole size $r_\perp$ drives the dipole amplitude closer to one (strong scattering). The small-$x$ evolution of the dipole amplitude and the corresponding evolution of the saturation scale are illustrated in Fig.\,\ref{fig:dipoleQs2__evolution}. It is worth mentioning that the small-$x$ evolution also changes the functional shape of the dipole; thus, it is not sufficient to simply parametrize the saturation scale. 

\section{Experimental signatures to date}\label{Sec:experiment} 

Having introduced the underlying theoretical principles and techniques that address a gluon saturated state, we now move on to an interpretation of experimental signatures at colliders from HERA to the LHC.  While our list is not meant to be exhaustive it does provide a summary of the published results  which are the pillars to Sections ~\ref{Sec:EIC} and ~\ref{Sec:conclusions}. The results described in this section have been tied to a number of phenomena including gluon saturation; throughout this document we will caution against competing mechanisms that can also explain the measurements without invoking saturation. As many of the competing mechanisms are process dependent,  we will address them case by case depending on the observable under consideration.

This section is organized as follows, first we will introduce structure functions and their historic relevance followed by diffraction and  semi-inclusive measurements. We apologize for the omission or superficial description of some the work of the scientists in our field for the sake of maintaining this manuscript of reasonable length.

\subsection{Structure functions}\label{subsec:structure functions}

One of the major achievements of the deep inelastic scattering experiments at HERA is the determination of the structure functions $F_2$ and $F_L$ of the proton ~\cite{ZEUS:1993ppj}, which can be determined from the total DIS cross-sections:
\begin{align}
    F_2(x,Q^2) &= \frac{Q^2}{4\pi \alpha_{em}} \left[\sigma^{\gamma*A}_T(x,Q^2) + \sigma^{\gamma*A}_L(x,Q^2) \right] \,, \\
    F_L(x,Q^2) &= \frac{Q^2}{4\pi \alpha_{em}} \sigma^{\gamma*A}_L(x,Q^2) \,.
\end{align}
From these objects it is possible to extract the PDFs. PDFs are universal parton densities  containing long-distance structure of hadrons and are independent of the colliding system (the same in DIS and proton-proton (pp) ). In the collinear framework, PDFs known at an initial scale $Q_0$ are evolved according to the DGLAP renormalization group equations to a different scale $Q$. At high energies, or equivalently small $x$, DGLAP evolution must be supplemented with BFKL dynamics which re-sums $\alpha_s \log(1/x)$ \cite{Lipatov:1976zz,Kuraev:1977fs,Balitsky:1978ic}. Compelling evidence of BFKL dynamics has been suggested in a recent analysis of HERA data with small-$x$ re-summation \cite{Ball:2017otu, Hentschinski:2013id}. Yet at even higher energies (or smaller $x$), the rapid rise of gluon densities cannot grow unchecked as it would violate unitarity, in other words probability conservation. It is expected that at high gluon densities, the nonlinear dynamics of QCD can result in the competing effect of gluon recombination taming growth of gluon distributions.

\begin{figure}[H]
    \centering
    \includegraphics[scale=0.5]{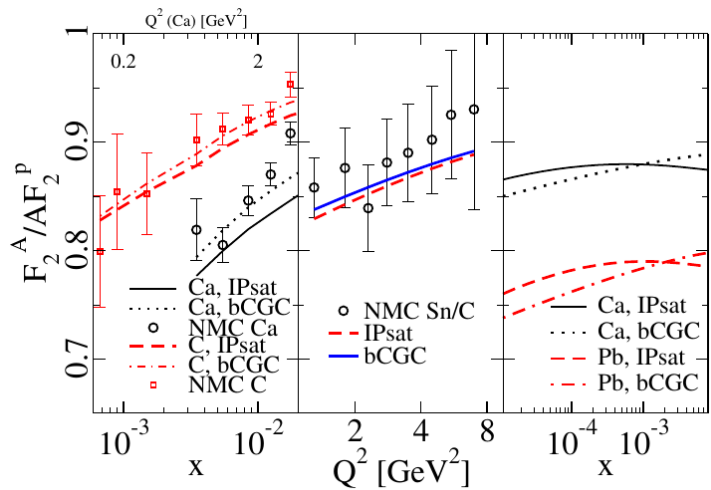}
    \caption{Predictions for shadowing compared to data from the New Muon Collaboration.  Center:  predictions for  Right: Predictions for Q$^{2}= 5$ GeV 2 as a function of $x$. Figure from ~\cite{Kowalski:2007rw}}
    \label{fig:NMC}
\end{figure}

Comparisons to New Muon Collaboration data (Fig.~\ref{fig:NMC}) and comprehensive analyses of the proton structure functions from HERA data have been performed in the saturation framework. The first comparisons to HERA data were performed over 20 years ago in \cite{Golec-Biernat:1998zce} using the Golec-Biernat Wusthoff (GBW) model, which assumed a simple parametrization of the dipole amplitude $D_x(r_\perp) = 1- \exp(-\frac{1}{4} r_\perp^2 Q_s^2(x))$. We point out that an additional parameter $\sigma_0/2$ that effectively accounts for the transverse area is required as this model does not capture impact parameter dependence.  Even so the GBW model had a reasonable agreement with data \footnote{Unfortunately, the GBW model fails to describe other observables such as single hadron inclusive spectra in pA due to its exponential tail, rather than the expected power law behavior.} and led to the observation that the total DIS cross-section can be described by a single variable $\tau = Q^2/Q_s^2(x)$. This phenomena is known as geometric scaling \cite{Stasto:2000er}.

Models that do incorporate the impact parameter $b_\perp$ dependence have been introduced e.g the IPsat model \cite{Kowalski:2003hm} and the the bCGC model \cite{Watt:2007nr}. Roughly speaking, these models incorporate the $b_\perp$ dependence of the saturation scale $Q_s^2(b_\perp) \sim T_p(b_\perp)$ by introducing the thickness function $T_p(b_\perp)$ which parametrizes the gluon density inside the proton. The impact parameter dependence is typically constrained by exclusive processes such as vector meson production, comparisons of these models with HERA data can be found in \cite{Rezaeian:2012ji,Rezaeian:2013tka}. A drawback of these frameworks is the lack of a rigorous treatment of small-$x$ evolution since it would require to incorporate significant contributions from confining physics. Some attempts to tackle this problem can be found in \cite{Berger:2011ew,Bendova:2019psy}. 

One of the most comprehensive studies of structure functions in the saturation framework was performed in \cite{Albacete:2010sy}. These studies used the running coupling BK equation supplemented with suitable initial conditions and accounted for contributions of heavy quarks. This is one of the first attempts at a rigorous description of HERA data using modern theoretical tools of the saturation framework. More recently studies have built upon this work incorporating collinear resummations in the BK equation \cite{Albacete:2015xza,Iancu:2015joa,Ducloue:2019jmy}.

\begin{figure}[H]
    \centering
    \includegraphics[scale=0.4]{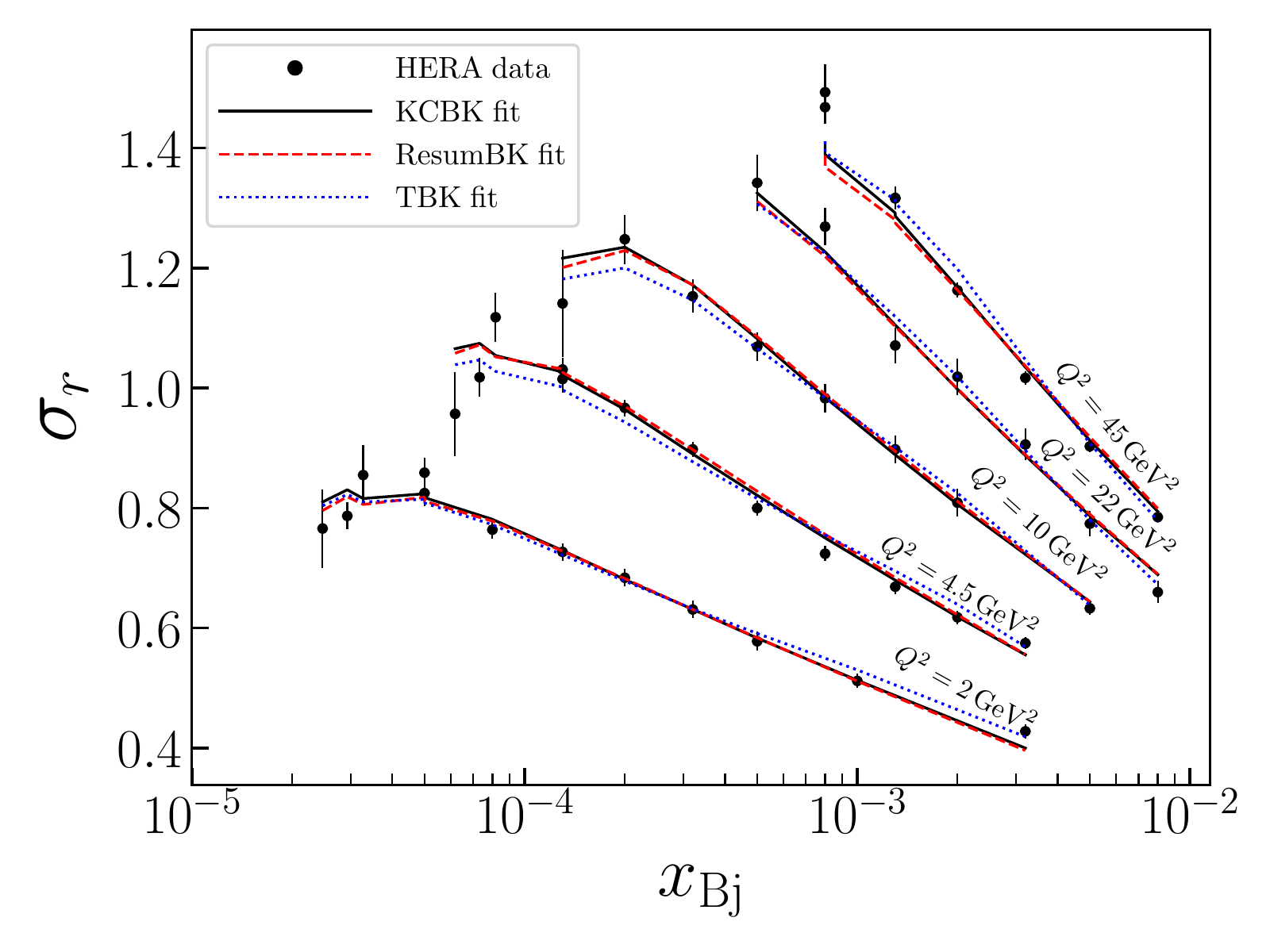}
    \includegraphics[scale=0.4]{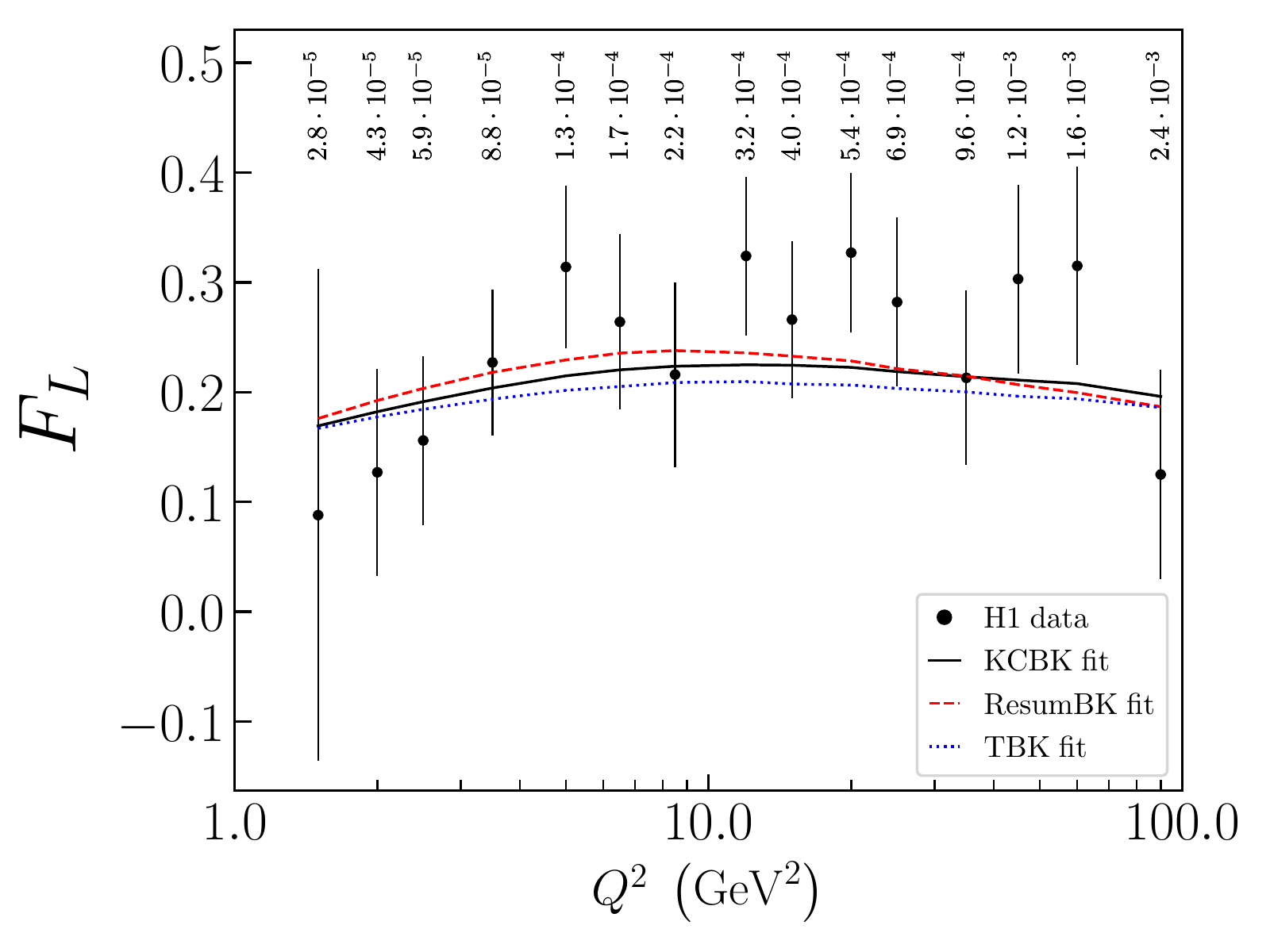}
    \caption{Comparison of the CGC at NLO compared to HERA data. Left: reduced cross-section at small-$x$. Right: $F_L$ structure function. Figure from \cite{Beuf:2020dxl}}
    \label{fig:HERA_CGCNLO}
\end{figure}

We conclude the discussion by highlighting that the state of the art was achieved in \cite{Beuf:2020dxl}. In this work, the authors compared HERA data to the predictions of the CGC at next-to-leading order (NLO) including impact factor and small-$x$ evolution equations. Their comparisons for the reduced cross-section (Eq.~\ref{Eq:reduced_xs}) and the longitudinal structure function are shown in Fig.\,\ref{fig:HERA_CGCNLO}.  As compared to the CGC leading order fits, the authors find that the evolution speed is naturally reduced by the NLO corrections without the need to introduce a large factor in the running coupling. It is worth mentioning that this study only included light-quark contribution as the computation for the impact factor for massive quark contributions is ongoing \cite{Beuf:2021qqa}. For completeness we include the reduced cross-section form which is given by
\begin{align}
    \sigma_r(x,y,Q^2) = F_{2}(x,Q^2) - \frac{y^2}{1+(1-y)^2} F_{L}(x,Q^2) \,,
\label{Eq:reduced_xs}\end{align}
where $y$ is inelasticity of the collision.

\subsubsection{Competing mechanisms in structure functions}

\begin{figure}[H]
    \centering
    \includegraphics[scale=0.32]{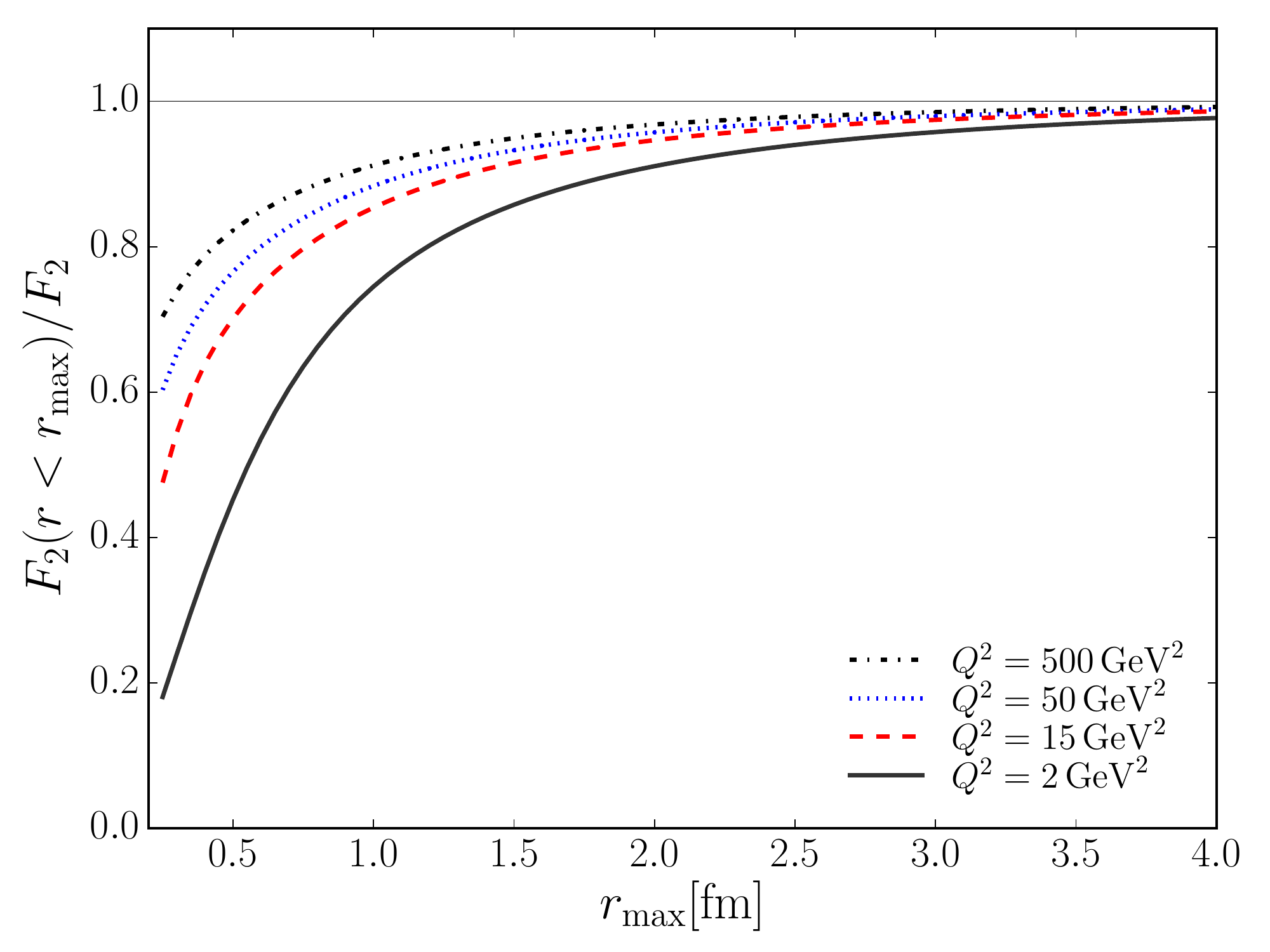}
    \includegraphics[scale=0.32]{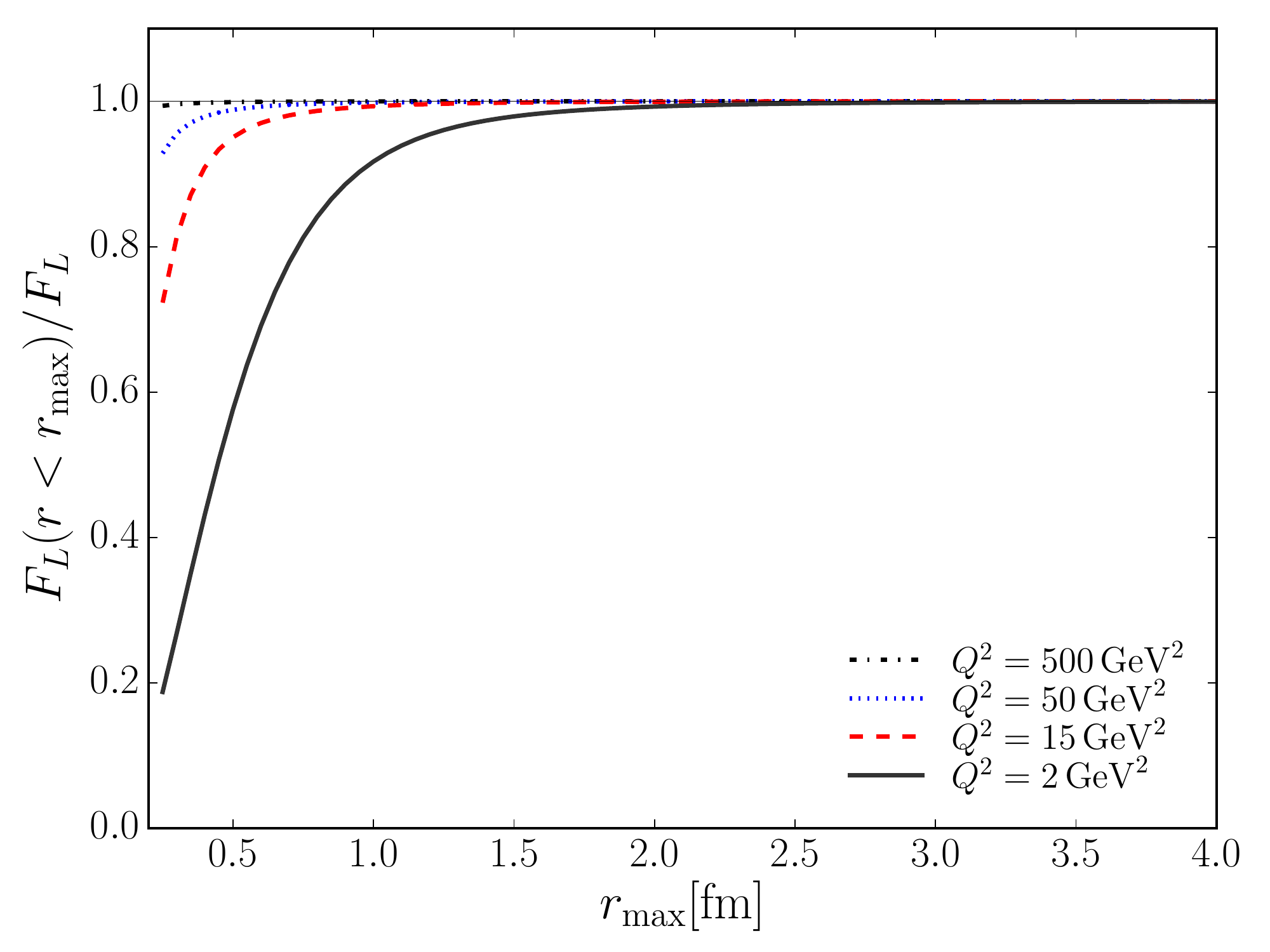}
    \caption{Contribution to the structure functions $F_2$ (left) and $F_L$ (right) from dipole sizes smaller than $r_{\rm max}$ at different photon virtualities. $F_L$ sensitivity to large dipoles is reduced compared to $F_2$ due to the different structure of light-cone wavefunctions between longitudinally and transversely polarized photons. Figure from \cite{Mantysaari:2018nng}.}
    \label{fig:dipole_size_contribution_F2FL}
\end{figure}

While structure functions  extracted from HERA and the NMC data  have  been influential, some aspects relevant to gluon saturation need to be confronted. A main feature to address is the impact of the non-linear phenomena of saturation for the description of the structure functions. This difficulty finds its roots in the large non-perturbative contributions to the determination of structure function at low to moderate $Q^2$. More specifically, structure functions $F_2$ and $F_L$ in the dipole picture can have a significant contribution from non-perturbatively large dipoles. This  has been demonstrated in \cite{Mantysaari:2018nng} and \cite{Mantysaari:2018zdd} where the authors study the contribution to $F_2$ and $F_L$ from large dipoles as shown in Fig.\,\ref{fig:dipole_size_contribution_F2FL}. Large dipole contributions arise from the so called \emph{aligned jet} configuration where either the quark or anti-quark carries most of the longitudinal momentum $(z \to 0,1)$. This configuration is more important for $F_2$ than 
$F_L$ due to the different structure of the light-cone wave-function between transversely and longitudinally polarized photons. Fig.\,\ref{fig:dipole_size_contribution_F2FL} shows that it is necessary to go to very large virtualities to suppress non-perturbatively large dipoles $(r_\perp \gtrsim \ 1.0\  \rm{GeV})$; however, at large $Q^2$ one expects less sensitivity to gluon saturation. This problem is ameliorated when studying charm structure functions as the mass of the quarks serve as an infrared cut-off.

\begin{figure}[H]
    \centering
    \includegraphics[scale=0.5]{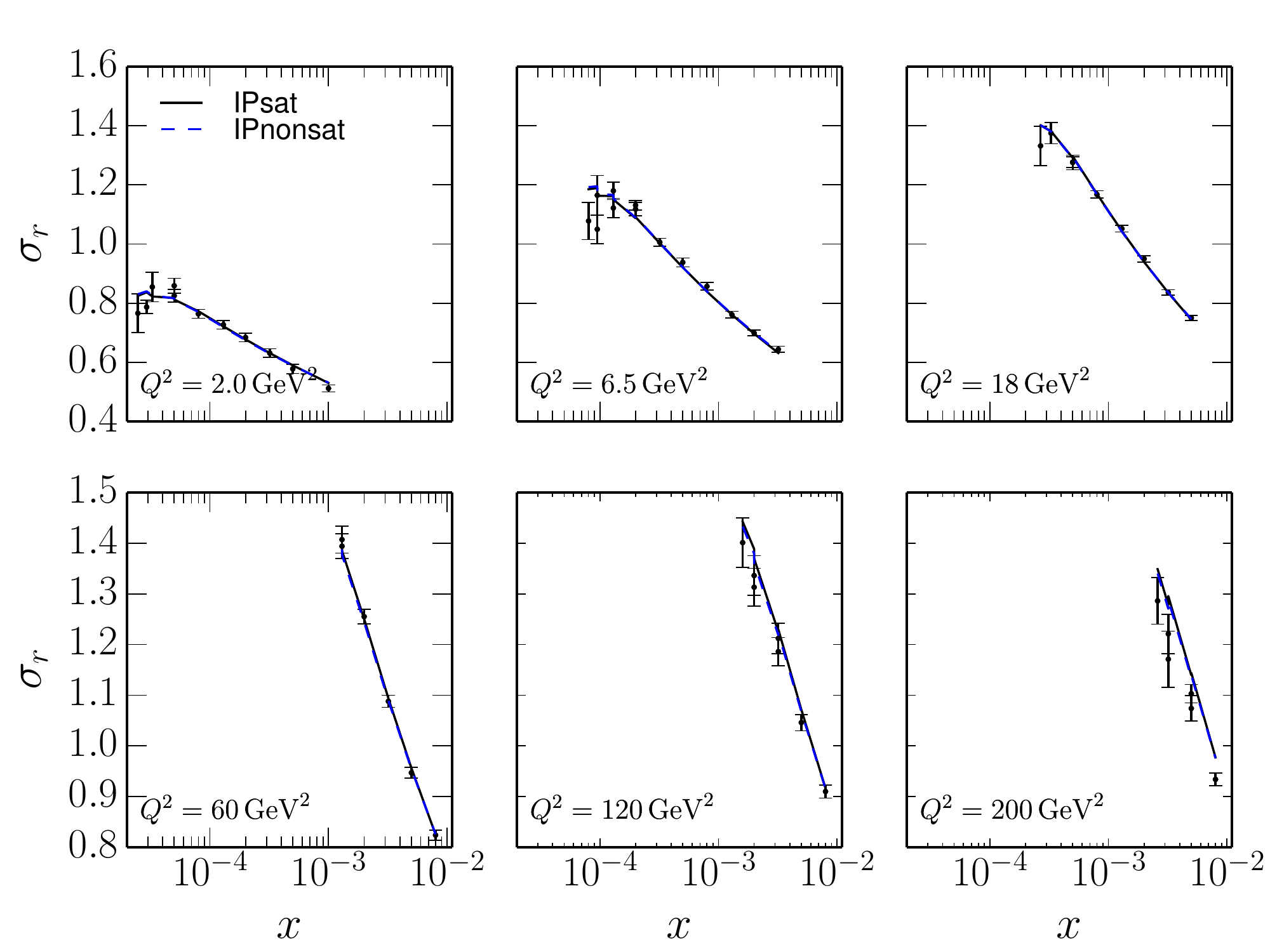}
    \caption{IPsat and IPnonsat (linearized IPsat) independent fits to inclusive reduced cross-section HERA data. Both fits result in almost indistinguishable results, hindering the extraction of a signal of gluon saturation at HERA. Figure from \cite{Mantysaari:2018nng}.}
    \label{fig:IPsat_vs_IPnonsat}
\end{figure}

Another difficulty in unambiguously determining the need of non-linear/gluon saturation effects in the description of HERA data structure functions is due to the parameter freedom allowed in the fits. To illustrate this flexibility, the authors in \cite{Mantysaari:2018nng} fitted structure functions using the IPsat model and its linearized version (IPnonsat) which is expected to exclude gluon  saturation dynamics. When independently fitted, both models result in almost indistinguishable results across a large phase space in $(x,Q^2)$, (see Fig.\,\ref{fig:IPsat_vs_IPnonsat}). This suggests that non-linear effects are not visible in the proton structure functions at HERA alone. It might be possible to reduce the freedom of these models by applying them to other physical processes. In Section ~\ref{Sec:EIC}, we will briefly discuss the improved potential to discover gluon saturation in the study of nuclear structure functions at future DIS collider experiments.

\subsection{Diffractive reactions}
\label{sec:diffractive_reactions}
Diffractive observables are characterized by a rapidity gap (the absence of particles produced in a given rapidity window) which originates from a color neutral exchange between the two colliding systems.  Since gluons carry color, this exchange requires at least two gluons which must be in the color singlet state. As consequence, diffractive measurements are sensitive to the "square" of the the gluon distribution (at lowest order in perturbation theory). Compared to inclusive measurements, this enhanced sensitivity to the gluon distribution makes diffractive observables excellent candidates for gluon saturation searches at small-$x$. In this section, we will mostly focus on diffractive production in the collision of a photon (virtual or real) with a proton or nucleus. This can be realized in deep inelastic scattering (DIS) and ultra-peripheral collisions (UPCs).

Diffractive DIS observables have been extensively studied at HERA. The first hints of saturation were observed in the analysis of inclusive and diffractive cross-sections using the GBW model in \cite{Golec-Biernat:1998zce,Golec-Biernat:1999qor}. The authors found that in the saturation framework the ratio of diffractive to inclusive events was almost constant with only a mild logarithmic dependence on the virtuality $Q^2$ and Bjorken-$x$. The results from these saturation models compared well to data, especially after they were furnished with DGLAP evolution \cite{Bartels:2002cj}. More refined studies for the description of the diffractive structure functions \cite{Marquet:2007nf} using the IPsat and bCGC model were carried out in \cite{Kowalski:2008sa}.

The exclusive production of vector particles (photons and vector mesons) are also powerful tools to study the gluon content of protons and nuclei. In addition to the energy and virtuality dependence of their production cross-sections, the momentum transfer squared $t$ dependence, and the dependence on the mass $M_{V}$ of the produced vector particle give more detailed insight into the gluon structure of nuclei at high energies. One can distinguish two cases: (i) coherent events in which the target (proton/nucleus) remains intact, and (ii) incoherent events in which the target (proton/nucleus) breaks up.

\begin{figure}[H]
    \centering
    \includegraphics[scale=0.5]{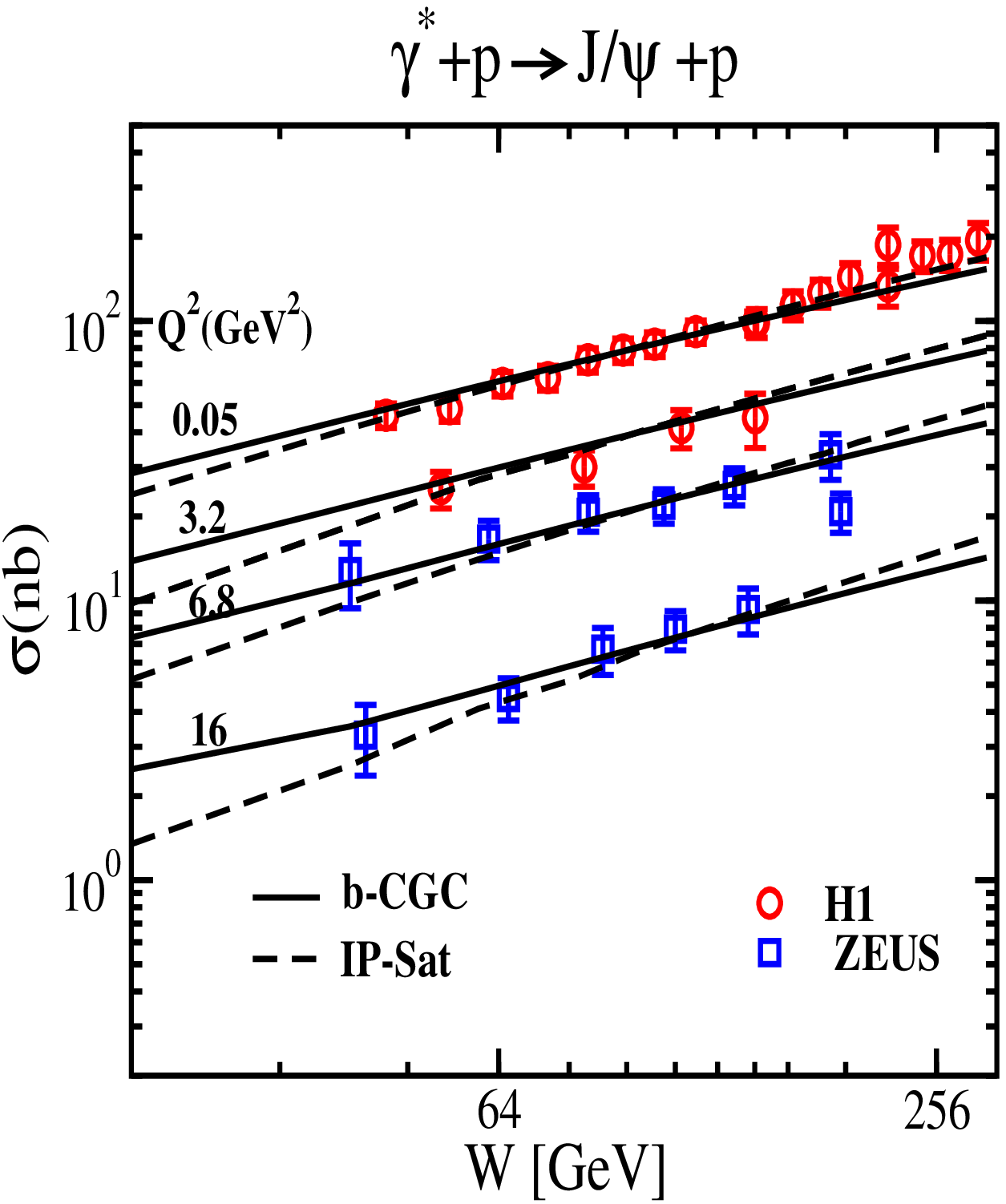}
    \includegraphics[scale=0.5]{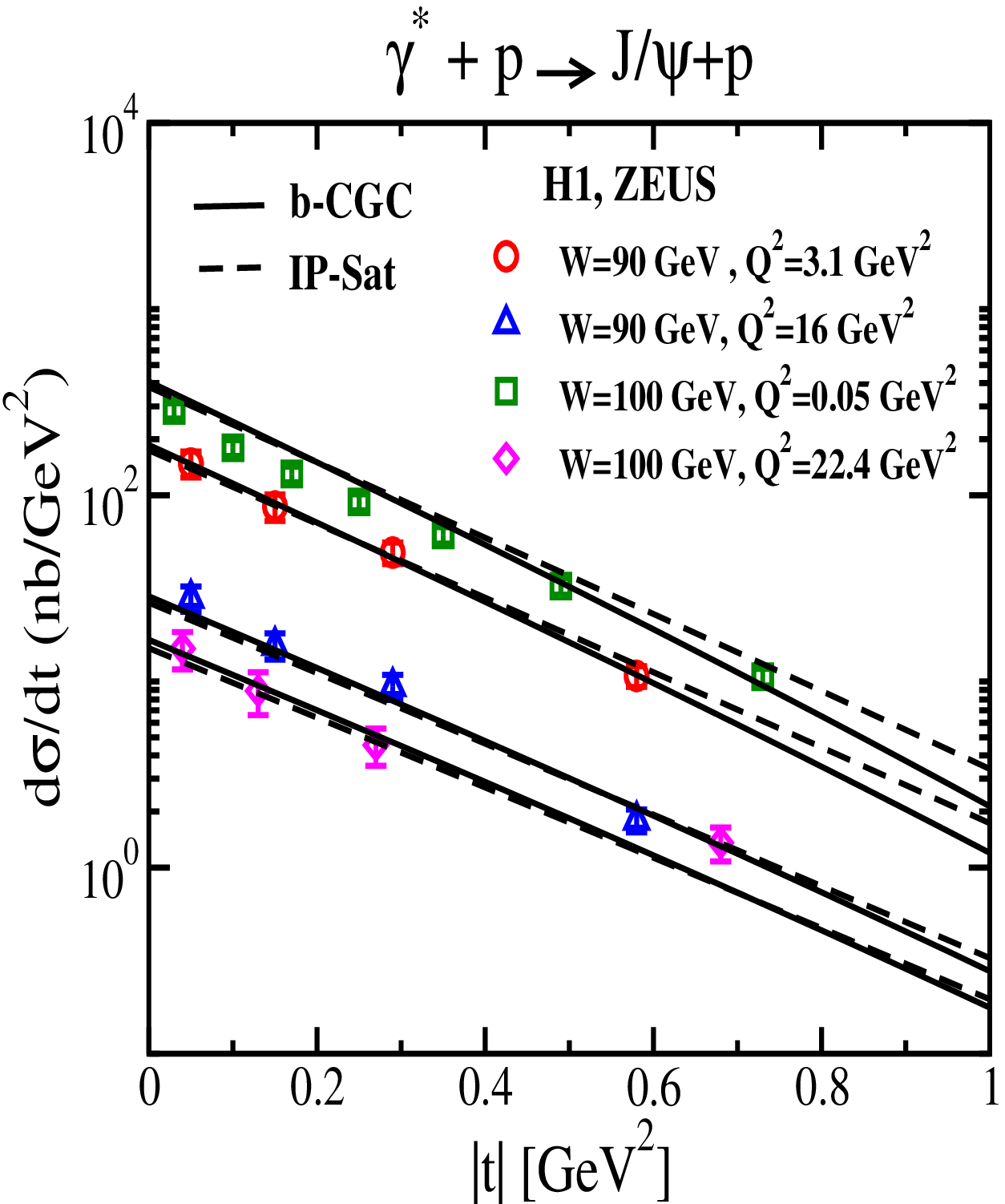}
    \caption{$J/\psi$ exclusive electroproduction data from HERA compared to saturation models (bCGC and IPsat). Left: energy dependence. Right: $|t|$ spectra. Figure from \cite{Rezaeian:2013tka}.}
    \label{fig:Jpsi_t_and_energydep}
\end{figure}

Coherent events are dominant at low $t \lesssim 1/R^2$, where $R$ is the size of the target, and they are sensitive to the average color density profile of the target. Both spectra and energy dependence for coherent vector meson production were compared to HERA data within a GBW model incorporating impact parameter dependence in \cite{Kowalski:2006hc} and later using the bCGC and IPsat models \cite{Watt:2007nr,Rezaeian:2012ji,Rezaeian:2013tka,Goncalves:2014wna} (see Fig.\ref{fig:Jpsi_t_and_energydep}). The impact parameter dependence of the dipole models is crucial as it is responsible for the cross-section not vanishing at non-zero momentum transfers. This dependence is typically modeled and the parameters are part of the fit. More complex studies using JIMWLK evolution have shown that the impact parameter dependence of the average color charge density evolves with energy: the gradients of color charge become smoother and the overall size of the profile grows \cite{Schlichting:2014ipa}.

Exclusive vector meson production can also occur in ultraperipheral nucleus-nucleus collisions, where either of the nuclei acts as a source of Weizsäcker-Williams real photons, which then interact with the other nucleus. These processes have been studied at RHIC and the LHC with the saturation models providing a good description of the data \cite{Armesto:2014sma,Goncalves:2014swa}. Recently, the energy evolution has been studied comparing models that incorporate either BFKL (linear) or BK (non-linear) evolution \cite{ArroyoGarcia:2019cfl}. The authors argue the onset of gluon saturation as they find the need for non-linear evolution to describe the vector meson photo-production data from HERA, RHIC and LHC.

\begin{figure}[H]
    \centering
    \includegraphics[scale=0.45]{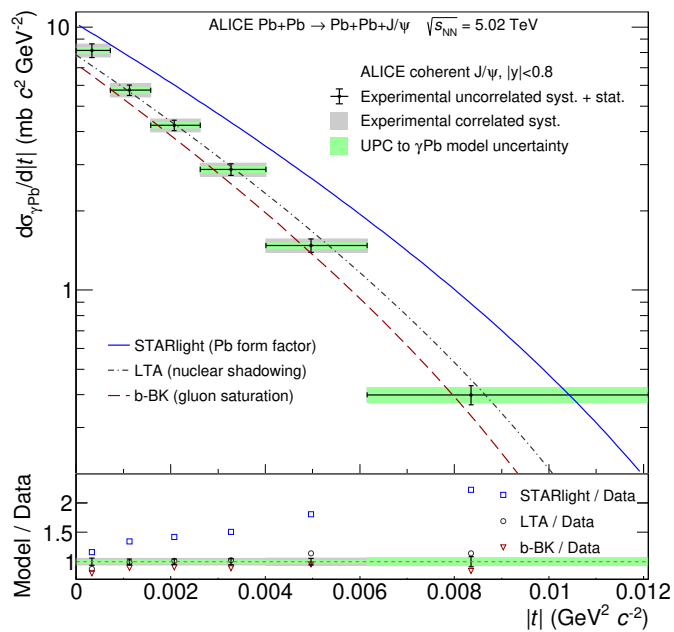}
    \caption{Photo-production of $J/\psi$ at very low values of $t$ compared to Starlight data \cite{Klein:2016yzr}, leading twist approach (LTA) \cite{Guzey:2016qwo}, and b-BK saturation model \cite{Bendova:2020hbb}. Figure from \cite{ALICE:2021tyx}.}
    \label{fig:very_lowt_Jpsi}
\end{figure}

One of the consequences of saturation is a steeper $t$-distribution compared to one obtained from the form factor or Fourier transform of the density profile~\cite{Klein:2016yzr}. This has recently received some attention \cite{Lappi:2021ieu} in light of the very precise ALICE measurements of $J/\psi$ photo-production at very low $t$ \cite{ALICE:2021tyx} as shown in Fig.\ref{fig:very_lowt_Jpsi}. Note that even after saturation is included as in the model in \cite{Bendova:2020hbb}, the spectrum is not steep enough to reproduce the lowest $t$ bin.

\begin{figure}[H]
    \centering
    \includegraphics[scale=0.42]{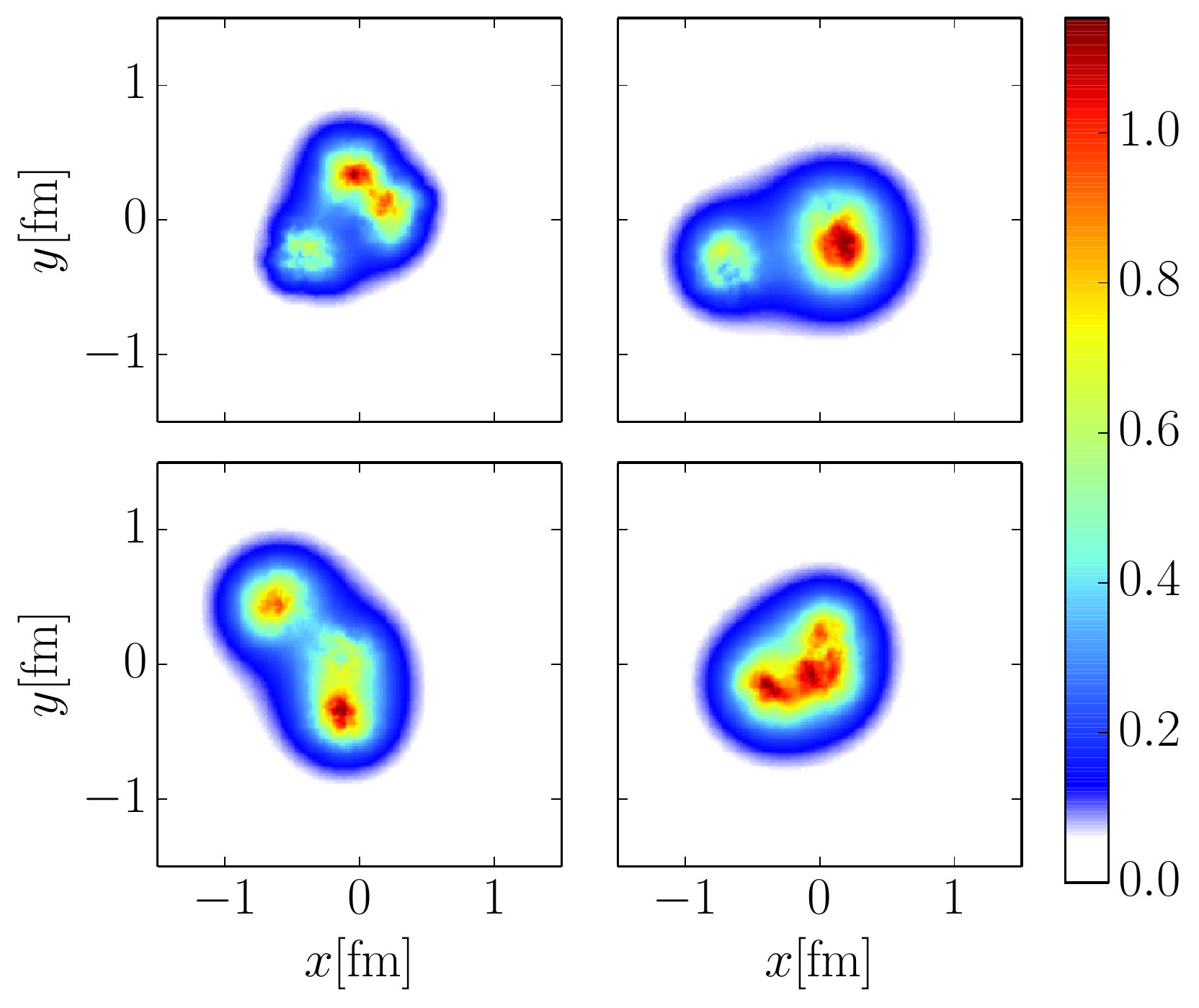}
    \includegraphics[scale=0.42]{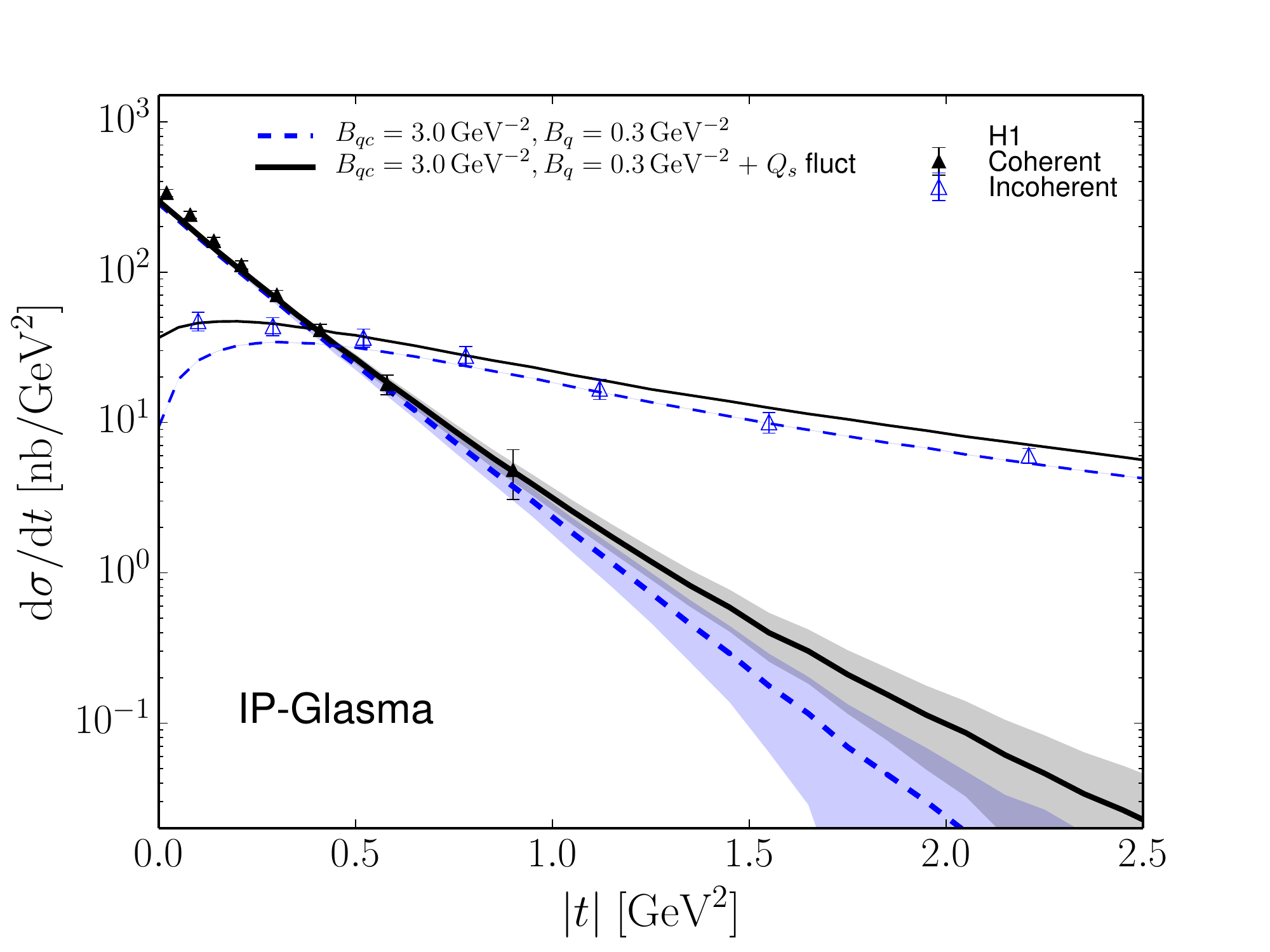}
    \caption{Left: color charge density for four different events. Right: Coherent and incoherent H1 data compared to predictions from CGC incorporating subnucleonic (and $Q_s$ saturation fluctuations).}\label{Fig:densityPlasma}
\end{figure}

In contrast to coherent production, incoherent events dominate at large values of $t$ and the spectrum is sensitive to event-by-event fluctuations \cite{Frankfurt:1993qi,Frankfurt:2008vi,Dominguez:2008aa,Lappi:2010dd}. In \cite{Mantysaari:2016ykx,Mantysaari:2016jaz} the authors found that in order to reproduce the data from HERA, it is necessary to incorporate subnucleonic fluctuations in terms of hotspots of color charge density (Fig.~\ref{Fig:densityPlasma}. These studies have been extended to UPCs at RHIC in \cite{Mantysaari:2017dwh} where the effect of fluctuations also significantly increases the distribution at large $t$. Furthermore, they have also been explored for LHC energies in \cite{Cepila:2016uku,Cepila:2018zky} where a model for the energy dependence of the number of hotspots has been introduced. For a comprehensive review on the subject of proton and nuclear shape fluctuations see \cite{Mantysaari:2020axf}).

It might also be possible to single out saturation effects by studying azimuthal correlations in the diffractive production of dihadron or dijets. This subject has been investigated recently for DIS and UPCs in \cite{Altinoluk:2015dpi,Mantysaari:2019csc,Salazar:2019ncp,Shi:2020djm,Boer:2021upt,ATLAS:2021jhn}.

\subsubsection{Competing mechanisms and systematic uncertainties}

A significant source of theoretical uncertainty in the production of vector mesons arises from the description of their light-cone wave-function and the model for the dipole amplitude. Different parametrizations for these objects have been recently compared when studying rapidity distributions \cite{Goncalves:2017wgg} where the authors find large systematics uncertainties. Recent developments on relativistic corrections to vector meson light-cone wave-functions can be found in  \cite{Lappi:2020ufv}.

As previously noted, the free parameters in the dipole models are typically chosen to reproduce HERA data, models are then used for predictions at RHIC and the LHC. However, saturation effects at HERA might be weak, as we argued in our discussion of the structure functions. The authors of \cite{Mantysaari:2018nng} find a similar description of HERA data when comparing fits with IPsat and IPnonsat, signaling that gluon saturation effects are weak; thus arguing for the need of nuclear DIS.

We close this section by mentioning that other compelling frameworks such as the leading twist approach \cite{Frankfurt:2011cs} based on QCD factorization theorems and nuclear shadowing can provide a good description of coherent vector meson photo-production \cite{Guzey:2013xba,Guzey:2013qza,Guzey:2016qwo,Guzey:2016piu} and diffractive dijet photo-production \cite{Guzey:2016tek}.

\subsection{Semi-inclusive reactions}\label{sec:semi_inclusive_reactions}
Semi-inclusive measurements defined where one or more particles are tagged, provide more detailed information about the dynamics of gluons than fully inclusive measurements such as structure functions. Experimental signatures of gluon saturation are expected to be imprinted in the transverse momentum and rapidity distributions of particles produced in hadronic collisions \cite{Kharzeev:2004bw}.  

\subsubsection{ Single inclusive production} 

Single inclusive particle production has been extensively studied in the saturation framework. The first ideas can be traced back to over 20 years ago, where inclusive forward gluon production \cite{Kovchegov:1998bi,Dumitru:2001jn,Kovner:2001vi} and inclusive forward quark production \cite{Dumitru:2002qt} in proton-nucleus where studied in the \emph{hybrid} formalism. In this framework incoming partons inside the proton are treated within the DGLAP collinear approximation and subsequently scatter eikonally off the strong field produced by the nucleus via correlators of light-like Wilson lines (see Sec.\,\ref{sec:DIS_to_pA}). These studies opened up the possibility to access the saturated gluon regime with semi-inclusive measurements in the collisions of a small dilute nucleus with a larger saturated nucleus. The conceptions soon capitalized in \cite{Kharzeev:2002pc,Kharzeev:2003wz}, where the authors argued that the high $p_\perp$ suppression observed in BRAHMS \cite{BRAHMS:2004xry} at forward rapidities in $d$-$Au$ collisions was a signature of the onset of gluon saturation. 

To characterize the suppression observed in these experiments, one defines the nuclear modification factor:
\begin{align}
    R_{A_1 A_2} = \frac{1}{N_{\mathrm{coll}}} \frac{\mathrm{d}  \sigma^{A_1 A_2 \to h X} }{\mathrm{d}^2 \boldsymbol{p}_\perp \mathrm{d} \eta} \Big / \frac{\mathrm{d}  \sigma^{pp \to h X} }{\mathrm{d}^2 \boldsymbol{p}_\perp \mathrm{d} \eta}\,, 
\end{align}
where $N_{\mathrm{coll}}$ is the number of binary collisions. This ratio is expected to be unity if nuclear collisions were a simple incoherent superposition of collisions with individual nucleons, while deviations from unity indicate coherent effects at play.

In the MV model, the presence of saturated gluons with typical momentum $k_\perp \sim Q_s$ induce a broadening of the transverse momentum distribution of the produced particles. More specifically,  the nuclear modification factor is suppressed for $k_\perp \lesssim Q_s$ and enhanced for $k_\perp \gtrsim Q_s$. This enhancement is  also called the Cronin peak \cite{Cronin:1974zm}. Indeed, in \cite{Kharzeev:2003wz} the authors explicitly show that not integrated gluon distributions (at small-$x$ there are two kinds: the dipole and the Weizsäcker-Williams type) obtained from the MV model satisfy a sum rule \footnote{The factor of $A$ in Eq.\,\eqref{eq:UGD_sum_rule_MV} arises from an $A^{2/3}$ overall area, and $A^{1/3}$ from the scaling of the saturation momentum.}:
\begin{align}
    \int \der^2 \pt \phi^{A}_{\rm MV}(p_\perp) = A \int \der^2 \pt \phi^{p}_{\rm MV}(p_\perp) \,.
    \label{eq:UGD_sum_rule_MV}
\end{align}
where $\phi^{p}_{\rm MV}(p_\perp)$ and $\phi^{A}_{\rm MV}(p_\perp)$ are the unintegrated gluon distributions for the proton and nucleus respectively.

\begin{figure}[H]
    \centering
    \includegraphics[scale=0.5]{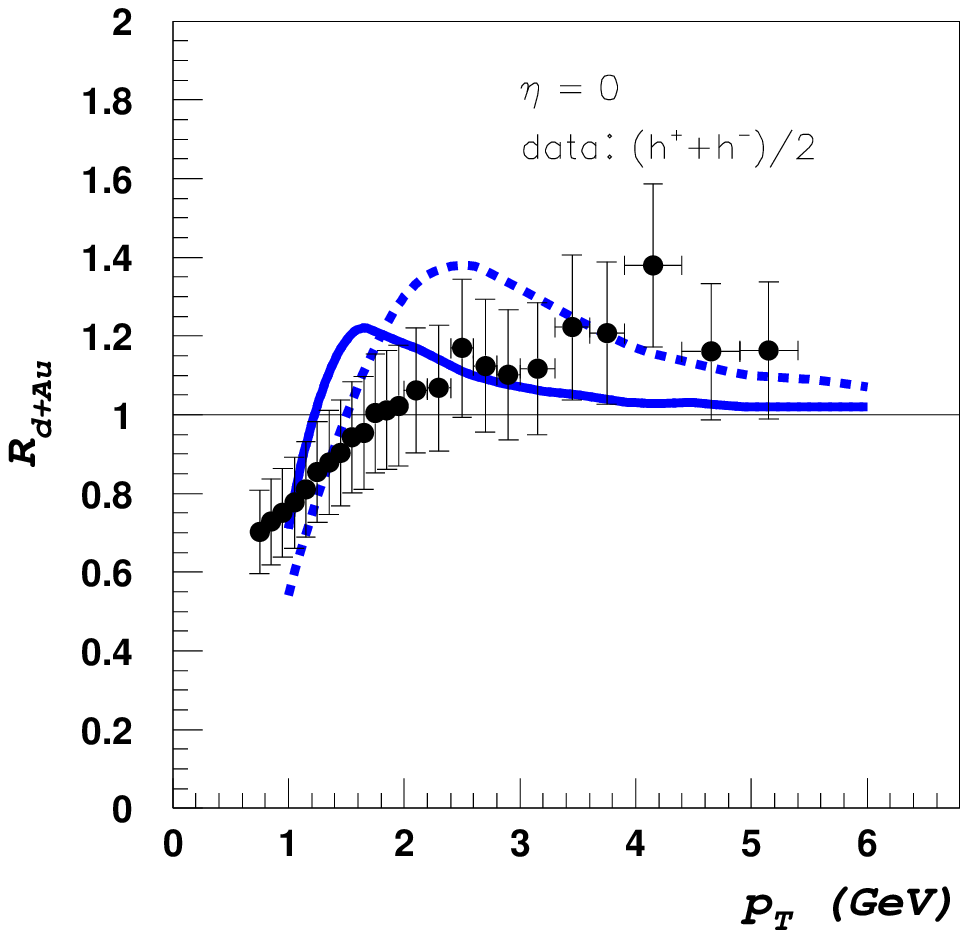}
    \includegraphics[scale=0.5]{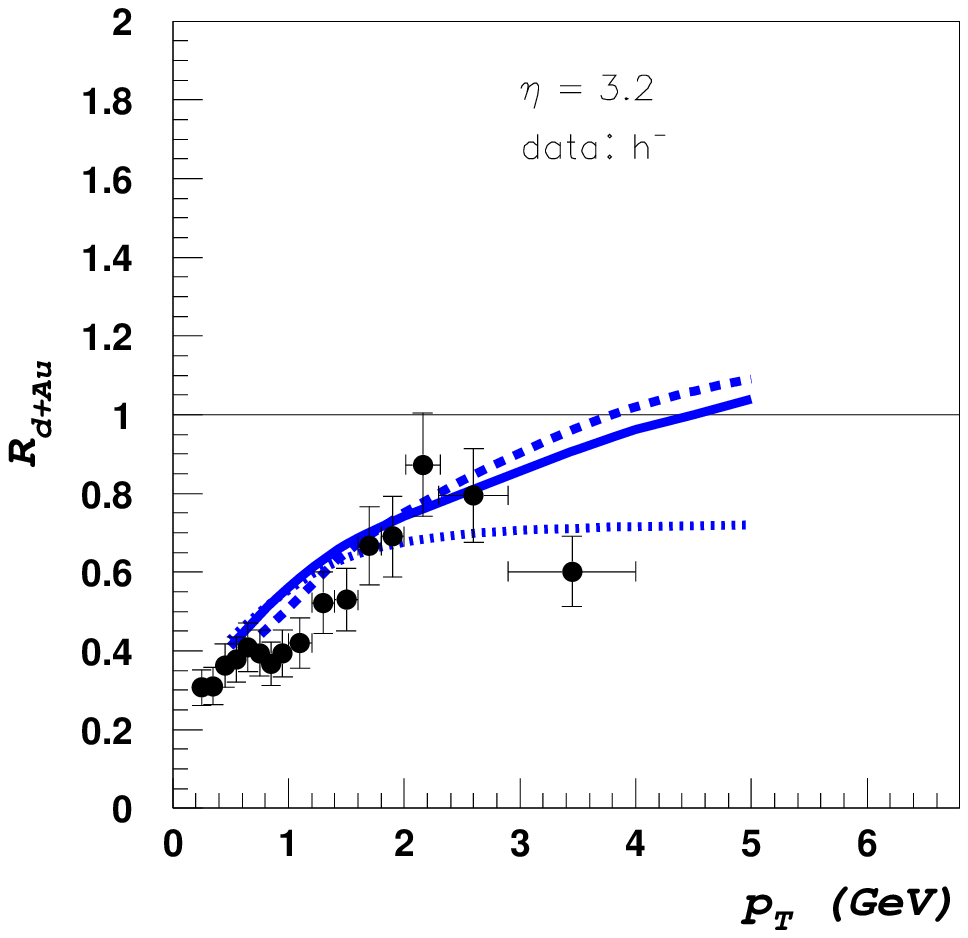}
    \caption{BRAHMS data on the nuclear modification factor $R_{dAu}$ of charged particles at midrapidity and forward rapidity (data from \cite{BRAHMS:2004xry}) compared with the saturation based results in \cite{Kharzeev:2004yx}. The midrapidity $p_\perp$ distribution is characterized by a Cronin peak, while at forward rapidities this peak is washed away.}
    \label{fig:nuclear_mod_factor_hadron_DAu}
\end{figure}

The authors also find by analytic arguments, that after sufficient quantum small-$x$ evolution, the sum rule in Eq.\,\eqref{eq:UGD_sum_rule_MV} is turned into an inequality: the $p_\perp$ integrated distribution is suppressed for larger nuclei.  These analytic arguments were confirmed in \cite{Albacete:2003iq} by evaluation of the corresponding unintegrated gluon distribution using numerical solutions of the BK equation. A quantitative comparison with RHIC data was carried out in \cite{Kharzeev:2004yx} (See Fig.\,\ref{fig:nuclear_mod_factor_hadron_DAu}). It is worth noting that in this work the authors shifted the nuclear saturation scale $Q_s^2 \to Q_s^2 + \kappa^2 A^{1/3}$ with $\kappa \sim 1.0 \ \rm{GeV}$ to describe the data, where they argued it was due to  non-perturbative low energy rescattering which is absent in the saturation framework. Nuclear suppression due to gluon saturation has also been studied at the LHC in \cite{Rezaeian:2012ye,Albacete:2012xq,Lappi:2013zma,Ducloue:2016ywt,Mantysaari:2019nnt}, e.g. $D$-meson production measurements~\cite{Ducloue:2016ywt} at the ALICE experiment\cite{ALICE:2018vhm} which was well described by CGC predictions. Notwithstanding, further studies are necessary as these data can also be well described without saturation see e.g. \cite{Eskola:2019bgf}.

Another opportunity to study gluon saturation is enabled by prompt or direct photon production in proton-nucleus collisions~\cite{Gelis:2002ki,Jalilian-Marian:2005qbq,Helenius:2014qla,JalilianMarian:2012bd,Benic:2016uku,Ducloue:2017kkq,Golec-Biernat:2020cah,SampaiodosSantos:2020raq}. Direct photons are differentiated from virtual or fragmentation photons as real photons originating from the electromagnetic vertex. High energy direct photons are a particularly interesting probe as they do not suffer from non-perturbative fragmentation.  Experimentally on the other hand they are challenging to identify. Photons are neutral final state particles which often rely on energy depositions in calorimeters or material conversions as part of the identification strategy. Their relative cross-sections are much smaller as compared to single hadron production which can be a competing background via photon decays. This difference in cross-sections is due to an additional factor of the electromagnetic structure constant $\alpha_{em}$.  General and beam related a backgrounds as will be discussed in the last Sec~.\ref{Sec:conclusions}  may also inundate and overlap in the detectors which can make a precise identification difficult in particular at moderate to high (or very low)  $p_\perp$ depending on the resolution limits of the detectors used. LHC results from the ALICE experiment have been measured~\cite{Acharya:2018dqe} at large $x$ and mid-rapidity. While there is little indication that the ALICE $x$ range/rapidity probed is ideal for a clean gluon saturation signal, sensitivity is anticipated, mainly, a suppression in the nuclear modification factor in pA collisions is expected featuring a Cronin peak at midrapidity and its disappearance at forward rapidities. Since the photon is colorless it is expected that its distribution will be less influenced by final state interactions.  There are number of active efforts at the LHC to measure similar photon observables at forward rapidity as well as with mid- and forward rapidity correlations  to investigate gluon saturation effects~\cite{ALICE:2020mso, Boettcher:2019kxa} with dedicated  calorimetry.

\label{Sec:quarkonia}
We wrap up the single inclusive discussion with quarkonia production. Quarkonium or hidden charm  in proton and heavy ion collisions also provides valuable opportunities to study gluon saturation at high scattering energies.
One of the advantages of quarkonia in terms of pQCD calculations is that the charm quark mass is larger than the typical QCD scale of $\Lambda_{QCD}$, making pQCD calculations meaningful via factorization. On the other hand the evolution into a bound state is intrinsically non perturbative due to the size of the $q\bar{q}$ system and the inverse of the binding energy not being  small enough.

A key consequence of the   saturation scale  defined as: $Q^{2}_{sat}\sim A^{1/3} x^{-0.3}$ with $A$ the mass number, is that the same scale for heavy nuclei in heavy ion colliders  is comparable with heavy quark mass. We borrow from a relevant formalism for quarkonia given in ~\cite{Watanabe:2016ert, Kharzeev:2005zr} for the arguments hereafter. In the case of pA collisions and in the nucleus (A) rest frame, the interaction time when proton scatters off the nucleus is characterized $\tau_{int}\sim R_{A}$. In a heavy ion collision a heavy quark pair ($q\bar{q}$) is produced over the time $\tau_{P}\sim \frac{E_{g}}{(2m_{q})^{2}}$, where $m_{q}$ is the quark mass and $E_{g}$ is the energy of the parent gluon ($E_{g}=x_{p}E_{p}$). The gluon  will ultimately split  into  the $q\bar{q}$ pair.  Momentum conservation for partonic scattering dictates that $\tau_{P}\sim \frac{1}{2x_{A}M_{N}}$, with $x_{a}$ representing the longitudinal momentum fraction of target nucleus carried by the incident gluons and $M_{A}$ being the mass of the nucleon. This latter indicates that in a \emph{proton going direction} or forward rapidity  the $q\bar{q}$ has a longer production time ($\tau_{P}$) than  $\tau_{int}$ owing to the Lorentz time dilation and the coherent interaction of the proton with the nucleus.  As the biding energy of quarkonium is smaller than its mass $m_{q}$~\cite{Kharzeev:2005zr, Kharzeev:1999bh}, quarkonium production is thus shorter than its formation time, $\tau_{F}\gg \tau_{P}\gg \tau_{int}$.
This tells us that the dynamics of the  $q\bar{q}$ bound state formation can be thus decoupled with cold nuclear effects thanks to its formation occurring well outside the target nucleus. 
A body of theoretical work exists which has compared 
low $p_{T}$ quarkonia production at LHC at forward rapidity using Color Evaporation Models (CEM), Color Singlet Models (CSM)  and  Non Relativistic QCD coupled with gluon saturation (NRQCD$+$CGC).

\begin{figure}[H]
    \centering
    \includegraphics[scale=0.4]{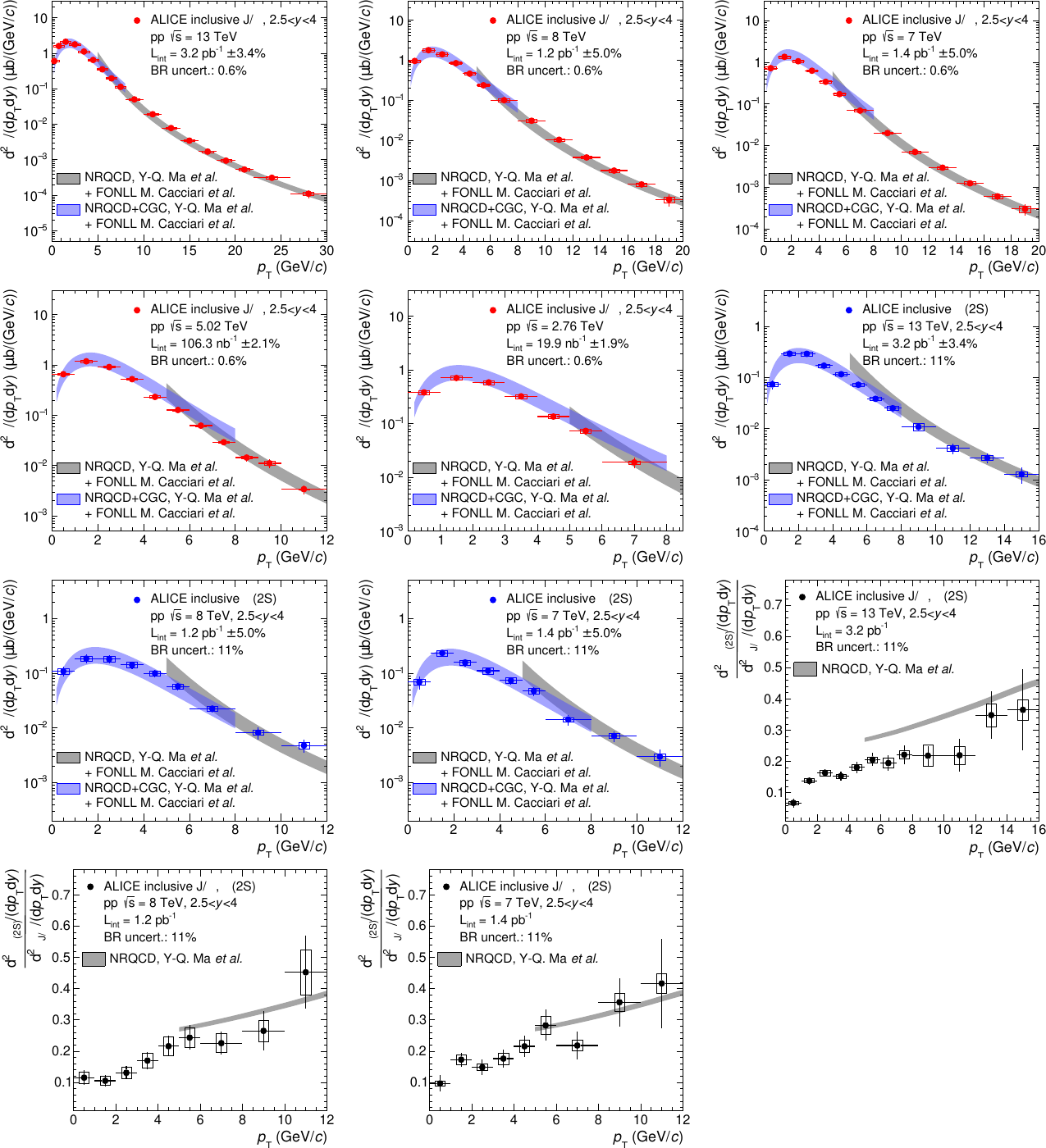}
    \caption{ $J/\psi$ and $\psi(2S)$ at five center of mass energies from the ALICE experiment  compared to summed CGC-NRQCD and FONLL
 calculations~\cite{Acharya:2017hjh}.
}
    \label{fig:Onia-ALICE}
\end{figure}

In the NRQCD$+CGC$ approach both Color Singlet and Color Octet state of the $c\bar{c}$ pairs are considered. The relative contribution of the states is parametrized using a finite set of universal long distance matrix elements (LDME), fitted to a subset of the data (e.g. Tevatron). Non-relativistic QCD is used to factorize the short-distance scale (annihilation), set by the heavy quark mass $M$, from the longer-distance scales (production). The short distance scales are expressed in terms of non-perturbative matrix elements of 4-fermion operators in non-relativistic QCD, with coefficients that can be computed using perturbation theory in the coupling constant $\alpha_{s}$.  The matrix elements are organized into a hierarchy according to their scaling with $v$, the typical velocity of the heavy quark~\cite{Bodwin:1994jh} making the heavy quark's velocity ($v$) and  $\alpha_{s}$ the two key parameters which are employed  to describe quarkonium production. 
Fig~\ref{fig:Onia-ALICE} illustrates a comprehensive comparison of $J/\psi$ and $\psi^{'}$ production at all center of energies that the LHC has collided protons and  at forward rapidity. These results are all well described by a number of independent CGC coupled with NRQCD calculations~\cite{Ma:2010jj, Ma:2014mri, Butenschoen:2010rq}.

\begin{figure}[H]
    \centering
    \includegraphics[scale=0.4]{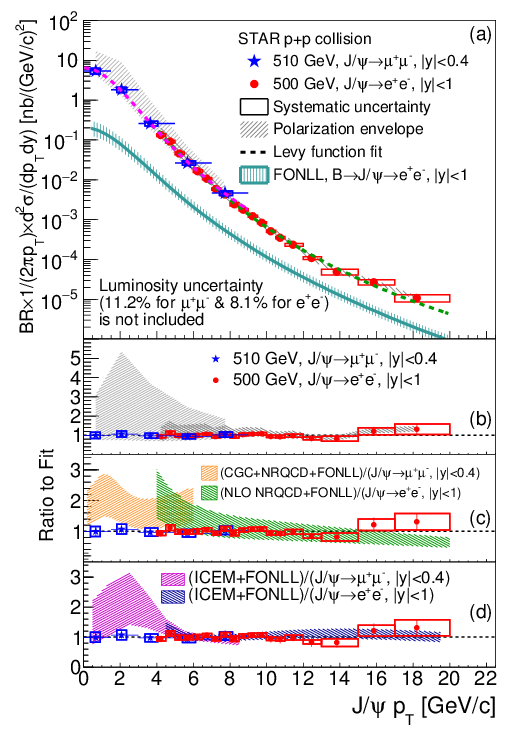}
    \caption{$J/\psi$ Production cross sections as a function of $p_\perp$  in pp collisions at $\sqrt{s}=$ 510 and 500 GeV measured through the $\mu^{+}\mu^{-}$ (blue stars) and $e^{+}e^{-}$ decay channels (red circles). From the STAR experiment~\cite{Adam:2019mrg}}
    \label{fig:StarHQ}
\end{figure}

\begin{figure}[H]
    \centering
    \includegraphics[scale=0.5]{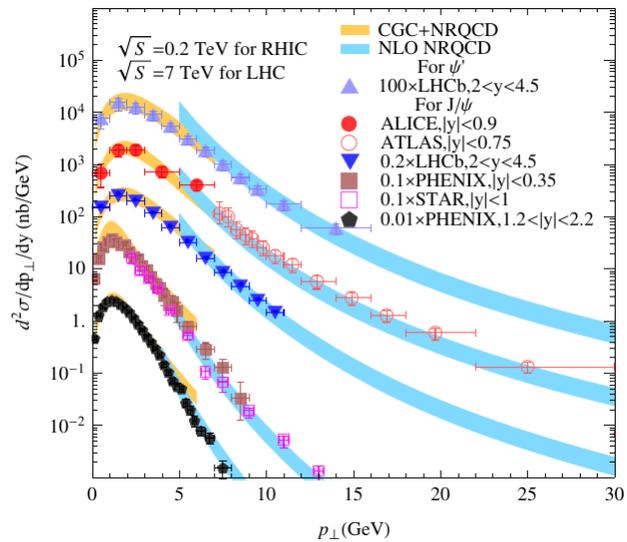}
    \caption{The $\psi^{'}$ (top curve) and $J/\psi$(other four curves) differential cross section as a function of p$_T$. Figure from~\cite{Ma:2014mri}.
}
    \label{fig:Onia}
\end{figure}

Figs.~\ref{fig:StarHQ}, ~\ref{fig:Onia} complete the global picture with experimental data at  the lower center of mass energies available at RHIC as well as more central rapidities both at the LHC and RHIC.  It is noted that CGC$+$NRQCD and CEM and their respective charmonium factorization approaches have been subject of many comparisons over the last two decades, one such comparison is documented in reference~\cite{Bodwin:2005hm}.

\subsubsection{Competing mechanisms in single inclusive production}

We begin by cautioning that the behavior of the nuclear modification factor shown in Fig.\,\ref{fig:nuclear_mod_factor_hadron_DAu} can also be described by other mechanisms:  the data at mid-rapidity can be reproduced well within the leading twist approach using nuclear PDFs \cite{Eskola:2009uj} and within Glauber like multiple scattering \cite{Accardi:2003jh}. At forward rapidities, it has been argued in \cite{Kopeliovich:2005ym} that the disappearance of the Cronin peak follows from energy-momentum conservation \footnote{Note that the partons in the dilute projectile (in this case the deuteron) involved in the forward production carry large momentum fraction $x$ close to the kinematic limit $x\sim 1$.}. Accordingly, the searches of saturation at RHIC using single hadron production need to be  supplemented by similar studies at the LHC, where the kinematic coverage in rapidity and transverse momentum is substantially larger and might allow for the distinction of these mechanisms and gluon saturation. 
For the quarkonia measurements, as it is the case for all results presented here, caution must be exercised  as number of other effects  may be at play and can also account for the experimental observations, this include and not are limited to collective effects (CE) and  Multiple Parton Interactions (MPI). For an interesting review on the subject of small systems and Cold Nuclear Matter effects such as MPI and CE, see ~\cite{2018EPJWC.17111001A}

\subsubsection{Double inclusive production}
\label{sec:double_inclusive}

We now move on to the study of inclusive two particle particle production in proton-nucleus or deuteron-nucleus collisions. For this process, it is very natural to study azimuthal angle correlations, which involve measuring the $\Delta\phi$ of particle pairs in an event. Typically, one observes two peaks in the distribution: a near side peak ($\Delta\phi \sim 0$) dominated by fragmentation of the leading jet, and an away side peak ($\Delta\phi \sim \pi$) produced by $2 \to 2$ back-to-back scatterings. It has been suggested that the emergence of gluon saturation might be studied in modifications to the away side peak by comparing pp to $p(d)$-$p(A)$ collisions as we will see shortly.

RHIC has measured a depletion of the  back-to-back peak in the production of forward dihadrons in $d$-$Au$ collisions when compared to the same distributions from pp collisions. This effect was predicted in the CGC formalism \cite{Marquet:2007vb} as a consequence of multiple scattering on the dense nucleus and the quantum evolution. If the two particles originate from the same parton, the collinear framework dictates that their transverse momenta must be (almost) back-to-back following momentum conservation. On the other hand, in the saturation framework (within the hybrid factorization) the scattered partons acquire a momentum imbalance from the dense nucleus with characteristic momentum scale $Q_s$. Since the saturation scale $Q_s$ grows with nuclear species, one expects a systematic enhancement of the suppression when the collision involves larger nuclei, higher energies or when the particles are produced at more forward rapidities.

 \begin{figure}[H]
     \centering
     \includegraphics[scale=0.35]{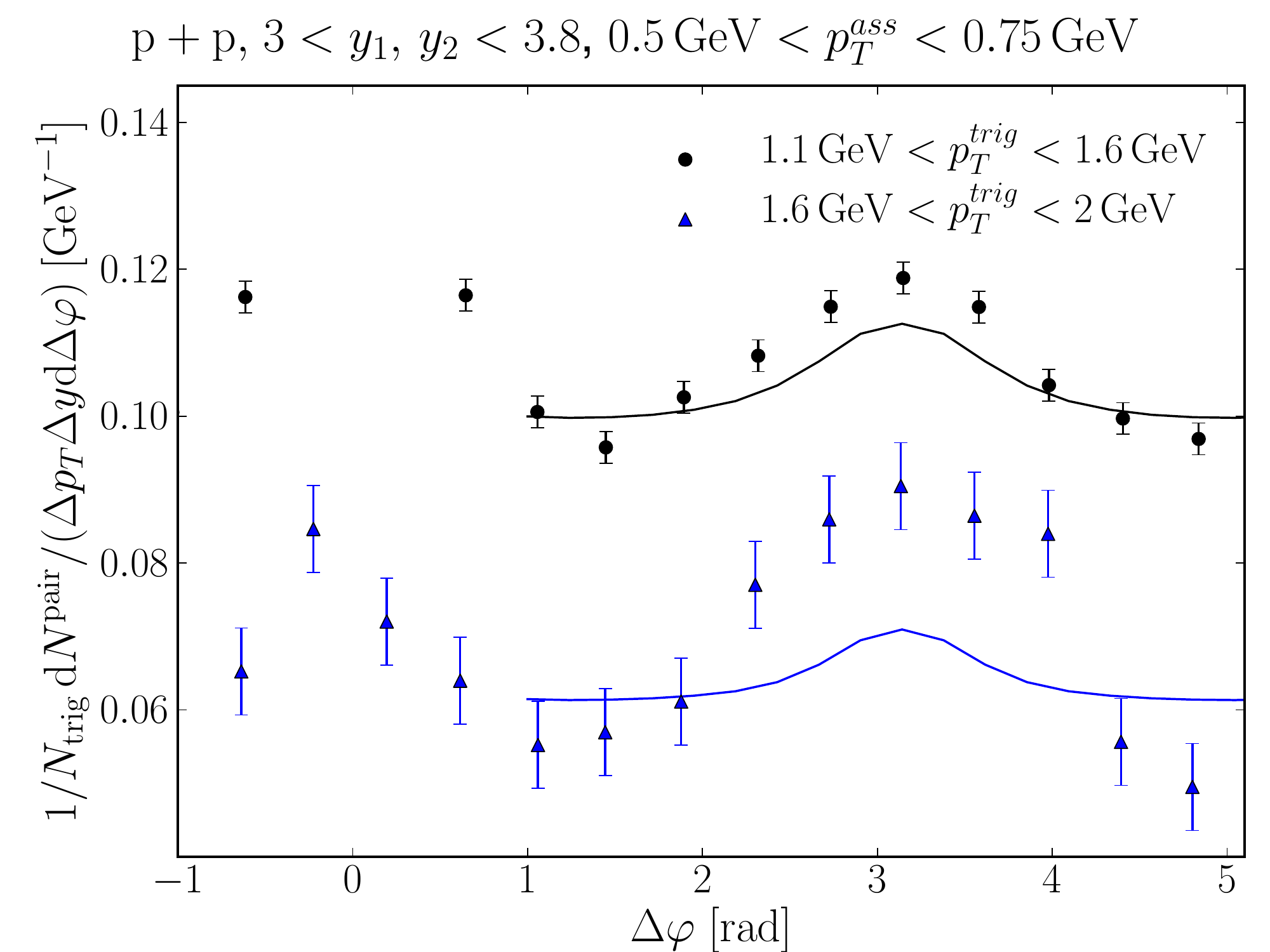}
     \includegraphics[scale=0.35]{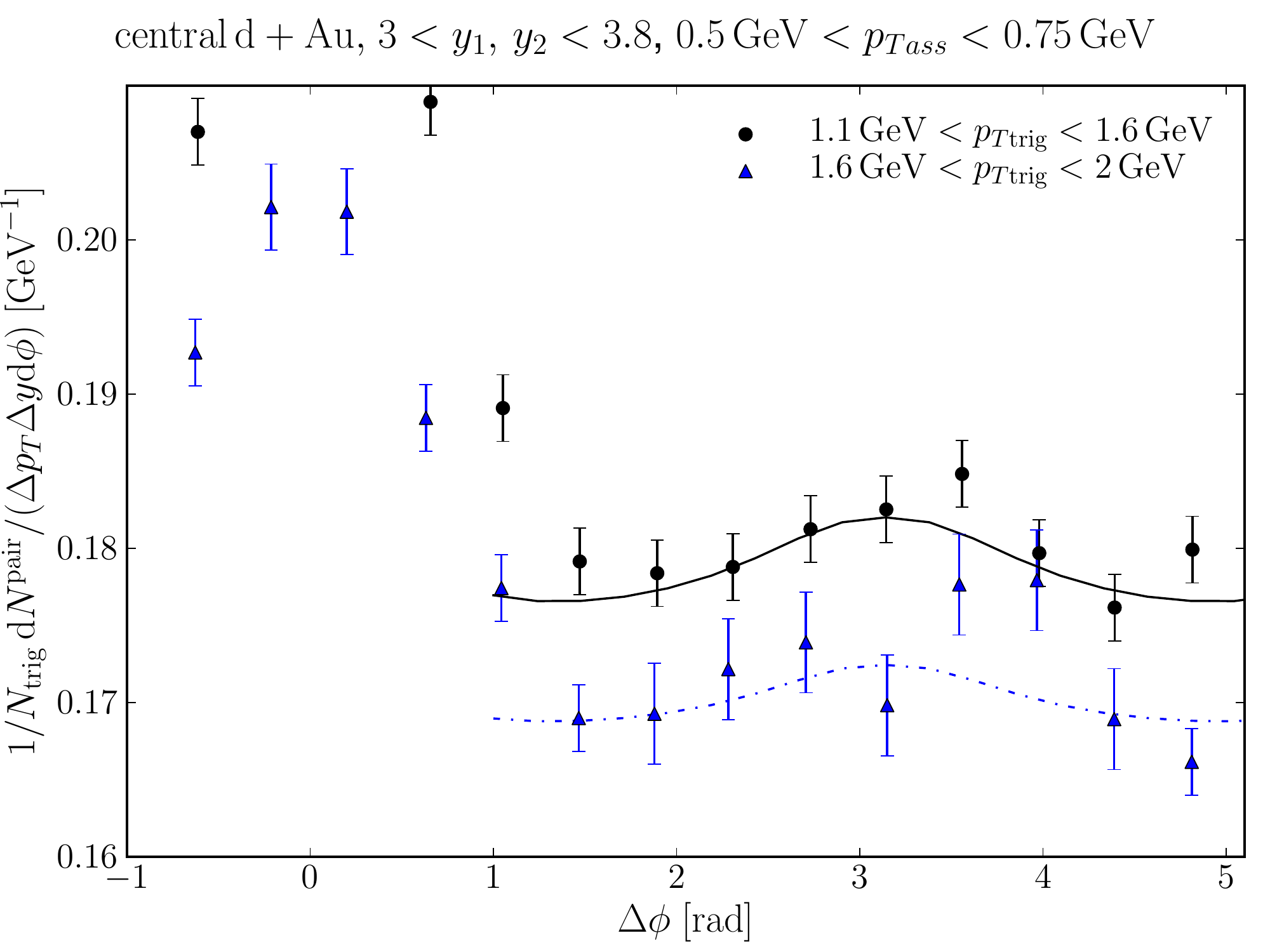}
     \caption{ Azimuthal correlation for $\pi^0$ production compared to PHENIX data \cite{PHENIX:2011puq}. Top: proton-proton collision. Bottom: central deuteron-gold collisions. Figures from \cite{Lappi:2012nh,Lappi:2012xe}. More modern version of this work both experimentally~\cite{Chu:2749297} and theoretically exist with forthcoming publications.}
     \label{fig:dAu_supression}
 \end{figure}
 
The first comparison of dihadron correlations at RHIC to the result of a CGC calculation was made in \cite{Albacete:2010pg}, while the important inelastic contribution (quadrupole) was included in \cite{Lappi:2012nh}. Both studies only considered the quark initiated channel (from deuteron), which is expected to be dominant at RHIC energies\footnote{At the LHC one has to include the gluon initiated channels as well.}. In addition, there is an angle independent contribution (pedestal) arising from double parton scattering \cite{Strikman:2010bg} that must be taken into account. Results show that gluon saturation qualitatively reproduces the systematics of suppression (see Fig.\,\ref{fig:dAu_supression}).  We note that a more modern experimental work from RHIC using pAl and pAu collisions has been recently presented in ~\cite{Chu:2749297} where a forthcoming publication is expected.

An important theoretical advancement was made in \cite{Dominguez:2011wm} where the authors established the connection between the CGC formalism and the TMD framework. These findings significantly simplified the theoretical computations allowing to include other channels (e.g. gluon initiated) and to incorporate higher order contributions. In the back-to-back limit the most important NLO contribution is the Sudakov factor derived in \cite{Mueller:2013wwa} which leads to a suppression of the back-to-back peak. In recent years, theoretical comparisons have been made using the TMD approximation: including both quark and gluon channels and rcBK evolution \cite{Albacete:2018ruq}  and including Sudakov resummation within the GBW model \cite{Stasto:2018rci}.

Similar studies have been carried out for dijet production at the LHC. Despite the fact that Sudakov resummation plays a dominant role (due to the higher $p_\perp$ required for insufficient jet reconstruction compared to hadron measurements), the results show that it is possible to distinguish this effect from gluon saturation. The first studies were performed without Sudakov in \cite{Kutak:2012rf,vanHameren:2014lna} and with Sudakov resummation in \cite{vanHameren:2014ala}. More recently, these results have been supplemented with kinematic power corrections within the so called Improved TMD (ITMD) framework \cite{Kotko:2015ura,vanHameren:2016ftb} extending the agreement with data to large non back-to-back configurations \cite{vanHameren:2019ysa} (see also for UPC studies \cite{Kotko:2017oxg}).

Finally, we point out that a similar depletion of the back-to-back peak was proposed in photon-hadron, photon-pion, and photon-jet correlations
at RHIC and the LHC \cite{Jalilian-Marian:2012wwi,Rezaeian:2012wa,Rezaeian:2016szi,Benic:2017znu,Goncalves:2020tvh}. It would be interesting to update these studies to include Sudakov resummation which is known to impact azimuthal correlations near the back-to-back peak \cite{Mueller:2013wwa}.

\subsubsection{Competing mechanisms in double inclusive production}
As discussed above a common challenge to uncover gluon saturation in these observables is to assess the impact of Sudakov resummation. Sudakov double logarithms can appear in these processes as a result of the incomplete cancellation between real and virtual contributions~\cite{Mueller:2012uf, Mueller:2013wwa} and are enhanced when the transverse momentum of the produced particles/jets is large. As a consequence, the searches for saturation are restricted to low to moderate transverse momentum phase space, where the Sudakov does not overwhelm the effects of gluon saturation but where jet reconstruction or effects of hadronization might obscure signals of saturation.

Other key physics mechanisms which have been used to explain the dihadron suppression observed at PHENIX and STAR are energy loss in the medium, final state radiation  and coherent power corrections~\cite{Kang:2011bp, Xing:2012ii}.

\subsection{High multiplicity and small systems}
\begin{figure}[H]
    \centering
    \includegraphics[scale=0.5]{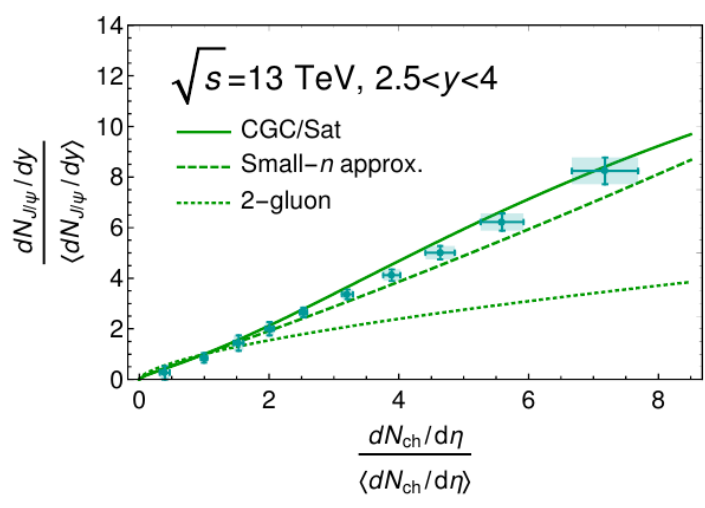}
    \caption{Multiplicity dependence of $J/\psi$ in $\sqrt{13}$ center of mass energies pp collisions at the ALICE experiment. Left figure corresponds to  ALICE data in the forward region~\cite{Khatun:2019slm, Acharya:2020pit}. Theory comparisons from ~\cite{Siddikov:2019xvf}}
    \label{fig:13TevLevin}
\end{figure}

We conclude with a brief note of recent observables that have been highlighted in \emph{small systems} and high energy collisions (pA, pp). These observables entail the classification of some of the results discussed through this document, into  \emph{high activity} or \emph{high multiplicity} environment classes. High activity has been recently used as a proxy to what is more commonly known and used in heavy ion collisions as  \emph{centrality}. Centrality has been traditionally used for describing and classifying the heavy ion collision system size  according to their impact parameter where the colliding nuclei are viewed as hard spheres with radius $R$.  In pp collisions this description is not as clear since until recently, the primary models used to describe centrality assumed the proton a point-like particle. In pA collisions it has additionally raised biases due to detector geometry and triggering effects potentially present in the experiments~\cite{Armesto:2015kwa, ALICE:2017svf, Connors:2017ptx}. 
Nevertheless a plethora of experimental results are currently available and a comprehensive paper that compared a number of  cold nuclear matter effects to LHC data in small systems was published in~\cite{Albacete:2017qng}. In this report among other subjects,  the multiplicity distributions of charged identified/unidentified hadrons in pp collisions at 7~TeV center of mas energies ~\cite{ALICE:2010mty,CMS:2012xvn} were compared to the IP-Glasma model~\cite{Schenke:2012wb, Schenke:2016lrs}. These comparisons showed an important milestone: the reproduction of overall mass ordering trends in   data over the whole multiplicity range. Recent quarkonia \emph{high multiplicity} measurements from the ALICE experiment ~\cite{Khatun:2019slm, Abelev:2012rz, Thakur:2018dmp, Acharya:2020pit} have been also compared recently to a CGC approach by E. Levin, I. Schmidt and  M. Siddikov~\cite{Siddikov:2019xvf} in which quarkonia data from p-p collisions and at forward rapidity  is successfully described by the CGC framework as shown in Fig.~\ref{fig:13TevLevin}.

\begin{figure}[H]
    \centering
    \includegraphics[scale=0.5]{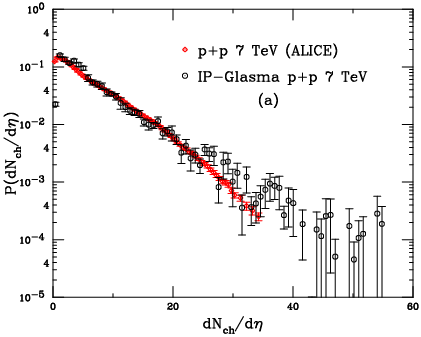}
    \includegraphics[scale=0.5]{IP-GlasmaCMS.pdf}

    \caption{IP-Glasma predictions for 7~TeV center of mass energies pp collisions for the multiplicity dependence of (top) unidentified charged hadrons, (center and bottom) identified hadrons. Figures from~\cite{Albacete:2017qng} and \cite{Schenke:2016lrs}}
    \label{fig:IPGlasma-Albacete}
\end{figure}

\subsubsection{A final note on competing mechanisms }
Single inclusive probes poses a number of challenges for evidencing  gluon saturation. A concern that is raised with the mid-rapidity ~$\sim 0$ region is the competing Glauber like mechanisms ~\cite{Accardi:2003jh} in addition to the energy scale which may be be less sensitive in the equivalent $x$ range. Energy conservation plays a non-negligible role which may be manifested in terms of energy loss mainly via radiation and hadronization effects. In addition as many of these results are produced at low $p_{T}$ where the bulk of particle production is \emph{soft}, in other words the  mean free path of the particles in the medium is very small and the phenomena can be explained with hydrodynamic evolution.

\section{A new generation of high energy  DIS  colliders}\label{Sec:EIC}
In 2020 the Department of Energy (DOE) granted approval for the Electron-Ion Collider (EIC) to be built in the USA~\cite{CD0} marking the beginning of a  new chapter for high-energy nuclear physics (HENP). The EIC is a key international facility that will collide electrons and protons/ions (e-p/A) at high energy with unprecedented luminosity. Amongst its rich physics program, the potential discovery of gluon saturation is one of the key missions of the EIC \cite{Boer:2011fh,Accardi:2012qut,Aschenauer:2017jsk,AbdulKhalek:2021gbh}. The prospect for the discovery for gluon saturation is facilitated thanks to its ability to collide electrons with large ions, where the saturation scale is enhanced $Q_s^2 \sim A^{1/3}$ as compared to the electron-proton collisions at HERA. Another future project of interest to our field is the Large  Hadron  electron  Collider  (LHeC). The LHeC is an ongoing accelerator study which would upgrade the existing LHC storage ring colliding an intense electron beam with a proton or ion beam from the High Luminosity–Large Hadron Collider (HL-LHC)~\cite{Agostini:2020fmq, Goncalves:2020ywm}. Among its rich and diverse physics capabilities, it would also present an opportunity to study gluon saturation in ep collisions while probing Bjorken-$x$ values as low as 10$^{-6}$.

The experimental discovery of gluon saturation will require comprehensive analyses  at future colliders. By the same token quantifying its characteristics will need an in-depth energy, $Q^2$ and mass number dependence scan on a number of quantities. Critical measurements that can help us in the discovery of a gluon saturated state can be classified once again into three main groups: structure functions, exclusive reactions, and semi-inclusive reactions. Below we will discuss the expected manifestations of gluon saturation for each of these processes while keeping an open tab to other mechanisms that could shroud its attributes.


\subsection{Structure functions}

In Sec.\,\ref{subsec:structure functions} we discussed the searches of gluon saturation in proton structure functions at HERA. Major obstacles for its clean extraction are: (i) the contribution from non-perturbatively large dipoles at low to moderate virtualities $Q^2$, and (ii) the non-perturbatively small size of the momentum saturation scale $Q_s^2$ accessed in measurements at HERA.

\begin{figure}[H]
    \centering
    \includegraphics[scale=0.5]{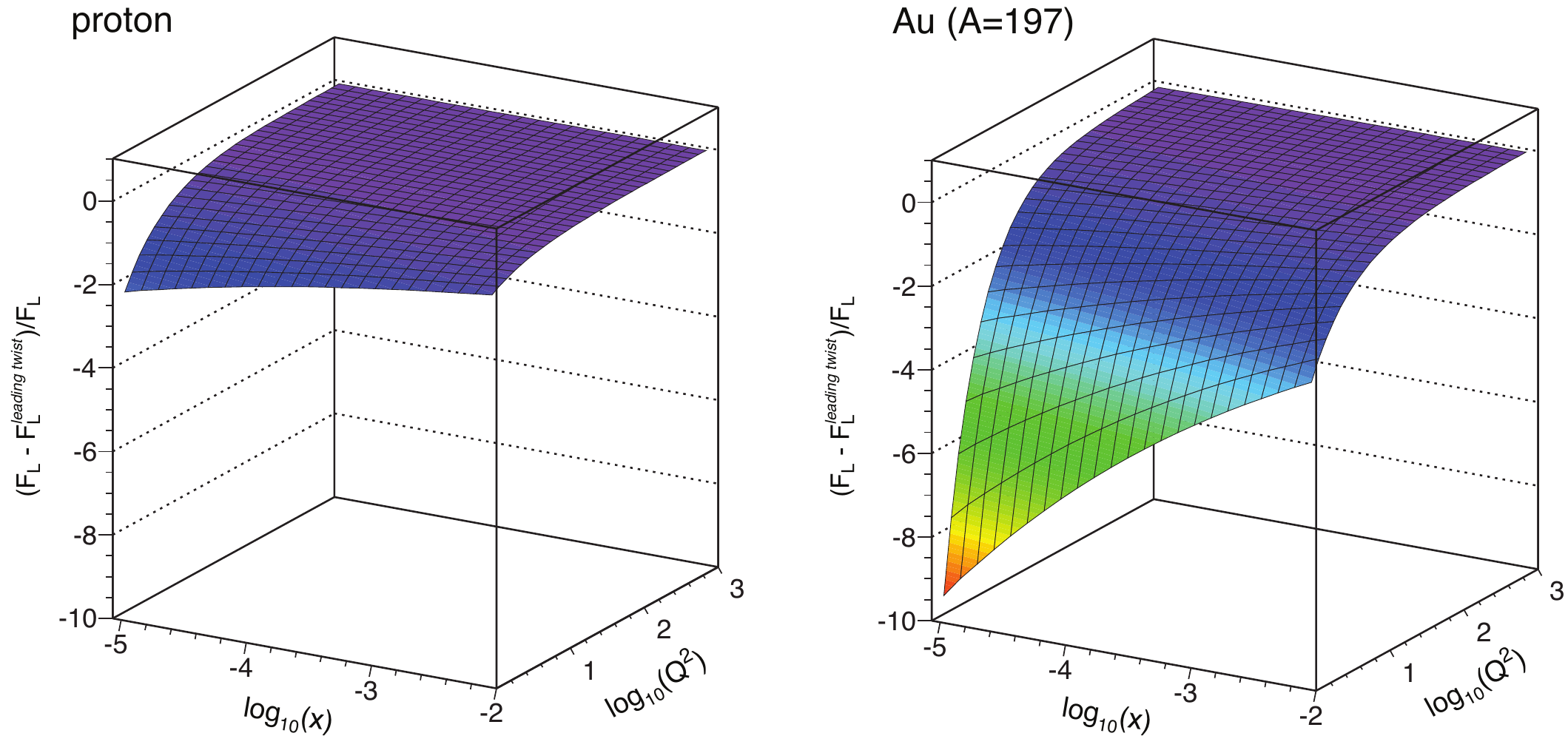}
    \caption{Relative difference between longitudinal structure function $F_{L}$ obtained from the saturation framework to that of the leading twist formalism for proton (left) and Gold (right) \cite{Bartels:2009tu}. Figure from \cite{Accardi:2012qut}.}
    \label{fig:FL_sat_vs_LT}
\end{figure}
While the expected top center of mass energy of the future EIC is lower than that at HERA, the possibility to collide electrons with large nuclei results in accessing larger values of the saturation scale as compared to electron-proton collisions. A comparison of the saturation scales accessible at HERA and the EIC was done in~\cite{Aschenauer:2017jsk}. 
The enhancement in the saturation scale is known as nuclear \emph{oomph} factor, and it is a result of the coherent interaction of the probe with partons along its path of propagation. Larger saturation scales at the EIC are expected to manifest as more pronounced differences between the saturation framework and the leading twist formalism when computing the structure functions. In  Fig.\,\ref{fig:FL_sat_vs_LT} we see the effects of gluon saturation, embodied in higher twist corrections, on the structure function $F_L$. As expected, the largest difference is observed in the small $(x,Q^2)$ corner.

\begin{figure}[H]
    \centering
    \includegraphics[scale=0.32]{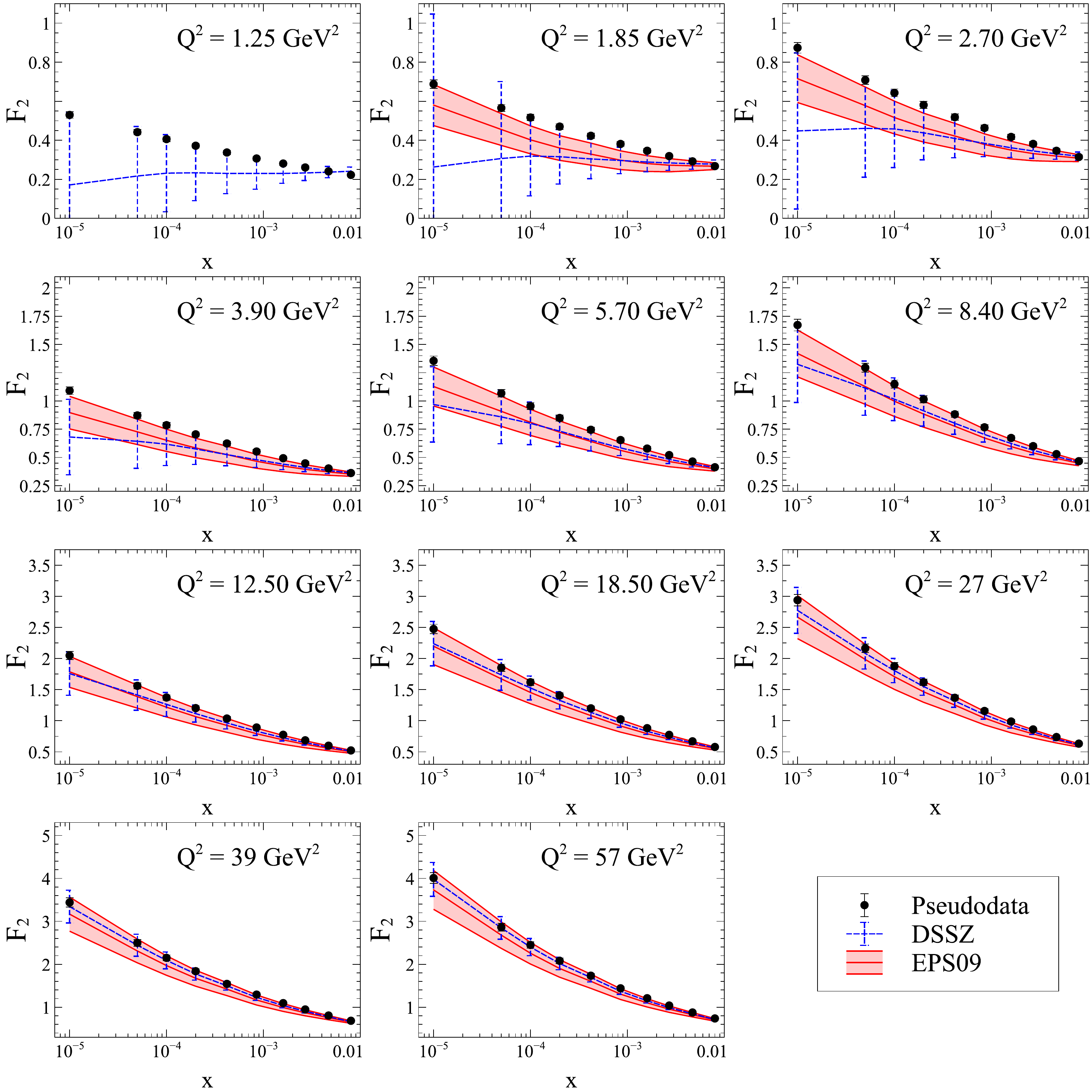}
    \caption{Comparison of the structure function $F_2$ obtained from the rcBK solutions to those from extrapolated nuclear PDFs. Figure from \cite{Marquet:2017bga}.}
    \label{fig:sat_vs_nuclearPDF_EIC}
\end{figure}
To more explicitly analyze the impacts of gluon saturation on the structure functions at the EIC, the authors of \cite{Marquet:2017bga} generated pseudodata for electron-gold collisions, using the running-coupling
Balitsky-Kovchegov evolution equation, and found tension between the compatibility of
these saturated pseudodata with extrapolations of the existing nuclear PDFs (see Fig.\,\ref{fig:sat_vs_nuclearPDF_EIC}). While this tension might result in a signature of gluon saturation, it remains possible that refittings of nPDFs could accommodate for these differences as nPDFs are not well constrained in the low $x$ regime. We have confidence that the high statistics of the EIC combined with forthcoming precise theoretical computations will distinguish both scenarios.

\begin{figure}[H]
    \centering
    \includegraphics[scale=0.38]{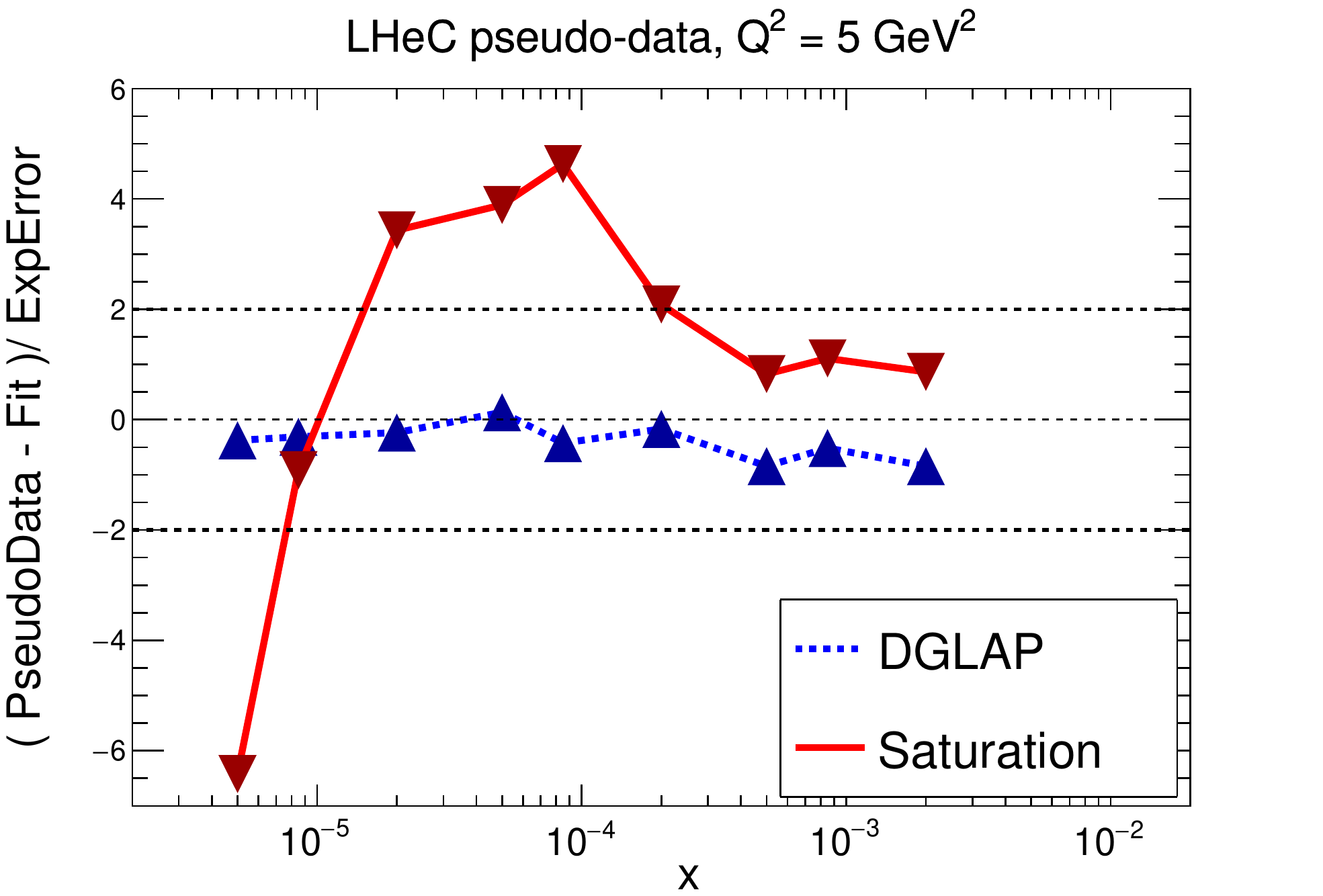}
    \includegraphics[scale=0.38]{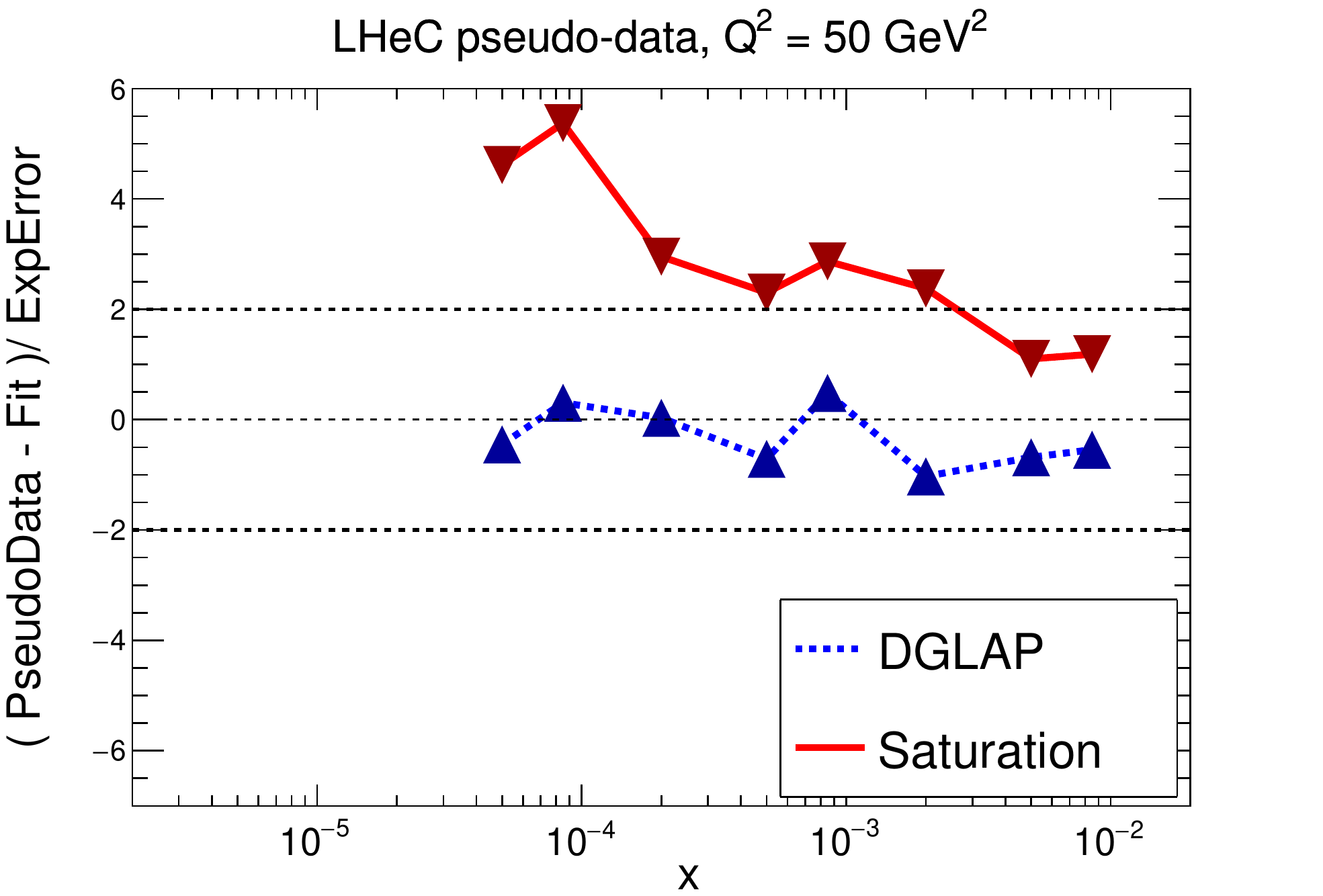}
    \caption{Pull (defined in Eq.\,\eqref{eq:pull_pseudodata_dglap}) between pseudodata for reduced cross-section and the fit based on DGLAP. The pseudodata has been generated either by DGLAP or by the GBW saturation model. Figure from \cite{Agostini:2020fmq}.}
    \label{fig:F2_LHeC_DGLAP_GBW}
\end{figure}

Complementary to the EIC, the LHeC will reach very low values of $x$, providing potential to discriminate linear DGLAP evolution from the non-linear QCD evolution via BK/JIMWLK equations. In a recent preliminary study that can be found in \cite{Agostini:2020fmq}, the authors generate pseudo data for the reduced cross-section using two models: (i) a DGLAP based model, and (ii) a saturation (GBW) based model. Subsequently, they fit each pseudodata set using a DGLAP calculation. As expected, the fit is excellent for the DGLAP generated pseudodata, while significant tension is observed when the DGLAP fit is applied to the pseudodata based on the saturation model. This is quantified by studying the pull:
\begin{align}
    P(x,Q^2) = \frac{\mathcal{F}_{\rm dat}(x,Q^2)-\mathcal{F}_{\rm fit}(x,Q^2)}{\delta_{\rm exp} \mathcal{F}(x,Q^2)} \,,
    \label{eq:pull_pseudodata_dglap}
\end{align}
where $\mathcal{F}_{\rm fit}$ is the central value of the result of the fit for the observable $\mathcal{F}$, $\mathcal{F}_{\rm dat}$ corresponds to the pseudodata generated by model (i) or (ii), and $\delta_{\rm exp} \mathcal{F}$ represented the experimental uncertainty. Their results are shown in Fig.\,\ref{fig:F2_LHeC_DGLAP_GBW}, where the magnitude of the pull is close to $0$ for the DGLAP pseudodata, and significantly larger than $0$ for the GBW (saturation) pseudodata. This tension suggests that if gluon saturation is present at LHeC, a DGLAP fit will not be able to  conceal it.

\subsection{Diffractive measurements}

Building upon the observables discussed in Sec.\,\ref{sec:diffractive_reactions}, we briefly discuss the potential for diffractive measurements at future colliders.

\begin{figure}[H]
    \centering
    \includegraphics[scale=0.4]{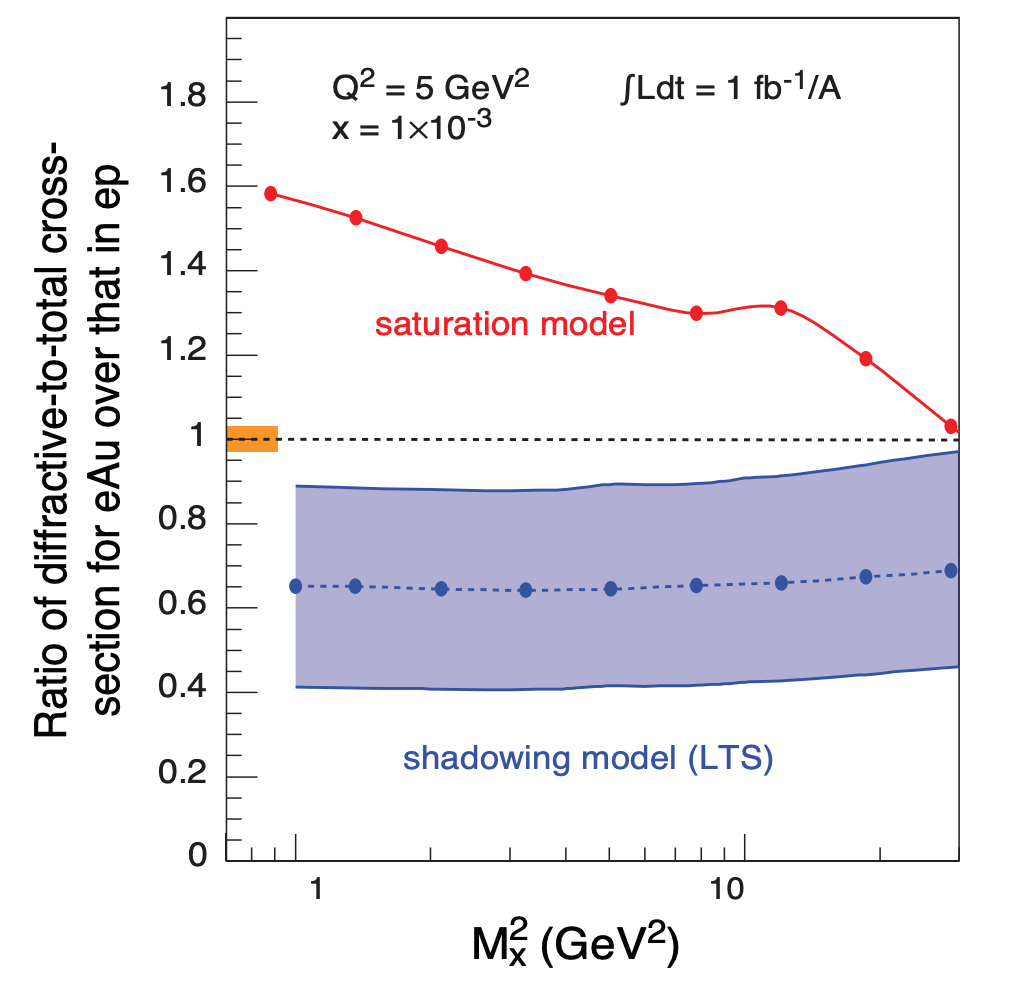}
    \caption{  Ratio of diffractive to inclusive DIS cross-section in eA normalized to pA (double ratio). Comparison between saturation predictions and the leading twist approach. Figure from \cite{Accardi:2012qut}.  }
    \label{fig:diffractive-to-inclusive}
\end{figure}

The ability to collide electrons with different nuclei opens up the possibility to study the nuclear modification factor for the production of diffractive events. Computations within the saturation framework result in the enhancement of nuclear diffractive structure functions when the invariant mass of the final state $M_{X}$ is small, which is dominated by the dipole Fock state. On the other hand, at large invariant masses the dominant state is that of a tripole ($q\bar{q}$+ gluon) which is absorbed more strongly in a denser target, resulting in a suppression of the nuclear structure function \cite{Kowalski:2008sa}. Another prediction of saturation models is that the number of diffractive events relative to all events is larger for nuclei than for protons, which can be quantified by a double ratio \cite{Kowalski:2008sa}   as shown in Fig.\,\ref{fig:diffractive-to-inclusive}. The leading twist formalism on the other hand predicts a slight suppression of diffractive events.

\begin{figure}[H]
    \centering
    \includegraphics[scale=0.3]{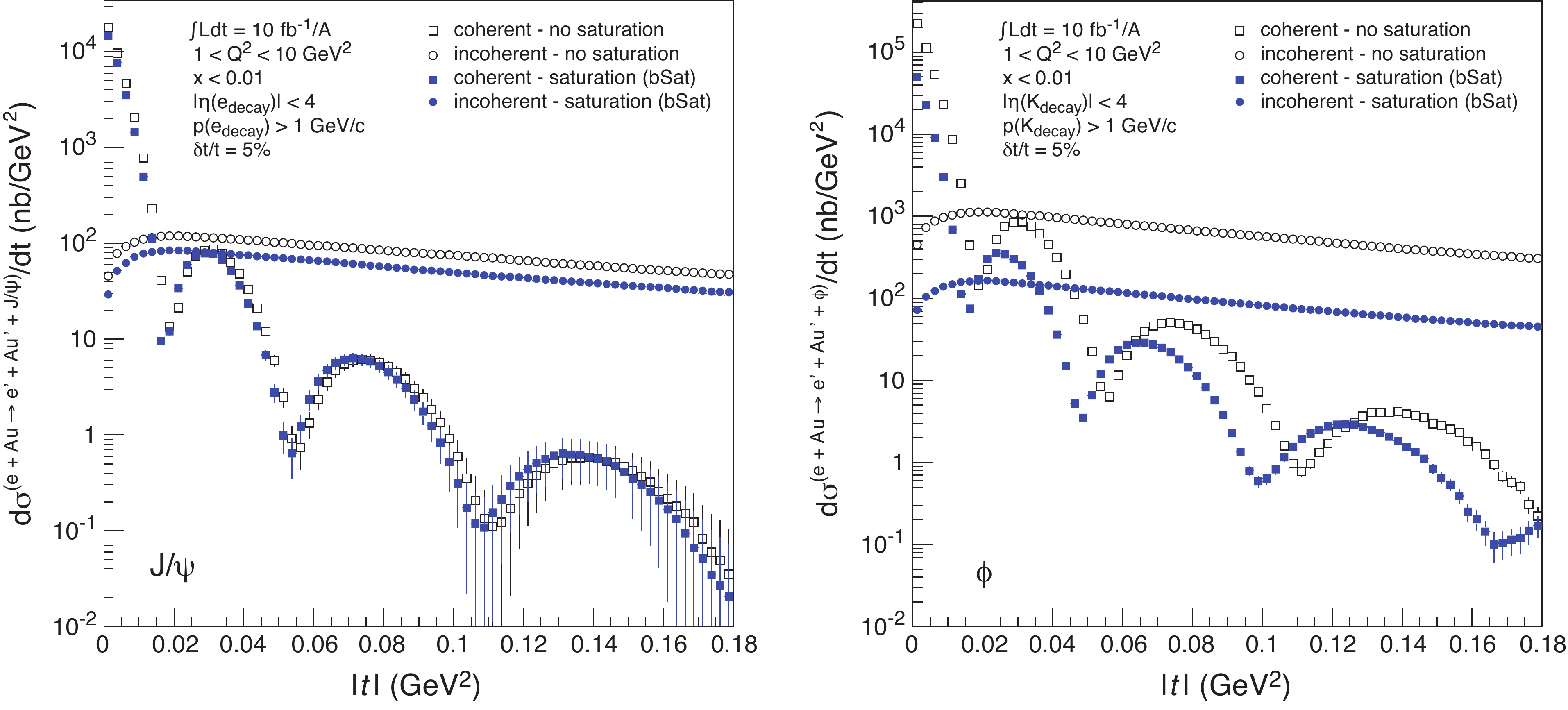}
    \caption{Transverse momentum spectra for the diffractive coherent production of vector mesons in electron-gold ion collisions. Left: $J/\psi$ production. Right: $\rho$ production. The figures show the comparison between models with and without saturation. Figure from \,\cite{Accardi:2012qut}. }
    \label{fig:VM_production_eA}
\end{figure}

Another possibility is the study of coherent vector meson electroproduction off nuclei. As in proton DIS, the spectrum of the exclusively produced particle in nuclear DIS provides a tomographic picture of the color charge density profile of the nuclear target. Predictions from the saturation framework are shown in Fig.\,\ref{fig:VM_production_eA} based on the calculations in \cite{Toll:2012mb,Toll:2013gda}. This figure shows that saturation results in spectra which deviate from the form factor (Fourier transform of the nuclear density profile). These deviations grow with energy and when the produced vector mesons are less massive\footnote{More massive vector mesons probe shorter distances where saturation effects are suppressed.}. To quantify saturation effects it is also  necessary to compare these results to predictions obtained from competing mechanisms such as the leading twist nuclear shadowing framework, where deviations from the form factor are also expected due to multiple scattering \cite{Frankfurt:2011cs}. Furthermore, theoretical control over the uncertainties for the light-cone wave-functions is necessary to distinguish saturation from non-saturation models. On the experimental side, enough statistics are necessary to resolve the peaks and dips of the spectra; particularly in the region in which the cross-section might be overwhelmed by incoherent (break-up of target) events or general beam-induced backgrounds as it will be discussed in Sec.~\ref{Sec:conclusions}.

Incoherent production is also interesting as one expects sensitivity to subnucleonic fluctuations of various kinds. At the EIC, in addition to incoherent events in heavy nucleus DIS, we will also be able to study DIS off light-nuclei and study the interplay of short range correlations and gluon saturation \cite{Miller:2015tjf,Mantysaari:2019jhh,Tu:2020ymk}. 

Finally we note that performing these measurements require detection  on the forward or backward region in a center of mass instrumentation design. Due to this a number of non-trivial experimental challenges are present that need to be considered for a statistically significant  and minimally biased measurement. These challenges include: control  of machine related backgrounds such as  synchrotron radiation and beam-gas interactions, magnetic field strengths,  Interaction Region (IR) designs, up-to-date adequate particle detection technologies including  designs which includes beam-line detectors capable to discern final state hadrons essential for identifying exclusive DIS events. 

\subsection{Semi-inclusive measurements}

In this section we revisit some of the observables discussed in Sec.\,\ref{sec:semi_inclusive_reactions} in the context of  Semi Inclusive Deep Inelastic (SIDIS) measurements. SIDIS at colliders offer several advantages over proton-proton and proton-nucleus collisions: 

(i) The kinematics of the electromagnetic probe (the virtual photon exchanged between the electron and the ion) can be fully reconstructed by measuring the scattered electron. In contrast to pp/pA collisions where the probes are quarks or gluons whose kinematics cannot be retrieved, but require convolutions with parton distribution functions. 

(ii) The number of mechanisms is less in electron-nucleus collisions as compared to proton-nucleus collisions, since in the former the probe is a virtual photon where in the latter one can have both quarks and gluons. 

(iii) The virtuality $Q^2$ of the exchanged photon can be used as a knob to scan between the non-linear saturated and linear QCD regimes.

We begin by discussing forward dihadron azimuthal correlations in ep, eA collisions. Motivated by the studies in pp and dAu collisions at RHIC and the LHC (c.f. Sec.\ref{sec:double_inclusive}), this process has received considerable attention in recent years and it is considered a promising channel for  gluon saturation searches at the EIC (see also \cite{Kolbe:2020tlq} for photon-hadron azimuthal correlations).

The away side peak in the azimuthal angle distribution of dihadron production is expected to be suppressed in nuclear DIS compared to proton DIS due to the momentum imbalance imparted by the saturated gluon inside the nucleus. At small-$x$ and in the TMD approximation this process involves only the Weizsäcker-Williams (WW) gluon distribution \cite{Dominguez:2011wm} given that the dominant partonic channel is virtual photon-gluon fusion\footnote{In electron-nucleus collisions there are no initial state interactions in the gauge links, in the language of TMDs, since the exchange photon is colorless. This is in contrast to proton-nucleus collisions, where the collinear quark or gluon to the proton carry color and thus initial interactions in the gauge links are present \cite{Mulders:2000sh,Dominguez:2011wm,Petreska:2018cbf}}. An advantage over proton-nucleus collisions is the absence of the pedestal arising from double parton scattering. 

\begin{figure}[H]
    \centering
    \includegraphics[scale=0.35]{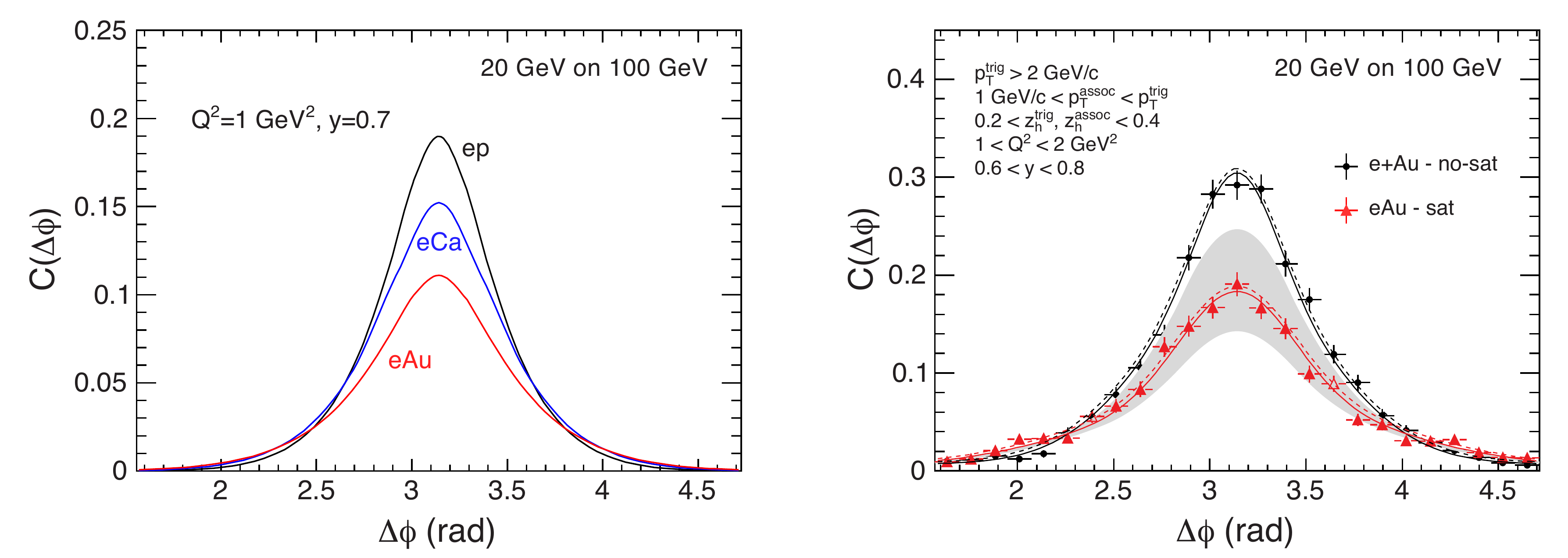}
    \caption{Left:  Dihadron correlation function in electron collisions with different nuclei showing a depletion of the back-to-back peak. Right: Comparison of the correlation function with and without saturation. The gray band is a result of varying the saturation scale. Figure from\,\cite{Accardi:2012qut}. }
    \label{fig:dihadron_suppresion_Adep}
\end{figure}

The first feasibility study for this process at the EIC has been carried out in \cite{Zheng:2014vka} employing a GBW model to compute the WW gluon TMD and including the Sudakov factor \cite{Mueller:2012uf,Mueller:2013wwa}. Their results for the correlation function (dihadron production normalized by single hadron production) are shown in the right panel of Fig.\,\ref{fig:dihadron_suppresion_Adep} showing a clear depletion of the back-to-back peak, while the left panel  shows the nuclear dependence of the suppression. 

To better interpret future results it would be necessary to update the predictions in Fig.~\ref{fig:dihadron_suppresion_Adep}~
to include a more realistic WW gluon distribution, e.g. obtained from the solution to the rcBK equation. Initial steps in this direction have been recently taken in~\cite{vanHameren:2021sqc} where the authors employ a model capturing rcBK evolution and also included kinematic power corrections~\cite{Altinoluk:2019fui,Altinoluk:2021ygv,Boussarie:2021lkb}.  It is also necessary to further investigate competing mechanisms which may deplete the away side peak due the momentum broadening; these can include cold nuclear matter energy loss and coherent power corrections as proposed in~\cite{Xing:2012ii}.

As for other observables that can be measured in DIS, recently, new signatures of gluon saturation have been proposed by studying single inclusive particle production~\cite{Marquet:2009ca,Iancu:2020jch}. These measurements are analogous to those at RHIC and the LHC, where the nuclear modification factor R$_{eA}$ develops a Cronin-like peak at mid rapidity which is then suppressed by saturation. Preliminary studies in~\cite{Iancu:2020jch} show very characteristic features of the transverse momentum distribution when studied at different rapidities and different virtualities and in DIS. The authors propose that this observable can be studied at perturbative virtualities $Q^2 \gtrsim 1 \ \mathrm{GeV}^2$ and that sensitivity to the saturated regime is enhanced for hadrons that carry a large longitudinal momentum fraction $z$.

\begin{figure}[H]
    \centering
    \includegraphics[scale=0.7]{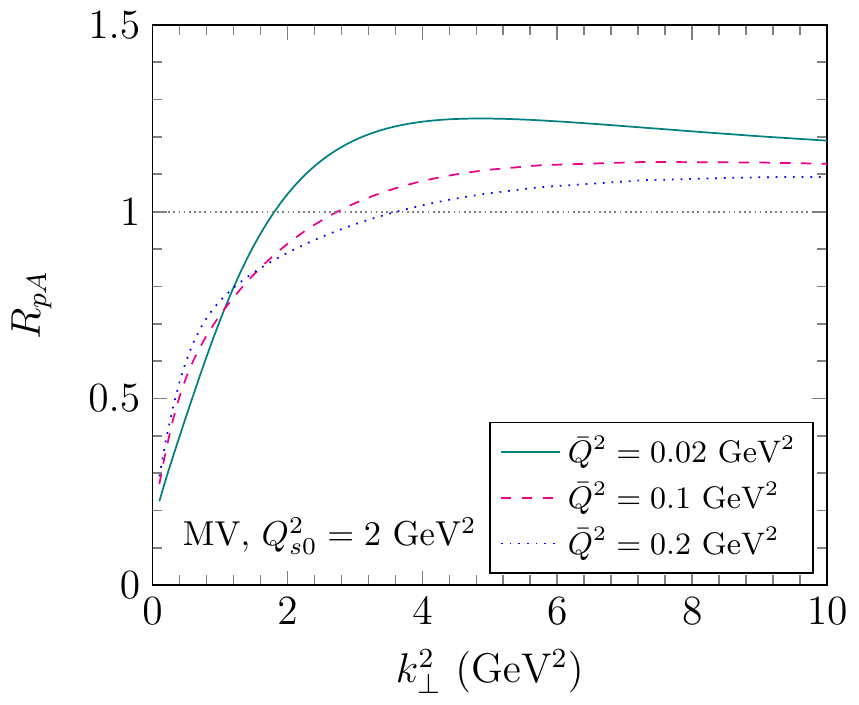}
    \includegraphics[scale=0.7]{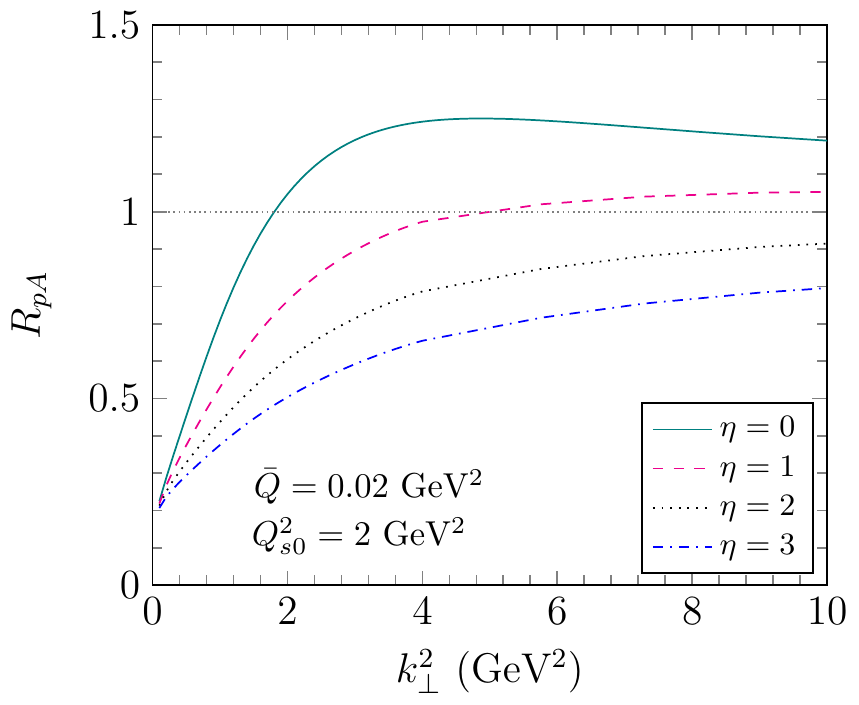}
    \caption{Nuclear modification factor for single semi-inclusive hadron production displaying a Cronin peak and its disappearance. Left: dependence on the virtuality. Right: dependence on rapidity of the produced particle. In this figures $\bar{Q}^2 = z (1-z) Q^2$, where $z$ is the longitudinal momentum fraction of the hadron relative to the virtual photon, and is chosen to be close to unity. Figure from\,\cite{Iancu:2020jch}. }
    
\end{figure}

\section{Discussion and concluding remarks}\label{Sec:conclusions}
\subsection{Theoretical advances}

In this document we have presented various observables that may pave the road for the discovery of gluon saturation at existing and future collider experiments. However, there are still significant sources of theoretical uncertainties that could complicate a systematic extraction from current and ensuing measurements if left unchecked. The CGC framework has entered a new era of theoretical developments which aim to push the precision of the saturation framework to the standards of collinear pQCD. In this section we briefly review recent advances in the field which align to a discovery direction.

Most of the observables presented  have been calculated in the CGC at leading order (LO) in the impact factor and with leading logarithmic (LL) small-$x$ evolution equations with running coupling corrections for the BK \cite{Kovchegov:2006vj,Balitsky:2006wa,Albacete:2007yr} and for the JIMWLK \cite{Lappi:2012vw}. Active efforts by the CGC theoretical community are being conducted to promote these observables to higher loop order accuracy. This requires the determination of the next-to-leading order (NLO) impact factors, and the numerical implementation of the next-to-leading logarithmic (NLL) small-$x$ evolution equations.  The NLL small-$x$ evolution equations have been derived in \cite{Balitsky:2008zza} for BK and in \cite{Balitsky:2013fea,Kovner:2013ona,Kovner:2014lca,Lublinsky:2016meo} for JIMWLK. Only the former has been implemented numerically in \cite{Lappi:2015fma}, where it was found that evolution is unstable for initial conditions of phenomenological interest. This issue of instability was resolved in \cite{Beuf:2014uia,Iancu:2015vea,Ducloue:2019ezk} by resummation of (anti-)collinear logarithms, with a numerical implementation realized in \cite{Lappi:2016fmu}. While a numerical implementation of the full NLL JIMWLK equation is not yet available, numerical codes exist \cite{Korcyl:2020orf,Cali:2021tsh} based on the collinearly improved JIMWLK equation proposed in \cite{Hatta:2016ujq}. 

To achieve precise computations of physical processes, one also needs high order computations of the corresponding impact factors. Significant progress has been made in this direction for a variety of processes at NLO in the CGC, which include: DIS structure functions \cite{Balitsky:2012bs,Beuf:2011xd,Beuf:2016wdz,Beuf:2017bpd,Lappi:2016oup,Hanninen:2017ddy,Ducloue:2017ftk,Beuf:2020dxl,Beuf:2021qqa}, exclusive dijet \cite{Boussarie:2016bkq,Boussarie:2016ogo,Boussarie:2019ero} and exclusive vector meson \cite{Mantysaari:2021ryb} in eA, inclusive di-jet+photon in in eA \cite{Roy:2019hwr}, single inclusive particle production in pA \cite{Chirilli:2011km,Chirilli:2012jd,Altinoluk:2014eka,Ducloue:2016shw}, partial results for inclusive dijet production in pA \cite{Iancu:2020mos},  and most recently inclusive dijet production in eA \cite{Caucal:2021ent}. Numerical results with NLO impact factor and NLL small-$x$ resummation have been obtained only for the structure functions \cite{Beuf:2020dxl}, and single inclusive particle production in pA \cite{Stasto:2013cha,Ducloue:2017mpb,Ducloue:2017dit,Liu:2020mpy} which compared well to data. We anticipate that in the next few years, more of these computations will be coupled to numerical routines and will provide precise quantitative results for the size of the NLO impact factors.

Another significant source of theoretical uncertainty lies in existing models used for the initial conditions of the small-$x$ evolution equations. The MV model \cite{McLerran:1993ni,McLerran:1993ka} is the most widely used framework to compute the initial conditions for the existing computations in the literature. This is in view of their simple numerical implementation relying on the Gaussian statistics of its color charge correlators \cite{Dumitru:2011vk,Iancu:2011nj}. Despite the fact that the MV model was conceived for the description of very large nuclei, it has been employed as the initial condition for protons and it has enjoyed great success in the phenomenological studies of HERA data (see also the variants MV$\gamma$ \cite{Albacete:2010sy} and MV$e$ \cite{Lappi:2013zma}).  It remains to be seen if it provides a good description of less inclusive observables where non-Gaussian effects might play a larger role \cite{Dumitru:2011ax,Dumitru:2011zz,Giannini:2020xme}. A recent alternative approach for the initial conditions has been taken in \cite{Dumitru:2018vpr,Dumitru:2020fdh,Dumitru:2020gla,Dumitru:2021tvw} where the authors follow a perturbative approach to find the two-point, three-point and four-point function of color charge correlators inside the proton from the light-cone wave-functions of its valence quarks. 

High energy Wilson line correlators are more naturally written in coordinate space, due to the diagonal nature of the scattering matrix in the eikonal approximation. This includes setting up their initial conditions as well as performing their small-$x$ evolution in coordinate space. However, most observables require their Fourier transform or convolutions to momentum space, making the relation between initial conditions and observables less transparent. Fortunately, in some special limits, it is possible to establish a clean factorization of the perturbatively calculable impact factors and the non-perturbative high energy correlators, which will make the connection of high energy correlators and their initial conditions more explicit. The paradigmatic example is the work in~\cite{Dominguez:2011wm} which established the connection between quadrupole and dipole operators to the WW gluon TMD and the dipole gluon TMD in the so called correlation limit. In the context of two particle production, this relation is physically realized in the near back-to-back production. This is an active area of research which has resulted in the development of improved TMD framework \cite{Kotko:2015ura,vanHameren:2016ftb,Petreska:2018cbf} and the CGC/TMD equivalence \cite{Altinoluk:2019fui,Altinoluk:2019wyu,Boussarie:2020vzf}. For comprehensive numerical studies both in pA and eA we refer the reader to \cite{Fujii:2020bkl,Boussarie:2021lkb}.

A further assumption of the saturation/CGC framework is the eikonal approximation, which is strictly valid at asymptotically high energies. Relaxing this assumption is necessary to access the physics of polarized measurements\cite{Kovchegov:2015pbl,Kovchegov:2016weo,Kovchegov:2016zex,Kovchegov:2017jxc,Kovchegov:2017lsr,Kovchegov:2018znm,Kovchegov:2020hgb,Adamiak:2021ppq,Cougoulic:2019aja,Cougoulic:2020tbc}. Efforts are also being carried out to study the effect of sub-eikonal contributions to various unpolarized observables \cite{Altinoluk:2014oxa,Altinoluk:2015gia,Altinoluk:2015xuy,Agostini:2019avp,Agostini:2019hkj,Altinoluk:2020oyd} with emphasis in non-trivial azimuthal correlations. These contributions might be relevant for precise EIC and RHIC phenomenology where the energies are much less compared to LHC or LHeC. 

Powerful techniques such as the glasma graph approximation \cite{Armesto:2006bv,Dumitru:2008wn,Dumitru:2010iy,Kovchegov:2012nd,Kovchegov:2013ewa} suited for collisions where both hadrons/nuclei are treated as dilute objects are being extended to account for asymmetric dilute-dense scenarios \cite{Altinoluk:2015uaa,Altinoluk:2018hcu,Altinoluk:2018ogz,Altinoluk:2020wpf}. On the other hand, it remains challenging to make progress in analytically understanding nuclear collisions where the saturation scales of both colliding objects are large, and one must resort to complex numerical evaluations of the full Yang-Mills evolution \cite{Krasnitz:2003jw,Lappi:2003bi,Blaizot:2010kh,Schenke:2015aqa}. Efforts to quantify the effect of saturation corrections to the dilute projectile on multi-particle production have been made in \cite{Schlichting:2019bvy} by comparing the dilute-dense approximation to the full Yang-Mills simulation. In addition, recent analytical work in \cite{Chirilli:2015tea,Kovchegov:2018jun,Li:2021yiv,Li:2021zmf} accounting for the effect of multiple scattering and saturation in the field of the proton has been carried out. It is imperative to assess the impact of these contributions in multiparticle production and azimuthal correlations at the RHIC and the LHC .

\subsection{Experimental requirements} 
The last twenty years have resulted in many experimental results which may have sensitivity to gluon saturation. Nonetheless, to this date we cannot unequivocally state that we have confirmed a gluon saturated state. 
One pre-requisite for the next generation of pertinent experimental publications is the need to  measure a diverse number of observables. As mentioned throughout this document, the capability to compare results at a number of  center of mass energies (energy, $Q^{2}$ scan)  as well as several atomic masses ($A$) is essential to obtain a clean physics picture at current and future colliders. Discovery of a gluon saturated state dictates better quantification of i)competing mechanisms, ii)regimes of validity and iii)transitions from dilute and dense gluon states to a truly gluon saturated state.   Appropriate detectors equipping finer kinematic areas (small-$x$) are needed to discern a set of  measurements sensitive to gluon saturation with better precision. All of these are vital requirements for the future experimental endeavours described in Section~\ref{Sec:EIC}.  

We highlight that the use of electron beams in the foreseeable future introduces a number of several machine induced backgrounds~\cite{SynRad:2004}  that need to be accounted for to ensure  successful data-taking and data-analyses campaigns. Past and current experiments~\cite{Seidel:2004rs,Belle-IISVD:2020wwc, Bartel:1985yd}  have  demonstrated 
  that beam-related backgrounds can shut down an accelerator, force detectors to reduce  bunch-crossing frequency and overall data rate to counteract beam-background interactions.   Mis-identification and tracking efficiency biases  using standard trigger, Monte Carlo and background rejection techniques can cripple even the cleanest theoretical probe.  The lessons learned from past and present experiments need to be carefully carried over to future experiments.  As of today there are a number of key experimental regions at the EIC and likely the LHeC which are susceptible to high backgrounds~\cite{AbdulKhalek:2021gbh}.  Additionally, many of the physics signatures we have outlined demand i) reconstruction of full DIS events ii) detection of small cross-sections very close to the collider beam pipe or in regions that may be bombarded by parasitic particles iii) scan of kinematic regions (e.g. $\eta,\,y,\,\phi$). Besides assuring we have the most up-to-date technologies and coverage to deal with the collision and data taking rates, it is imperative that we evolve our particle detection techniques to cope with a new generation of diverse measurements. In the last decade, analyses of large data from high-energy nuclear experiments have been rapidly evolving to Artificial Intelligence techniques.  Collider experiments located at the European Organization for Nuclear Research (CERN) have begun a massive effort to introduce and develop AI techniques in all of their physics experiments.  The USA's Department of Energy  (DOE) and other national science organizations (NSF, APS) have made interdisciplinary AI research, including at future colliders, a key effort that will support the nation’s long-term economic and national security ~\cite{DOE:AI}, it is becoming clear that the next standard in high energy nuclear physics computing will be based on AI~\cite{Feickert:2021ajf, Kasieczka:2021xcg, Amoroso:2020lgh}. Some of the techniques which AI could help in the long term to unambiguously measure  gluon saturation can include: 
 
 \begin{enumerate}
   \item Accurate description of  physics and machine induced  backgrounds. This requires an effort of open-sourced, cross-collaboration simulation packages that include theory, phenomenology studies as well as up-to-date machine background knowledge. Two principal machine backgrounds that we can learn from past experiments are synchrotron radiation and beam-gas interactions. Synchrotron radiation occurs when the trajectory of a charged particle is bent, synchrotron photons are emitted tangential to the particle’s path. More concretely, these backgrounds can affect tracking detectors and calorimeters by  depositing energy leading to detector \emph{hits}. Ultimately this can also lead to a large number of ghost tracks and large detector occupancy effects. Beam-gas interactions on the other hand occur when proton or ion beam particles collide with residual gas. Ion beam interactions with gas cause beam particle losses and halo, which can reach  the detectors.  Addition of these backgrounds in future simulations is needed for detector design or AI-based data  training  techniques;  as such these should be included in the next generation of DIS experiments.~\cite{Feickert:2021ajf}
   
    \item Improved jet tagging capabilities which can  disentangle jets that come from quarks, gluons,  gluon-dense vs saturated gluon signatures.
    Jet tagging refers to  the reconstruction of streams of particles coming from the collision or displaced vertices with the flexibility of a loose event selection requirement. The classification of jets depends on the kinematic variables such as transverse momentum  ($p_\perp$), pseudorapidity (rapidity) $\eta$($y$), azimuthal angle $\phi$, number of tracks, energy ($E$). We remind the reader that jets can be contaminated by many soft processes that are not correlated to the jet.  We often rely on classification/regression tasks which give us an approximation of the background. A potential AI application  which should build upon existing experiments and further developed could be to extract and study list of features using kinematic variables from  simulations. The list of features could be used to form jet images or graphs in $\eta-\phi$ plane which will be used as an input of various AI-related algorithms to classify jet events from background events~\cite{Guest:2018yhq}.

  \item Precisely identify particles: open and hidden charm mesons, direct photons, electrons all while minimizing biases. While standard cut and slice techniques have done a excellent job when the detectors are adequate and production cross-sections are large, many rare resonances  or small cross-sections have suffered from these same methods and have yet reached statistical significance.  While machine learning techniques are currently implemented  for identification of rare particles in certain physics cases of nuclear experiments at accelerators, AI is at its infancy and has not replaced or considerably complemented  standard particle identification methods at  high energy nuclear experiments.  Applying Machine Learning algorithms can give advantages in the signal to background ratios as strict cuts and slices on the variables are minimized or eliminated altogether. This however requires a dedicated computing effort to go beyond the standard ML methods used so far. 
\end{enumerate}

In this document we have outlined a number of existing challenges, potential solutions as well as the high gains/rewards in unambiguously identifying a gluon saturated state. These can be summarized into being able to disentangle the features that characterize a truly saturated state from ones that may indicate the presence of competing effects or a gluon density which may be large but not necessarily saturated. This challenge is particularly well-suited for a new generation of tools and techniques at the theory and experimental level. It is also well suited for a new generation of cross-experiment and cross-field collaborations.

\section{Acknowledgments}\label{Sec:Ack}
We would like to thank Bj\"{o}rn Schenke,  Ernst Sichtermann, Raju Venugopalan, for providing elucidating comments to our project. Astrid Morreale and Farid Salazar are respectively supported by i)the U.S. Department of Energy, Office of Science, Office of Nuclear Physics under contract No.~${\rm KB-01-01-02\_2}$OPE. ii) the U.S. Department of Energy, Office of Science, Office of Nuclear Physics under contract No.~DE-SC0012704, and the joint Brookhaven National Laboratory-Stony Brook University Center for Frontiers in Nuclear Science (CFNS).

\clearpage
\vspace{6pt} 





\abbreviations{The following abbreviations are used in this manuscript:\\

\noindent 
\begin{tabular}{@{}ll}
MDPI & Multidisciplinary Digital Publishing Institute\\
DOAJ & Directory of open access journals\\
TLA & Three letter acronym\\
LD & Linear dichroism
\end{tabular}}

\appendixtitles{no} 

\end{paracol}
\reftitle{References}



\end{document}